\documentclass[12pt,openany]{book}
\usepackage{textcomp}
\usepackage{hyperref}
\usepackage{graphicx}
\usepackage{amsmath}
\usepackage[round]{natbib}

\newcommand{\myfigures}[4]{
\begin{figure}[!htb]
  \centering
  \includegraphics[scale=#4]{#1}
  \caption{\bf #2}
  \label{#3}
\end{figure}
}

\newcommand{\myfigure}[3]{
\myfigures{#1}{#2}{#3}{1.0}
}
\newcommand{\degree}{$^\circ$}
\newcommand{\frontsection}[1]{
\newpage
\begin{center}
{\bf \huge #1}
\end{center}
}
\newcommand{\signatureline}[2]{
\vspace{3.5ex}
\hspace{2.5in}\rule{2.5in}{.1pt}\\
\hspace{2.5in}#1\\
\hspace{2.5in}#2\\
}
\begin{document}
\frontmatter

\title{Pebble bed pebble motion: Simulation and Applications}
\author{Joshua J. Cogliati} 
\date{A dissertation\\ 
submitted in partial fulfillment of\\ 
the requirements for the degree of\\ 
Doctor of Philosophy in Applied Science and Engineering\\ 
Idaho State University\\ 
October 2010 } 

\maketitle 

\frontsection{Copyright}

This work is copyright 2010 by Joshua J. Cogliati.  This work is licensed
under the Creative Commons Attribution 3.0 Unported License. To view a
copy of this license, visit
\url{http://creativecommons.org/licenses/by/3.0/} or send a letter to
Creative Commons, 171 Second Street, Suite 300, San Francisco,
California, 94105, USA.  As specified in the license, verbatim copying
and use is freely permitted, and properly attributed use is also allowed.

\frontsection{Committee Approval Page}

\noindent
To the graduate faculty:

The members of the Committee appointed to examine the thesis of Joshua J. Cogliati find it satisfactory and recommend that it be accepted.

\begin{flushleft}
\signatureline{Dr. Abderrafi M. Ougouag,}{Major Advisor, Chair}
\signatureline{Dr. Mary Lou Dunzik-Gougar,}{Co-Chair}
\signatureline{Dr. Michael Lineberry,}{Committee Member}
\signatureline{Dr. Steve C. Chiu,}{Committee Member}
\signatureline{Dr. Steve Shropshire,}{Graduate Faculty Representative}
\end{flushleft}

\newpage 

\ \\

\newpage

\frontsection{Acknowledgments}

Thanks are due to many people who have provided information, comments
and insight. I apologize in advance for anyone that I have left out.
The original direction and technical ideas came from my adviser
Abderrafi Ougouag as well as continued encouragement and discussion
throughout its creation.  Thanks go to Javier Ortensi for the
encouragement and discussion as he figured out how use the earthquake
data and I figured out how to generate it.  At INL the following
people assisted: Rob Bratton and Will Windes with the graphite
literature review, Brian Boer with discussion and German translation,
Hongbin Zhang with encouragement and Chinese translation and Suzette
J. Payne for providing me with the Earthquake motion data.  The
following PBMR (South Africa) employees provided valuable help
locating graphite literature and data: Frederik Reitsma, Pieter Goede
and Alastair Ramlakan.  The Juelich (Germany) people Peter Pohl,
Johannes Fachinger and Werner von Lensa provided valuable assistance
with understanding AVR.  Thanks to Professor Mary Dunzik-Gougar for
introducing me to many of these people, as well as encouragement and
feedback on this PhD and participating as co-chair on the dissertation
committee.  Thanks to the other members of my committee, Dr. Michael
Lineberry, Dr. Steve C. Chiu and Dr. Steve Shropshire, for providing
valuable feedback on the dissertation.  Thanks to Gannon Johnson for
pointing out that length needed to be tallied separately from the
length times force tally for the wear calculation (this allowed the
vibration issue to be found).  Thanks to Professor Jan Leen
Kloosterman of the Delft University of Technology for providing me the
PEBDAN program used for calculating Dancoff factors.  The work was
partially supported by the U.S. Department of Energy, Assistant
Secretary for the office of Nuclear Energy, under DOE Idaho Operations
Office Contract DEAC07-05ID14517.  The financial support is gratefully
acknowledged.

This dissertation contains work that was first published in the
following conferences: Mathematics and Computation, Supercomputing,
Reactor Physics and Nuclear and Biological Applications, Palais des
Papes, Avignon, France, September 12-15, 2005; HTR2006: 3rd
International Topical Meeting on High Temperature Reactor Technology
October 1-4, 2006, Johannesburg, South Africa; Joint International
Topical Meeting on Mathematics \& Computation and Supercomputing in
Nuclear Applications (M\&C + SNA 2007) Monterey, California, April
15-19, 2007; Proceedings of the 4th International Topical Meeting on
High Temperature Reactor Technology, HTR2008, September 28-October 1,
2008, Washington, DC USA and PHYSOR 2010 - Advances in Reactor
Physics to Power the Nuclear Renaissance Pittsburgh, Pennsylvania,
USA, May 9-14, 2010.  

My mother helped me edit papers I wrote in my pre-college years, and
my father taught me that math is useful in the real world.  For this
and all the other help in launching me into the world, I thank my parents.  

Last, but certainly not least, thanks go to my wife, Elizabeth Cogliati
for her encouragement and support.  This support has provided me with
the time needed to work on the completion of this dissertation.  This
goes above and beyond the call of duty since I started a PhD the same
time we started a family.  Thanks to my son for letting my skip my
first Nuclear Engineering exam by being born, as well as asking all
the really important questions.  Thanks to my daughter for working
right beside me on her dissertation on the little red toy laptop.

\tableofcontents
\listoffigures 
\listoftables

\chapter{Abstract}

Pebble bed reactors (PBR) have moving graphite fuel pebbles.  This unique
feature provides advantages, but also means that simulation of the
reactor requires understanding the typical motion and location of the
granular flow of pebbles.

This dissertation presents a method for simulation of motion of the
pebbles in a PBR. A new mechanical motion simulator, PEBBLES,
efficiently simulates the key elements of motion of the pebbles in a
PBR.  This model simulates gravitational force and contact forces
including kinetic and true static friction.  It's used for a variety
of tasks including simulation of the effect of earthquakes on a PBR,
calculation of packing fractions, Dancoff factors, pebble wear and the
pebble force on the walls.  The simulator includes a new differential
static friction model for the varied geometries of PBRs.  A new static
friction benchmark was devised via analytically solving the mechanics
equations to determine the minimum pebble-to-pebble friction and
pebble-to-surface friction for a five pebble pyramid.  This pyramid
check as well as a comparison to the Janssen formula was used to test
the new static friction equations.

Because larger pebble bed simulations involve hundreds of thousands of
pebbles and long periods of time, the PEBBLES code has been
parallelized.  PEBBLES runs on shared memory architectures and
distributed memory architectures.  For the shared memory architecture,
the code uses a new O(n) lock-less parallel collision detection
algorithm to determine which pebbles are likely to be in contact.  The
new collision detection algorithm improves on the traditional
non-parallel O(n log(n)) collision detection algorithm.  These
features combine to form a fast parallel pebble motion simulation.

The PEBBLES code provides new capabilities for understanding and
optimizing PBRs.  The PEBBLES code has provided the pebble motion data
required to calculate the motion of pebbles during a simulated
earthquake.  The PEBBLES code provides the ability to determine the
contact forces and the lengths of motion in contact. This information
combined with the proper wear coefficients can be used to determine
the dust production from mechanical wear.  These new capabilities
enhance the understanding of PBRs, and the capabilities of the code
will allow future improvements in understanding.


\mainmatter

\chapter{Introduction}

\section{Pebble Bed Reactors Introduction}

Pebble bed nuclear reactors are a unique reactor type that have been
proposed and used experimentally. Pebble bed reactors were initially
developed in Germany in the 1960s when the AVR demonstration reactor
was created. The 10 megawatt HTR-10 reactor achieved first criticality
in 2000 in China and future reactors are planned. In South Africa,
Pebble Bed Modular Reactor Pty.~Ltd.~was designing a full scale pebble
bed reactor to produce process heat or electricity.

Pebble bed nuclear reactors use graphite spheres (usually about 6 cm
in diameter) for containing the fuel of the reactor. The graphite
spheres encase smaller spheres of TRistructural-ISOtropic (TRISO)
particle fuel. Unlike most reactors, the fuel is not placed in an
orderly static arrangement.  Instead, the graphite spheres are dropped
into the top of the reactor, travel randomly down through the reactor
core, and are removed from the bottom of the reactor.  The pebbles are
then possibly recirculated depending on the amount of burnup of the
pebble and the reactor's method of operation.

The first pebble bed reactor was conceived in 1950s in the West
Germany using helium gas-cooling and spherical graphite fuel
elements. Construction on the Arbeitsgemeinschaft Versuchsreaktor
(AVR) 15 MWe reactor was started in 1959 at the KFA Research Centre
J\"ulich.  It started operation in 1967 and continued operation for 21
years until 1988.  The reactor operated with an outlet temperature of
950\degree C.  The AVR demonstrated the potential for the pebble bed
reactor concept.  Over the course of its operation, loss-of-coolant
experiments were successfully performed.

The second pebble bed reactor was the Thorium High Temperature Reactor
(THTR).  This reactor was built in West Germany for an electric
utility.  It was a 300 MWe plant that achieved full power in September
1986.  In October 1988, when the reactor was shutdown for maintenance,
35 bolt heads were found in the hot gas ducts leading to the steam
generators.  The determination was made that the plant could be
restarted, but funding difficulties prevented this from occurring and
the reactor was decommissioned
\citep{SummaryGascooledReactorPrograms}.

The third pebble bed reactor to be constructed and the only one that
is currently operable is the 10 MWt High Temperature Reactor (HTR-10).
This reactor is at the Tsinghua University in China.  Construction was
started in 1994 and reached first criticality in December 2000.  This
reactor is helium cooled and has an outlet temperature of 700\degree C
\citep{DesignFeaturesOfHTR10,OverviewOfHTR10}.


The use of high temperature helium cooled graphite moderated reactors
with TRISO fuel particles have a number of advantages. A TRISO
particle consists of spherical fuel kernel (such as uranium oxide)
surrounded by four concentric layers: 1) a porous carbon buffer layer
to accommodate fission product gases which limits pressure on the
outer layers, 2) an interior pyrolytic carbon layer, 3) a layer of
silicon carbide, and 4) an outer layer of pyrolytic carbon. The
pyrolytic layers shrink and creep with irradiation, partially
offsetting the pressure from the fission products in the interior as
well as helping contain the fission gases.  The silicon carbide acts
as a containment mechanism for the metallic fission
products.\citep{MillerPettiMaki2002} These layers provide an in-core 
containment structure for the radioactive fuel and fission products.

The high temperature gas reactors have some advantages over
conventional light water reactors.  First, the higher outlet
temperatures allow higher Carnot efficiency to be
obtained\footnote{The outlet temperatures for pebble bed reactors have
  ranged from 700 \degree C to 950 \degree C, compared to typical
  outlet temperatures on the order of 300\degree C for light water
  reactors, so the intrinsic Carnot efficiency is higher. }.  Second,
the higher temperatures can be used for process heat, which can reduce
the use of methane.  Third, the high temperature under which TRISO
particles can operate allows for the exploitation of the negative
temperature coefficient to safely shutdown the reactor without use of
control rods.\footnote{Control rods are needed for a cold shutdown,
  however.}  Fourth, the higher temperature is above the annealing
temperature for graphite, which safely removes Wigner
energy\footnote{The accumulation of Wigner energy led to the
  Windscale fire in that lower temperature graphite reactor.}. These
are advantages of both prismatic and pebble bed high temperature
reactors.\citep{GougarExternal,DesignFeaturesOfHTR10}

Pebble bed reactors, unlike most other reactors types, have moving
fuel.  This provides advantages but complicates modeling the reactors.
A key advantage is that pebble bed reactors can be refueled online,
that is, reactor shutdown is not needed for refueling.  As a
consequence, the reactors have low excess reactivity, as new pebbles
can be added or excess pebbles removed to maintain the reactor at
critical.  The low excess reactivity removes the need for burnable
poisons.  A final advantage is that the moving fuel allows the pebble
bed to be run with optimal moderation, where both increases and
decreases in the fuel-to-moderator ratio cause reduction in
reactivity.  \citet{optimalmoderation} discuss the advantages of
optimal moderation including improved fuel utilization.  However,
because the fuel is moving, many traditional methods of modeling
nuclear reactors are inapplicable without a method for quantifying the
motion.  Hence, there is a need for development of methods usable for
pebble bed reactor modeling.

\section{Dissertation Introduction}




This dissertation describes a computer code, PEBBLES, that is designed
to provide a method of simulating the motion of the pebbles in a
pebble bed reactor.  

Chapter \ref{mechanics_model} provides the details of how the
simulation works.  Chapter \ref{static_friction_model} has a new
static friction model developed for this dissertation.  

Several checks have been made of the code.  Figure \ref{bb_comparison}
compares the PEBBLES simulation to experimentally determined radial
packing fractions.  Section \ref{pyramid_test} describes a new
analytical benchmark that was used to test the static friction model
in PEBBLES.  Section \ref{janssen_test} uses the Janssen model to test
the static friction in a cylindrical vat.  

Motivating all the above are the new applications, including Dancoff
factors (\ref{dancoff_factors_ss}), calculating the angle of repose
(\ref{angle_of_repose}) and modeling an earthquake in section
\ref{earthquake}.  

\chapter{Motivation}


Most nuclear reactors have fixed fuel, including typical light water
reactors.  Some reactor designs, such as non-fixed fuel molten salt
reactors, have fuel that is in fluid flow.  Most designs for pebble bed
reactors, however, have moving granular fuel.  Since this fuel is
neither fixed nor easily treatable as a fluid, predicting the behavior
of the reactor requires the ability to understand the characteristics
of the positions and motion of the pebbles.  For example, predicting
the probability of a neutron leaving one TRISO's fueled region and
entering another fueled region depends on the typical locations of the
pebbles.  A second example is predicting the effect of an earthquake
on the reactivity of the pebble bed reactor.  This requires knowing
how the positions of the pebbles in the reactor change from the forces
of the earthquake.  Accurate prediction of the typical features of
the flow and arrangement of the pebbles in the pebble bed reactor would be
highly useful for their design and operation.

The challenge is to gain the ability to predict the pebble flow and
pebble positions for start-up, steady state and transient pebble bed
reactor operation.

The objective of the research presented in this dissertation is to
provide this predicting ability.  The approach used is to create a
distinct element method computer simulation.  The simulation
determines the locations and velocities of all the pebbles in a pebble
bed reactor and can calculate needed tallies from this data.  Over the
course of creating this simulation, various applications of the
simulation were performed.  These models allow the operation of the
pebble bed reactor to be better understood.

\chapter{Previous work}

Because the purpose of this dissertation is to produce a high fidelity
simulation that can provide predictions of the pattern and flow of
pebbles, previous efforts to simulate granular methods and packing
were examined. A variety of simulations of the motion of discrete
elements have been created for different
purposes. \citet{LuAbdouandYing2001} applied a discrete element method
(DEM) to determine the characteristics of packed beds used as fusion
reactor blankets. \citet{JullienPavlovitchandMeakin1992} used a DEM to
determine packing fractions for spheres using different non-motion
methods. \citet{Soppe1990} used a rain method to determine pore
structures in different sized spheres. The rain method randomly
chooses a horizontal position, and then lowers a sphere down until it
reaches other existing spheres. This is then repeated to fill up the
container. \citet{Freundetal2003} used a rain method for fluid flow in
chemical processing.

The use of non-motion pebble packing methods provide an approximation
of the positions of the pebble. Unfortunately, non-motion methods will
tend to either under pack or over pack (sometimes both in the same
model). For large pebble bed reactors, the approximately ten-meter
height of the reactor core will result in different forces at the
bottom than at the top. This will change the packing fractions between
the top and the bottom, so without key physics, including static
friction and the transmittal of force, non-motion physics models will
not even be able to get correct positional information. Non-physics
based modeling can not be used for predicting the effect of changes in
static friction or pebble loading methods even if only the position
data is required.

The initial PEBBLES code for calculation of pebble positions minimized
the sum of the gravitational and Hookes' law potential energies by
adjusting pebble positions. However, that simulation was insufficient
for determining flow and motion parameters and simulation of
earthquake packing. 


Additional references addressing full particle motion simulation were
evaluated.  \citet{Kohring1995} created a 3-D discrete element method
simulation to study diffusional mixing and provided detailed
information on calculating the kinetic forces for the simulation. The
author describes a simple method of calculating static friction.
\citet{Haile1997} discusses both how to simulate hard spheres and soft
spheres using only potential energy.  The soft sphere method in Haile
proved useful for determining plausible pebble positions, but is
insufficient for modeling the motion.  Hard spheres are simulated by
calculating the collision results from conservation laws.  Soft
spheres are simulated by allowing small overlaps, and then having a
resulting force dependent on the overlap.  Soft spheres are similar to
what physically happens, in that the contact area distorts, allowing
distant points to approach closer than would be possible if the
spheres were truly infinitely hard and only touched at one
infinitesimal point.  Hard spheres are impractical for a pebble bed
due to the frequent and continuous contact between spheres so soft
spheres are used instead.  The dissertation by \citet{Ristow1998}
describes multiple methods for simulation of granular materials.  On
Ristow's list of methods was a model similar to that used as the
kernel of the work supporting this dissertation.  Ristow's
dissertation mentioned static friction and provided useful references
that will be discussed in Section
\ref{static_friction_simulation_review}.  

To determine particle flows, \citet{Wait2001} developed a discrete
element method that included only dynamic friction. Concurrently with
this dissertation research, \citet{Rycroftetal2006} used a discrete
element method, created for other purposes, to simulate the flow of
pebbles through a pebble bed reactor.  

Multiple other discrete element codes have been created and PEBBLES is
similar to several of the full motion models. For most of the
applications discussed in this dissertation, only a model that
simulates the physics with high fidelity is useful.  The PEBBLES
dynamic friction model is similar to the model used by Wait or
Rycroft, but the static friction model incorporates some new
improvements that will be discussed later.


In addition to simulation by computer, other methods of determining
the properties of granular fluids have been
used. \citet{BedenigRauschandSchmidt1968} used a scale model to
experimentally determine residence spectra (the amount of time that
pebbles from a given group take to pass through a reactor) for
different exit cone angles. \citet{KadakandBazant2004} used scale
models and small spheres to estimate the flow of pebbles through a
full scale pebble bed reactor. These researchers also examined the
mixing that would occur between different radial zones as the pebbles
traveled downward. \citet{BernalMasonandScott1960} carefully lowered
steel spheres into cylinders and shook the cylinders to determine both
loose and dense packing fractions. The packing fraction and boundary
density fluctuations were experimentally measured by
\citet{BenenatiandBrosilow1962}. The Benenati and Brosilow data have
been used to verify that the PEBBLES code was producing correct
boundary density fluctuations (See Figure \ref{bb_comparison}). Many
experiments were performed in the designing and operating of the AVR
reactor to determine relevant properties such as residence times and
optimal chute parameters \citep{AVR1990}. These experiments provide
data for testing the implementation of any computational model of
pebble flow.

\myfigures{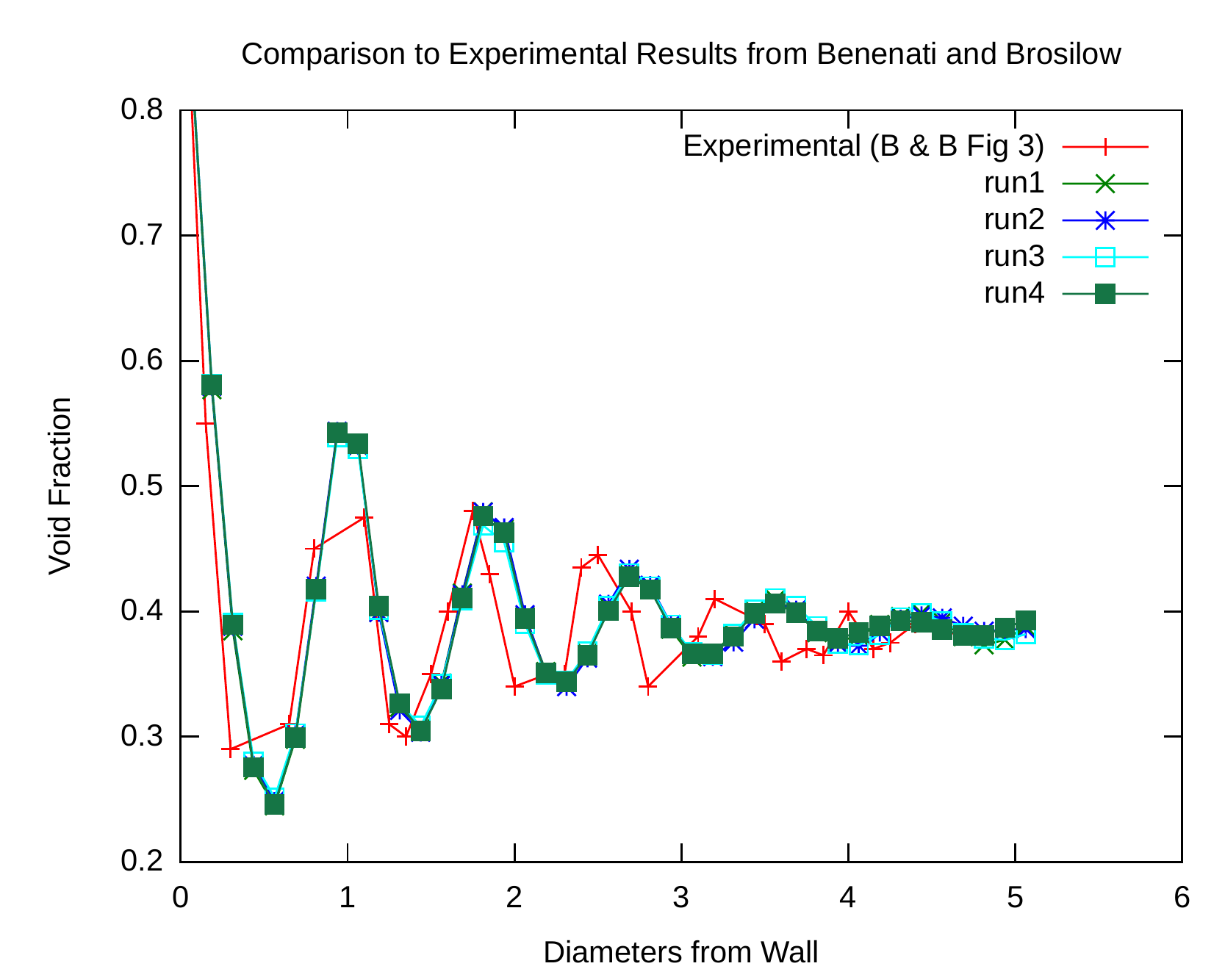}{Comparison Between PEBBLES Outputs and Benenati and Brosilow Data}{bb_comparison}{0.8}

The PEBBLES simulation uses elements from a number of sources and uses
standard classical mechanics for calculating the motion of the pebbles
based on the forces calculated. The features in PEBBLES have been
chosen to implement the necessary fidelity required while allowing run
times small enough to accommodate hundreds of thousands of
pebbles. The next sections will discuss handling static friction.

\section{Static Friction Overview}

Static friction is an important effect in the movement of pebbles and
their locations in pebble bed reactors. This section briefly reviews
static friction and its effects in pebble bed reactors.  Static
friction is a force between two contacting bodies that counteracts
relative motion between them when they are moving sufficiently
slowly\citep{MarionandThornton2004}. Macroscopically, the maximum
magnitude of the force is proportional to the normal force with the
following equation:

\begin{equation}
\label{static_friction_equation}
|\mathbf{F}_{s}|\le \mu |\mathbf{F}_{\perp }|
\end{equation}

where $\mu$ is the coefficient of static friction,
$\mathbf{F}_{s}$ is the static friction force and $\mathbf{F}_{\perp}$
is the normal (load) force.

Static friction results in several effects on granular materials.
Without static friction, the angle of the slope of a pile of a
material (angle of repose) would be zero\citep{Duran1999}.  Static
friction also allows `bridges' or arches to be formed near the outlet
chute.  If the outlet chute is too small, the bridging will be stable
enough to clog the chute.  Static friction will also transfer force
from the pebbles to the walls.  This will result in lower pressure on
the walls than would occur without static
friction\citep{Sperl2006,Walker1966}.

For an elastic sphere, static friction's counteracting force is the
result of elastic displacement of the contact point.  Without static
friction, the contact point would \emph{slide} as a result of relative
motion at the surface.  With static friction, the spheres will
experience local shear that distorts their shape so that the contact
point remains constant.  This change will be called \emph{stuck-slip},
and continues until the counteracting force exceeds $\mu
\mathbf{F}_{\perp}$.  When the counteracting force exceeds that value,
the contact point changes and slide occurs.  The mechanics of this
force with elastic spheres were investigated by
\citet{MindlinandDeresiewicz1953}.  Their work created exact formulas
for the force as a function of the past relative motion and force.


\section{Simulation of Mechanics of Granular Material}
\label{static_friction_simulation_review}

Many simulations of granular materials incorporating static friction
have been devised.  \citet{cundall_and_strack} developed an early
distinct element simulation of granular materials that incorporated a
computationally efficient static friction approximation.  Their method
involved integration of the relative velocity at the contact point and
using the sum as a proxy for the current static friction force.  Since
their method was used for simulation of 2-D circles, adaptation was
required for 3-D granular materials.  One key aspect of adaptation is
determining how the stuck-slip direction changes as a result of
contacting objects' changing orientation.

\citet{VuQuocandZhang1999} and \citet{vu_quoc} developed a 3-D
discrete-element method for granular flows.  This model was used for
simulation of particle flow in chutes.  They used a simplification of
the Mindlin and Deresiewicz model for calculating the stuck-slip
magnitude, and project the stuck-slip onto the tangent plane each
time-step to rotate the stuck-slip force direction.  This correctly
rotates the stuck-slip, but requires that the rotation of the
stuck-slip be done as a separate step since it not written in a
differential form.

\citet{granular_flow} and \citet{confined_packings} describe a 3-D
differential version of the Cundall and Strack method.  The literature
states that particle wall interactions are done identically.  The
amount of computation of the model is less than the Vu-Quoc, Zhang and
Walton model.  This model was used for modeling pebble bed
flow\citep{rycroft,Rycroftetal2006}.  This model however, does not
specify how to apply their differential version to modeling curved
walls.

\chapter{Mechanics Model}
\label{mechanics_model}

The PEBBLES simulation calculates the forces on each individual
pebble.  These forces are then used to calculate the subsequent motion
and position of the pebbles.  

\section{Overview of Model}

The PEBBLES simulation tracks each individual pebble's velocity,
position, angular velocity and static friction loadings.  The
following classical mechanics differential equations are used for
calculating the time derivatives of those variables:

\begin{equation}
\frac{d{\mathbf{v}}_{i}}{dt}=\frac{m_{i}\mathbf{{g}}+\sum _{i\neq
    j}\mathbf{{F}}_{ij}+\mathbf{{F}}_{ci}}{m_{i}}
\end{equation}

\begin{equation} 
\frac{d{\mathbf{p}}_{i}}{dt}={\mathbf{v}}_{i}
\end{equation}

\begin{equation}
\frac{d\mathbf{\omega }_{i}}{dt}=\frac{\sum _{i\neq
    j}\mathbf{{F}}_{\parallel ij}\times
  r_{i}\hat{\mathbf{{n}}}_{ij}+\mathbf{F}_{\parallel ci}\times
  r_{i}\hat{\mathbf{n}}_{ci}}{I_{i}}
\end{equation}

\begin{equation}
\frac{d\mathbf{s}_{ij}}{dt} = \mathbf{S}(\mathbf{F}_{\perp ij},\mathbf{v}_i,\mathbf{v}_j,\mathbf{p}_i,\mathbf{p}_j,\mathbf{s}_{ij})
\end{equation}
where $\mathbf{F}_{ij}$ is the force from pebble $j$ on pebble $i$,
$\mathbf{F}_{ci}$ is the force of the container on pebble $i$,
$\mathbf{g}$ is the gravitational acceleration constant, $m_{i}$ is
the mass of pebble $i$, $\mathbf{v}_{i}$ is the velocity of pebble
$i$, $\mathbf{p}_{i}$ is the position vector for pebble $i$,
$\omega_{i}$ is the angular velocity of pebble \textit{i},
$\mathbf{F}_{\parallel ij}$ is the tangential force between pebbles
$i$ and $j$, $\mathbf{F}_{\perp ij}$ is the perpendicular force
between pebbles $i$ and $j$, $r_{i}$ is the radius of pebble $i$,
$I_{i}$ is the moment of inertia for pebble \textit{i},
$\mathbf{F}_{\parallel ci}$ is the tangential force of the container
on pebble \textit{i}, $\hat{\mathbf{n}}_{ci}$ is the unit vector
normal to the container wall on pebble \textit{i},
$\mathbf{\hat{n}}_{ij}$ is the unit vector pointing from the position
of pebble $i$ to that of pebble $j$, $\mathbf{s}_{ij}$ is the current
static friction loading between pebbles $i$ and $j$, and $\mathbf{S}$
is the function to compute the change in the static friction
loading. The static friction model contributes to the
$\mathbf{F}_{\parallel ij}$ term which is also part of the
$\mathbf{F}_{ij}$ term.  Figure \ref{vectors} illustrates the
principal vectors with pebble $i$ going in the $\mathbf{v}_i$
direction and rotating around the $\mathbf{\omega}_i$ axis, and pebble
$j$ going in the $\mathbf{v}_j$ direction and rotating around the
$\mathbf{\omega}_j$ axis.

\myfigures{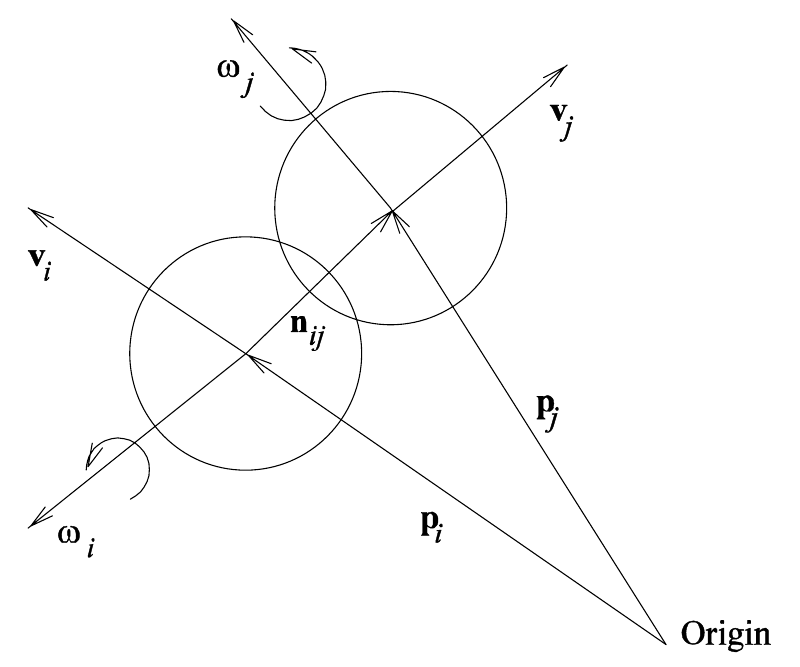}{Principle Vectors in the Interaction of Two Pebbles}{vectors}{1.0}

The mass and moment of inertia are calculated
assuming spherical symmetry with the equations: 

\begin{equation}
m=\frac{4}{3}\pi \left[\rho _{c}r_{c}^{3}+\rho
_{o}(r_{o}^{3}-r_{c}^{3})\right]
\end{equation}

\begin{equation}
I=\frac{8}{15}\pi \left[\rho _{c}r_{c}^{5}+\rho
_{o}(r_{o}^{5}-r_{c}^{5})\right]
\end{equation}

where  $r_{c}$ is the radius of inner (fueled) zone of the pebble, 
$r_{o}$ is the radius of whole pebble,  $\rho_{c}$ is the average
density of center fueled region and  $\rho_{o}$ is the average density
of outer non-fueled region.

The dynamic (or kinetic) friction model is based on the model
described by \citet{Wait2001}. Wait's and PEBBLES model calculate the
dynamic friction using a combination of the relative velocities and
pressure between the pebbles, as shown in Equations (\ref{F_perpij}) and
(\ref{F_dparallelij}):

\begin{equation}
\label{F_perpij}
\mathbf{F}_{\perp ij}=hl_{ij}\hat{\mathbf{n}}_{ij}-C_{\perp}\mathbf{v}_{\perp ij},l_{ij}>0
\end{equation}

\begin{equation}
\label{F_dparallelij}
\mathbf{F}_{d\parallel ij}=-min(\mu
|\mathbf{F}_{\perp ij}|,C_{\parallel}|\mathbf{v}_{\parallel ij}|)\hat{\mathbf{v}}_{\parallel ij},l_{ij}>0
\end{equation}

where $C_{\parallel}$ is the tangential dash-pot constant, $C_{\perp}$
is the normal dash-pot constant, $\mathbf{F}_{\perp ij}$ is the normal
force between pebbles \textit{i} and \textit{j},
$\mathbf{F}_{d\parallel ij}$ is the tangential dynamic friction force
between pebbles \textit{i} and \textit{j}, $h$ is the normal Hooke's
law constant, $l_{ij}$ is the overlap between pebbles \textit{i} and
\textit{j}, $\mathbf{v}_{\parallel ij}$ is the component of the
velocity between two pebbles perpendicular to the line joining their
centers, $\mathbf{v}_{\perp ij}$ is the component of the velocity
between two pebbles parallel to the line joining their centers,
$\mathbf{v}_{ij}$ is the relative velocity between pebbles \textit{i}
and \textit{j} and $\mu $ is the kinetic friction coefficient. Equations
(\ref{F_ij}-\ref{v_ij}) relate supplemental variables to the primary
variables:

\begin{equation} 
\label{F_ij}
\mathbf{F}_{ij}=\mathbf{F}_{\perp ij}+\mathbf{F}_{\parallel ij}
\end{equation}

\begin{equation}
\mathbf{v}_{\perp ij}=(\mathbf{v}_{ij}\cdot
\hat{\mathbf{n}}_{ij})\mathbf{\hat{n}}_{ij}
\end{equation}

\begin{equation} 
\mathbf{v}_{\parallel ij}=\mathbf{v}_{ij}-\mathbf{v}_{\perp ij}
\end{equation}

\begin{equation} 
\label{v_ij}
\mathbf{v}_{ij}=(\mathbf{v}_{i}+\mathbf{\omega
}_{i}\times
r_{i}\mathbf{\hat{n}}_{ij})-(\mathbf{v}_{j}+\mathbf{\omega
}_{j}\times r_{j}\mathbf{\hat{n}}_{ji})
\end{equation}

The friction force is then bounded by the friction coefficient and the
normal force, to prevent it from being too great:

\begin{equation}
\mathbf{F}_{f\parallel ij} = \mathbf{F}_{s\parallel ij} + \mathbf{F}_{d\parallel ij}
\end{equation}

\begin{equation}
\label{F_bound}
\mathbf{F}_{\parallel ij} = min(\mu|\mathbf{F}_{\perp
  ij}|,|\mathbf{F}_{f\parallel ij}|){\hat{\mathbf{F}}_{f\parallel ij}}
\end{equation}

where $\mathbf{F}_{s \parallel ij}$ is the static friction force
between pebbles $i$ and $j$, $\mathbf{F}_{d \parallel ij}$ is the
kinetic friction force between pebbles $i$ and $j$, $h_{s}$ is the
coefficient for force from slip, $\mathbf{s}_{ij}$ is the slip
distance perpendicular to the normal force between pebbles $i$ and
$j$, ${v_{\mathrm{max}}}$ is the maximum velocity under which static
friction is allowed to operate, and $\mu$ is the static friction
coefficient when the velocity is less than $v_{\mathrm{max}}$ and the
kinetic friction coefficient when the velocity is greater. These
equations fully enforces the first requirement of a static friction
method, ${|\mathbf{F}_{s}|\le \mathit{{\mu}}|\mathbf{F}_{\perp}|}$.

\section{Integration}

When all the position, linear velocity, angular velocity and slips are
combined into a vector $\mathbf{y}$, the whole computation can be
written as a differential formulation in the form:

\begin{align}
&\mathbf{y}' = \mathbf{f}(t,\mathbf{y}) \\
&\mathbf{y}(t_0) = \mathbf{y}_0
\end{align}

This can be solved by a variety of methods with the simplest being
Euler's method:

\begin{equation}
\mathbf{y}_1 = \mathbf{y}_0 + \Delta t \mathbf{f}(t, \mathbf{y}_0)
\end{equation}

In addition, both the Runge-Kutta method and the Adams-Moulton method
can be used for solving this equation.  These methods improve the
accuracy of the simulation.  However, they do not improve the
wall-clock time at the lowest stable simulation, since the additional
time required for computation negates the advantage of being able to
use somewhat longer time-steps.  In addition, when running on a
cluster, more data needs to be transferred since the methods allow
non-contacting pebbles to affect each other in a single `time-step
calculation'.

\section{Geometry Modeling}

For any geometry interaction, two things need to be calculated, the
overlap distance $l$ (or, technically, the mutual approach of distant
points) and the normal to the surface $\hat{n}$.  The input is the
radius of the pebble $r$ and the position of the pebble, $\mathbf{p}$
with components $\mathbf{p}_x$, $\mathbf{p}_y$, and $\mathbf{p}_z$

For the floor contact this is:

\begin{align}
l &= (\mathbf{p}_z - r) - floor\_location \\
\hat{n} &= \hat{z}
\end{align}

For cylinder contact on the inside of a cylinder this is:

\begin{align}
pr &= \sqrt{\mathbf{p}_x^2+\mathbf{p}_y^2} \\
l &= (pr + r) - cylinder\_radius \\
\hat{n} &= \frac{-\mathbf{p}_x}{pr}\hat{x}+\frac{-\mathbf{p}_y}{pr}\hat{y}
\end{align}

For cylinder contact on the outside of a cylinder this is:

\begin{align}
pr &= \sqrt{\mathbf{p}_x^2+\mathbf{p}_y^2} \\
l &= cylinder\_radius + r - pr \\
\hat{n} &= \frac{\mathbf{p}_x}{pr}\hat{x}+\frac{\mathbf{p}_y}{pr}\hat{y}
\end{align}

For contact on the inside of a cone defined by the $radius = m z+b$:

\begin{align}
pr &= \sqrt{\mathbf{p}_x^2+\mathbf{p}_y^2} \\
z_c &= \frac{m(pr-b)+z}{m^2+1} \\
r_c &= m z_c+b \\
x_c &= (r_c/pr) \mathbf{p}_y \\
y_c &= (r_c/pr) \mathbf{p}_x \\
\mathbf{c} &= x_c \hat{x}+y_c \hat{y} + z_c \hat{z}  \\
d &= \mathbf{p} - \mathbf{c} \\
l &= |d| + r, r_c < pr \\
\hat{n} &= -\hat{d}, r_c < pr \\
l &= r - |d|, r_c >= pr \\
\hat{n} &= \hat{d}, r_c >= pr 
\end{align}

These equations are derived from minimizing the distance between the
contact point $\mathbf{c}$ and the pebble position $\mathbf{p}$. 


For contact on a plane defined by $ax+by+cz+d=0$ where the equation
has been normalized so that $a^2 + b^2 + c^2 = 1$, the following is used:

\begin{align}
dp &= a\mathbf{p}_x+b\mathbf{p}_y+c\mathbf{p}_z+d\\
l &= r - dp\\
\hat{n} &= a\hat{x}+b\hat{y}+c\hat{z}
\end{align}

Combinatorial geometry operations can be done.  Intersections and
unions of multiple geometry types are done by calculating the overlaps
and normals for all the geometry objects in the intersection or union.
For an intersection, where there is overlap on all the geometry
objects, then the smallest overlap and associated normal are kept,
which may be no overlap.  For a union, the largest overlap and its
associated normal are kept.

For testing that a geometry is correct, a simple check is to fill up
the geometry with pebbles using one of the methods described in
Section \ref{packing_methods}, and then make sure that linear and
angular energy dissipate.  Many geometry errors will show up by
artificially creating extra linear momentum.  For example, if a plane
is only defined at the top, but it is possible for pebbles to leak
deep into the bottom of the plane, they will go from having no overlap
to a very high overlap, which will give the pebble a large force.
This results in extra energy being added each time a pebble encounters
the poorly defined plane, which will show up in energy tallies.

\section{Packing Methods}
\label{packing_methods}

The pebbles are packed using three main methods.  The simplest creates
a very loose packing with an approximately 0.15 packing fraction by
randomly choosing locations, and removing the overlapping ones. These
pebbles then allowed to fall down to compact to a realistic packing
fraction.

The second is the PRIMe method developed by
\citet{KloostermanandOugouag2005}.  In this method large numbers of
random positions (on the order of 100,000 more than will fit) are
generated.  The random positions are sorted by height, and starting at
the bottom, the ones that fit are kept.  Figure \ref{prime}
illustrates this process.  This generates packing fractions of
approximately 0.55.  Then they are allowed to fall to compact.  This
compaction takes less time than starting with a 0.15 packing fraction.

\myfigures{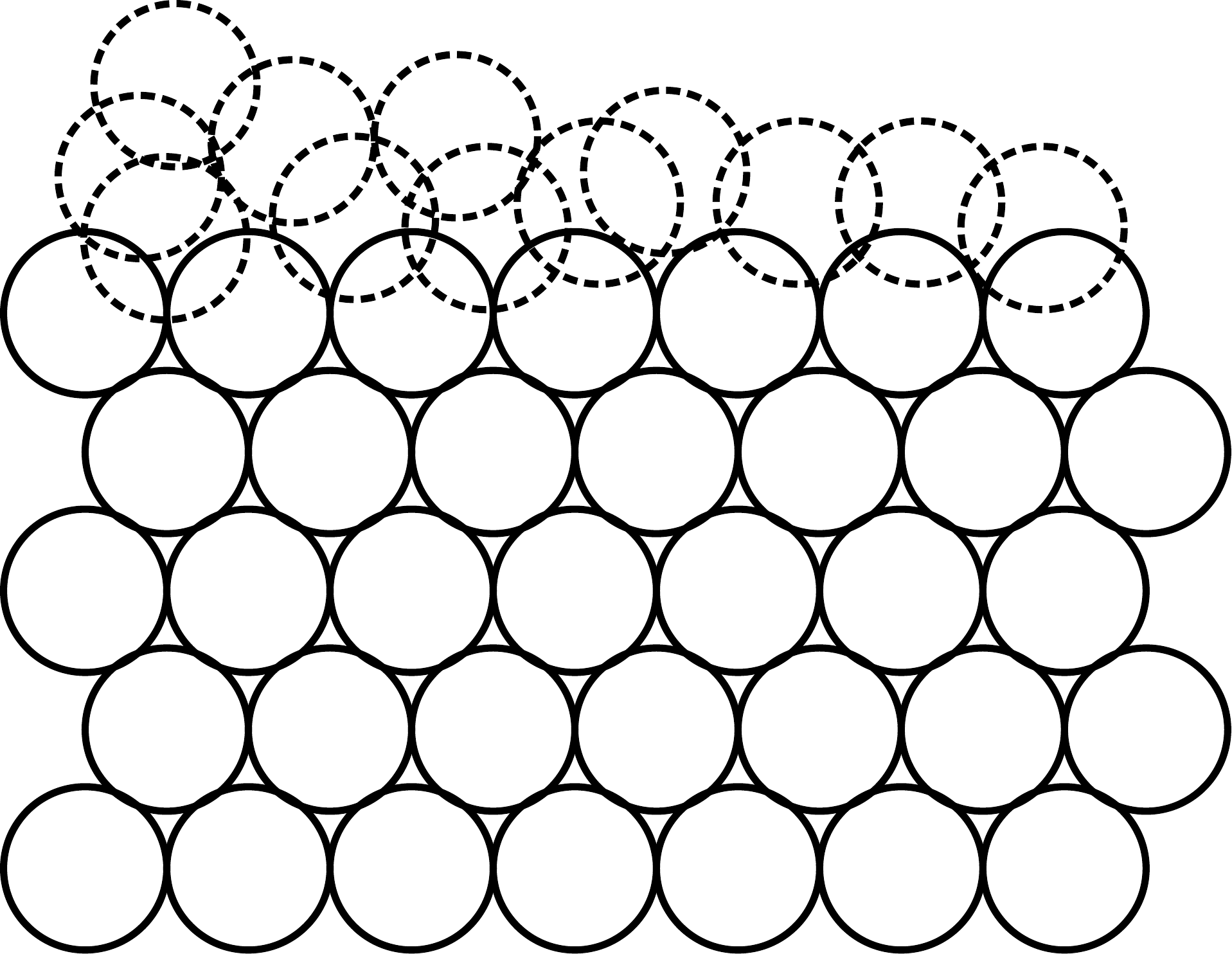}{PRIMe Method Illustration}{prime}{0.3}

The last method is to automatically generates virtual chutes above the
bed where the actual inlet chutes are, and then loads the pebbles into
the chutes.  Figure \ref{virtual_chute} shows this in progress.  This
allows locations that have piles where the inlet chutes are, but can
be done quicker than a recirculation. The other two methods generate
flat surfaces at the top, which is unrealistic, since the surface of a
recirculated bed will have cones under each inlet chute.  

\myfigures{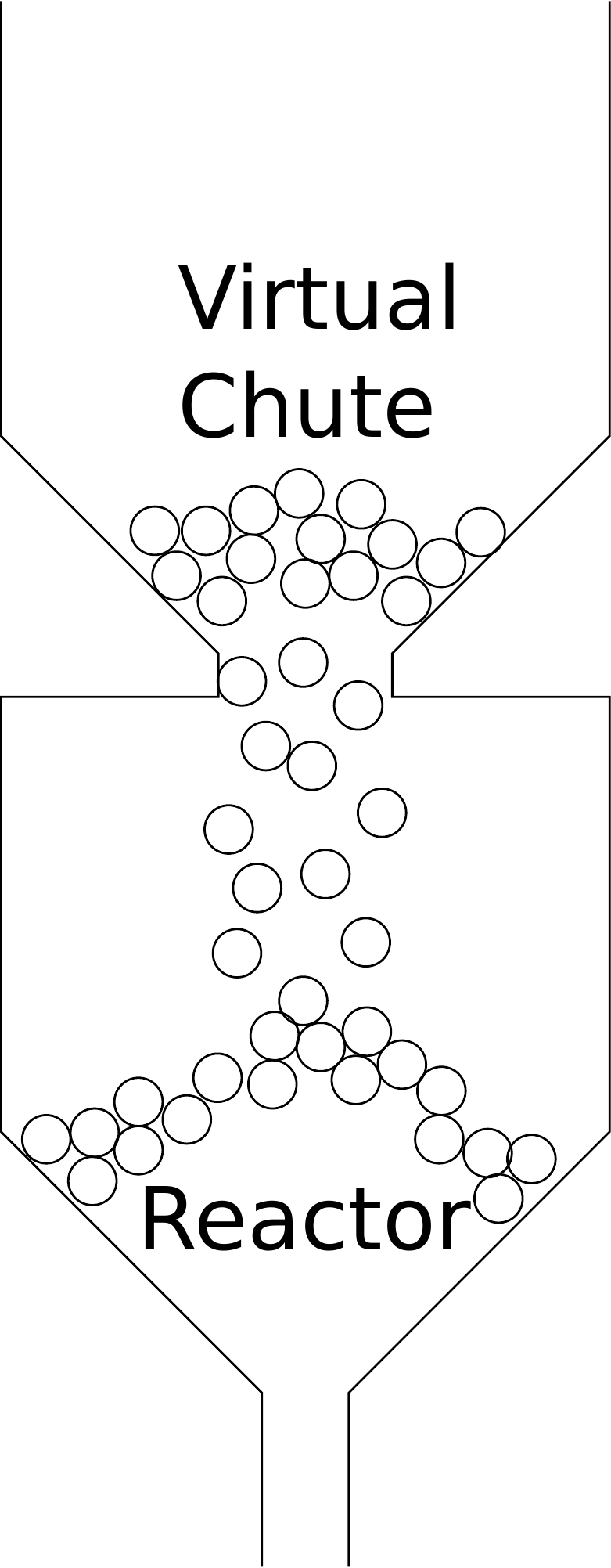}{Virtual Chute Method}{virtual_chute}{0.4}

\section{Typical Parameters}

The typical parameters used with the PEBBLES code are described in
Table \ref{typical_constants}.  Alternative numbers will be described
when used.


\begin{table}[!htb]
  \centering
  \caption{\bf Typical Constants used in Simulation}
  \label{typical_constants}
  \begin{tabular}{l|l}
    Constant & Value \\
    \hline
    Gravitational Acceleration $g$ & 9.8 m/s$^2$ \\
    radius of pebbles $r$ & 0.03 m\\
    Mass of Pebble $m$ & 0.2071 kg\\
    Moment of Inertia $I$ & 7.367e-05 kg m$^2$\\
    Hooke's law constant $h$ & 1.0e6 \\
    Dash-pot constants $C_\parallel$ and $C_\perp$ & 200.0\\
    Kinetic Friction Coefficient $\mu$ & 0.4 or sometimes 0.25\\
    Static Friction Coefficient $\mu_s$ & 0.65 or sometimes 0.35\\
    Maximum static friction velocity $v_{max}$ & 0.1 m/s\\
  \end{tabular}
\end{table}

\chapter{A New Static Friction Model}
\label{static_friction_model}

The static friction model in PEBBLES is used to calculate the force
and magnitude of the static friction force.  Other models have been
created before to calculate static friction, but the PEBBLES model
provides the combination of being a differential model (as opposed to
one where the force is rotated as a separate step) and being
able to handle the type of geometries that exist in pebble bed
reactors.

The static friction model has two key requirements.  First, the force from
stuck-slip must be updated based on relative motion of the pebbles.  Second,
the current direction of the force must be calculated since the
pebbles can rotate in space.  

\subsection{Use of Parallel Velocity for Slip Updating}

For elastic spheres, the true method of updating the stuck-slip force
is to use the method of \citet{MindlinandDeresiewicz1953}.  This
method requires computationally and memory intensive calculations to
track the forces.  Instead, a simpler method is used to approximate
the force.  This method, described by \citet{cundall_and_strack} uses
the integration of the parallel relative velocity as the displacement.
The essential idea is that the farther the pebbles have stuck-slipped
at the contact point, the greater the counteracting static friction
force needs to be.  This is what happens under more accurate models
such as Mindlin and Deresiewicz.  There are two approximations imposed
by this assumption. First, the amount the force changes is independent
of the normal force.  Second, the true hysteretic effects that are
dependent on details of the loading history are ignored.  For
simulations where the exact dynamics of static friction are important,
these could potentially be serious errors.  However, since static
friction only occurs when the relative speed is low, the dynamics of
the simulation usually are unimportant.  Thus, for most circumstances,
the following approximation can be used for the rate of change of the
magnitude and non-rotational change of the stuck-slip:

\begin{equation}
\label{basic_slip_change}
\frac{d\mathbf{s}_{ij}}{dt}=\mathbf{v}_{\parallel ij}
\end{equation}

\subsection{Adjustment of Oversize Slips}

The slips can build up to unrealistically large amounts, that is,
greater than $\mu|\mathbf{F}_{\perp}|$; equation
\ref{basic_slip_change} places no limit on the maximum size of the
slip.  The excessive slip is solved at two different locations. First,
when the frictions are added together to determine the total friction
they are limited by $\mu|\mathbf{F}_{\perp}|$ in equation
(\ref{F_bound}).  This by itself is insufficient, because the slip is
storing potential energy that appears anytime the normal force
increases.  This manifests itself by causing vibration of the pebbles
to continue for long periods of time.  Two methods for fixing the
hidden slip problem are available in PEBBLES.  The simplest drops any
slip that exceeds the static friction threshold (or an input parameter
value somewhat above the static friction threshold so small vibrations
do not cause the slip to disappear).

The second method used in PEBBLES is to decrease the slip that is over
a threshold value.  If the slip is too great, a derivative that is the
opposite as the current slip is added as an additional term in the
slip time derivative.  This occurs in the following additional term:

\begin{equation}
\label{dls_eqn}
\frac{d\mathbf{s}_{ij}}{dt}=-\mathrm{R}(|\mathbf{s}_{ij}|-s_{sd}\mu|\mathbf{F}_{\perp ij}|)\hat{s_{ij}}
\end{equation}

In this $\mathrm{R}(x)$ is the ramp function (which is $x$ if $x > 0$
and 0 otherwise), $s_{sd}$ is a constant to select how much the slip
is allowed to exceed the static friction threshold (usually 1.1 in
PEBBLES).  This derivative adder is used in most PEBBLES runs since it
does allow vibrational energy to decrease, yet does not cause the
pyramid benchmark to fail like complete removal of too great slips
does.

When using non-Euler integration methods, the change in slip is
calculated multiple times.  Each time it is calculated, it might be
set to be zeroed.  In the PEBBLES code, if any of the added up slips
for a given contact were set to be zeroed, the final slip is zeroed.
This is not an ideal method, but it works well enough.

\subsection{Rotation of Stuck-Slip}

The static friction force must also be rotated so that it is in the
plane of contact between the two pebbles. When there is a difference
between the pebbles' center velocities, which changes in the relative
pebble center location, change in the direction in the stuck-slip
occurs. That is:

\begin{equation}
\label{position_difference}
\mathbf{p}_{in+1}-\mathbf{p}_{jn+1}\approx
\mathbf{p}_{in}-\mathbf{p}_{jn}+(\mathbf{v}_{in}-\mathbf{v}_{jn})\Delta
t
\end{equation}



First, let $\mathbf{n}_{ijn}=\mathbf{p}_{i}-\mathbf{p}_{j}$ and
$d{\mathbf{n}}_{ijn}=\mathbf{v}_{i}-\mathbf{v}_{j}$.  The cross
product $-d{\mathbf{n}}_{ijn}\times \mathbf{n}_{ijn}$ is perpendicular
to both $\mathbf{n}$ and $d\mathbf{n}$ and signed to create the axis
around which $\mathbf{s}$ is rotated in a right-handed
direction. Then, using the cross product of the axis and $\mathbf{s}$,
$-(d\mathbf{n}_{ij}\times \mathbf{n}_{ijn})\times \mathbf{s}_{ijn}$
gives the correct direction that $\mathbf{s}$ should be increased.

Next, determine the factors required to make the differential the
proper length. By cross product laws,

\begin{equation}
|-(d\mathbf{n}_{ij}\times
\mathbf{n}_{ijn})\times
\mathbf{s}_{ijn}|=|d\mathbf{n}_{ij}||\mathbf{n}_{ijn}||\mathbf{s}_{ijn}|\sin
\theta \sin \phi 
\end{equation}

where ${\theta}$ is the angle between $\mathbf{n}_{\mathbf{ijn}}$ and
$d\mathbf{n}_{\mathbf{ij}}$ and $\phi$ is the angle between
$d\mathbf{n}_{ij}\times \mathbf{n}_{ijn}$ and $\mathbf{s}_{ijn}$.

The relevant vectors are shown in figure \ref{static_friction_vectors}.

\myfigure{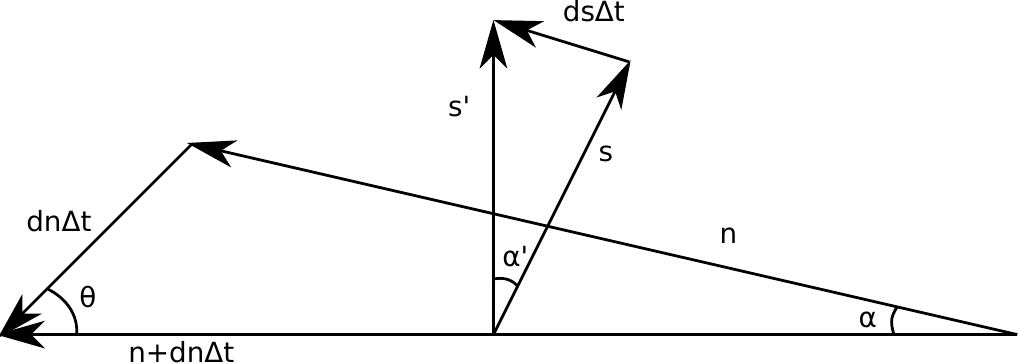}{Static Friction Vectors}{static_friction_vectors}


The goal is to rotate $\mathbf{s}$ by angle $\alpha\prime$ which is
the `projection' into the proper plane of the angle $\alpha$ that
$\mathbf{n}$ rotates by.  Since the direction has been determined, for
simplicity the figure leaves the indexes off, and concentrates on
determining the lengths.  In \figurename~\ref{static_friction_vectors},
$\mathbf{s}$ is the old slip vector, $\mathbf{s}\prime$ is the
new slip vector, $\mathbf{n}$ is the vector pointing from one pebble to
another.  The vector $d\mathbf{n}\Delta t$ is added to $\mathbf{n}$
to get the new $\mathbf{n}\prime$,
$\mathbf{n}+d\mathbf{n}{\Delta} t$.  The initial condition is that
$\mathbf{s}$ and $\mathbf{n}$ are perpendicular. The final conditions
are that $\mathbf{s}\prime$ and $\mathbf{n}\prime$
are perpendicular, and that $\mathbf{s}$ and
$\mathbf{s}\prime$ are the same length and that
$\mathbf{s}\prime$ is the closest vector to $\mathbf{s}$ as it
can be while satisfying the other conditions.  There is no
requirement that $\mathbf{s}$ or $\mathbf{s}\prime$ are
coplanar with $d\mathbf{n}{\Delta} t$ (otherwise
${\alpha}\prime$ would be equal to ${\alpha}$).  From the
law of sines we have:

\begin{equation}
\frac{|d\mathbf{n}\Delta t|}{\sin \alpha
}=\frac{|\mathbf{n}|}{\sin \theta }
\end{equation}
so

\begin{equation}
\sin \alpha =\frac{|d\mathbf{n}\Delta t|\sin \theta
}{|\mathbf{n}|}
\end{equation}

In \figurename~\ref{seq:refIllustration1} the projection to the
correct plane occurs.  First by using $\phi$ the length of
$\mathbf{s}$ is projected to the plane.  Note that $\phi$ is the
angle both to $\mathbf{s}$ and to $\mathbf{s}\prime$.  So, the length
of the line on the $d\mathbf{n}\times\mathbf{n}$ plane is $|\mathbf{s}| \sin
{\phi}$, and the length of the straight line at the end of the
triangle is $|\mathbf{s}| \sin {\phi} \sin {\alpha}$ (note that the
chord length is $|\mathbf{s}| (\sin {\phi}){\alpha}$, but as
${\Delta} t$ approaches 0 the other can be used).  From these
calculations, the length of the $d\mathbf{s}{\Delta} t$ can be
calculated with the following formula:

\begin{equation}
d\mathbf{s}\Delta t=\frac{|s|\sin \phi
|d\mathbf{n}\Delta t|\sin \theta }{|\mathbf{n}|}
\end{equation}

Since   $|-(d\mathbf{n}_{ij}\times
\mathbf{n}_{ijn})\times
\mathbf{s}_{ijn}|=|d\mathbf{n}_{ij}||\mathbf{n}_{ijn}||\mathbf{s}_{ijn}|\sin
\theta \sin \phi $ the formula for the rotation is:

\begin{equation}
\mathbf{s}_{ijn+1}=-\frac{(d{\mathbf{n}}_{ijn}\times
  \mathbf{n}_{ijn})\times \mathbf{s}_{ijn}}{\mathbf{n}^{2}}\Delta
t+\mathbf{s}_{ijn}
\end{equation}

\myfigures{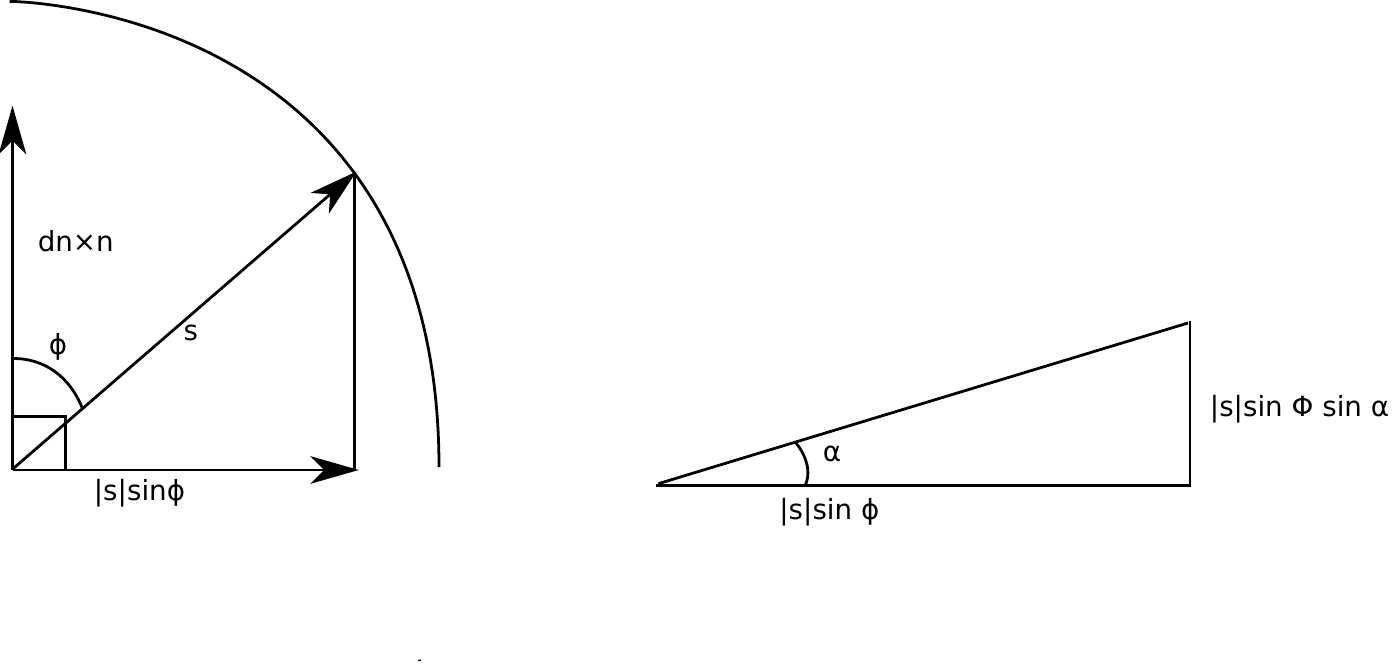}{Projections to ds}{seq:refIllustration1}{0.9}

As a differential equation this is:

\begin{equation}
\frac{d\mathbf{s}_{ij}}{dt}=-\frac{\left[((\mathbf{v}_{i}-\mathbf{v}_{j})\times
(\mathbf{p}_{i}-\mathbf{p}_{j}))\times
\mathbf{s}_{ij}\right]}{|\mathbf{p}_{i}-\mathbf{p}_{j}|^{2}}
\end{equation}

By the vector property $a \times (b \times c) = b(a \cdot c) - c(a
\cdot b)$ and since $(\mathbf{p}_{i}-\mathbf{p}_{j}) \cdot
\mathbf{s}_{ij} = 0$, this can be rewritten as the version in
\citet{granular_flow}:

\begin{equation}
\frac{d\mathbf{s}_{ij}}{dt} =- \frac{(\mathbf{p}_{i}-\mathbf{p}_{j}) (\mathbf{s}_{ij} \cdot (\mathbf{v}_{i}-\mathbf{v}_{j}))}{|\mathbf{p}_{i}-\mathbf{p}_{j}|^{2}}
\end{equation}

\subsection{Differential Equation for Surface Slip Rotating}

It might seem that the wall interaction could be modeled the same way
as the pebble-to-pebble interaction.  For sufficiently simple wall
geometries this may be possible, but actual pebble bed reactor
geometries are more complicated, and violate some of the assumptions
that underpin the derivation.  For a flat surface, there is no
rotation, so the formula can be entirely dropped.  For a spherical
surface, it would be possible to measure the curvature at pebble to
surface contact point in the direction of relative velocity to the
surface.  This curvature could then be used as an effective radius in
the pebble-to-pebble formulas.  

The pebble reactor walls have additional features that violate
assumptions made for the derivation.  For surfaces such as a cone, the
curvature is not in general constant, because the path can follow
elliptical curves.  As well, the curvature has discontinuities where
different parts of the reactor join together (for example, the
transition from the outlet cone to the outlet chute).  At these
transitions, the assumption that the slip stays parallel to the
surface fails because the slip is parallel to the old surface, but the
new surface has a different normal.

Because of the complications with using the pebble to pebble
interaction, PEBBLES uses an approximation of the ``rotation delta.''
This is similar to the \citet{VuQuocandZhang1999} method of adjusting
the slip so that it is parallel to the surface each time.  Each time
when the slip is used, a temporary version of the slip that is
properly aligned to the surface is computed and used for calculating
the force.  As well, a rotation to move the slip more parallel to the
surface is also computed.

The rotation is computed as follows.  Let the normal direction of the
wall at the point of contact of the pebble be $\mathbf{n}$, and the
old stuck-slip be $\mathbf{s}$.  Let $a$ be the angle between
$\mathbf{n}$ and $\mathbf{s}$.  $\mathbf{n}\times \mathbf{s}$ is
perpendicular to both $\mathbf{n}$ and $\mathbf{s}$ and so this cross
product is the axis that needs to be rotated around.  Then
$(\mathbf{n}\times \mathbf{s})\times \mathbf{s}$ is perpendicular to
this vector, so it is either pointing directly towards $\mathbf{n}$ if
$a$ is acute or directly away from $\mathbf{n}$ if $a$ is obtuse.  To
obtain the correct direction, this vector is multiplied by the scalar
$\mathbf{s}\cdot \mathbf{n}$ which has the correct sign from $\cos a$.
The magnitude of $(\mathbf{s}\cdot \mathbf{n})[(\mathbf{n}\times
  \mathbf{s})\times \mathbf{s}]$ needs to be determined for
reasonableness.  We define the angle $b$, which is between
$(\mathbf{n}\times \mathbf{s})$ and $\mathbf{s}$.  By these
definitions the magnitude is $(|\mathbf{s}||\mathbf{n}|\cos
a)[(|\mathbf{n}||\mathbf{s}|\sin a)|\mathbf{s}|\sin b]$.  $b$ is a
right angle since $\mathbf{n}\times \mathbf{s}$ is perpendicular to
$\mathbf{s}$, so $\sin b = 1$.  Collecting terms gives the magnitude
as $|\mathbf{s}|^{3}|\mathbf{n}|^{2}\cos a\sin a$ which is divided by
$|\mathbf{n}\times \mathbf{s}||\mathbf{n}||\mathbf{s}|$ to give the
full term the magnitude $|\mathbf{s}|\cos a$.  This is the length of
the vector that goes from $\mathbf{s}$ to the plane perpendicular to
$\mathbf{n}$.  This produces equation \ref{wall_rotation}, which can
be used to ensure that the wall stuck-slip vector rotates towards the
correct direction.


\begin{equation}
\label{wall_rotation}
\frac{d\mathbf{s}}{dt}= (\mathbf{s}\cdot
\mathbf{n})\frac{[(\mathbf{n}\times \mathbf{s})\times
    \mathbf{s}]}{|\mathbf{n}\times
  \mathbf{s}||\mathbf{n}||\mathbf{s}|}
\end{equation}

\myfigures{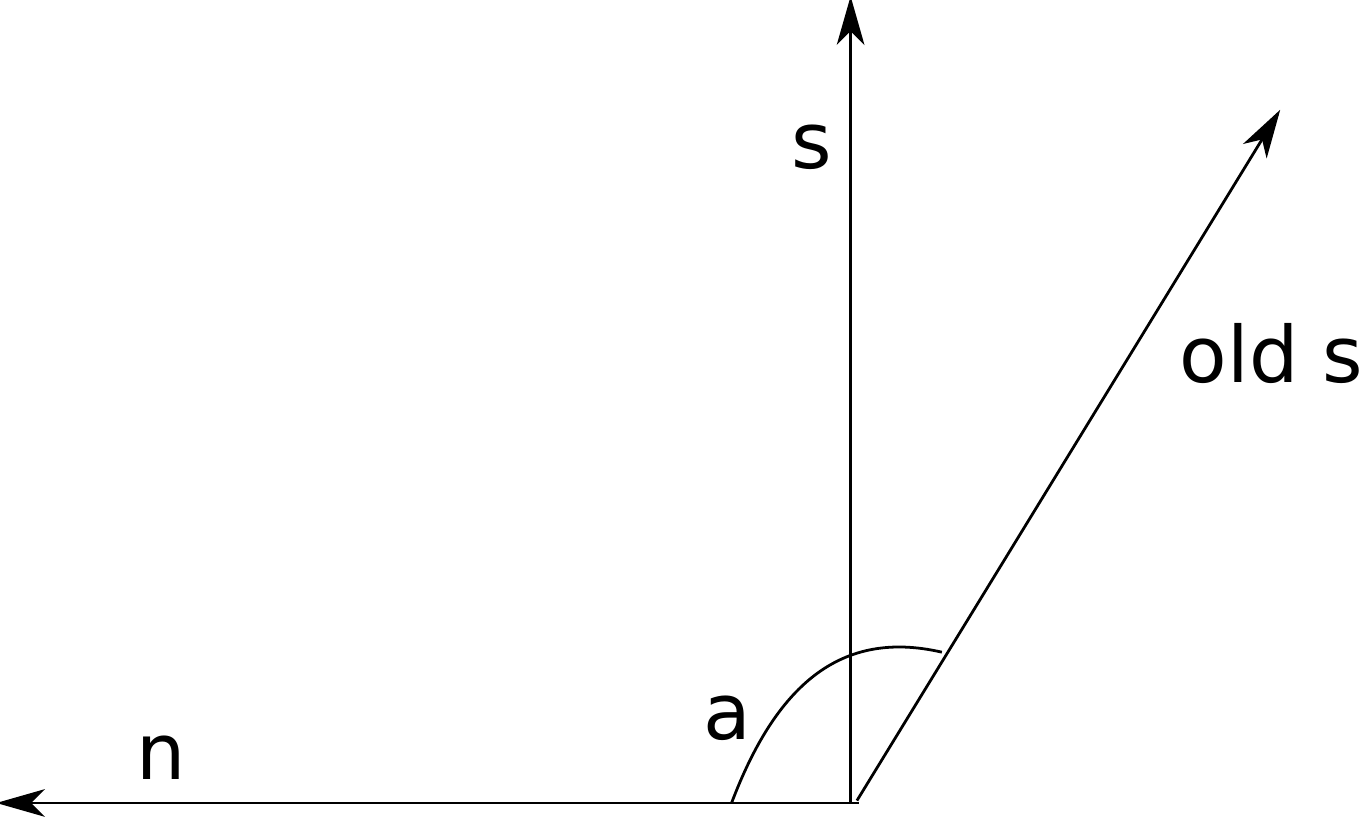}{Static Friction Vectors for Wall}{wall_static}{0.4}








\section{Testing of Static Friction Model With Pyramid Test}
\label{pyramid_test}

Static friction is an important physical feature in the implementation
of mechanical models of pebbles motion in a pebble bed, and checking
its correctness is important. A pyramid static friction test model was
devised as a simple tool for verifying the implementation of a static
friction model within the code.  The main advantages of the pyramid
test are that the model test is realistic and that it can be modeled
analytically, providing an exact basis for the comparison. The test
benchmark consists of a pyramid of five spheres on a flat surface.
This configuration is used because the forces acting on each pebble
can be calculated simply and the physical behavior of a model with
only kinetic friction is fully predictable on physical and
mathematical grounds: with only kinetic friction and no static
friction, the pyramid will quickly flatten. Even insufficient static
friction will result in the same outcome.  The four bottom spheres are
arranged as closely as possible in a square, and the fifth sphere is
placed on top of them as shown in Fig.~\ref{pyramid_spheres}.

\myfigures{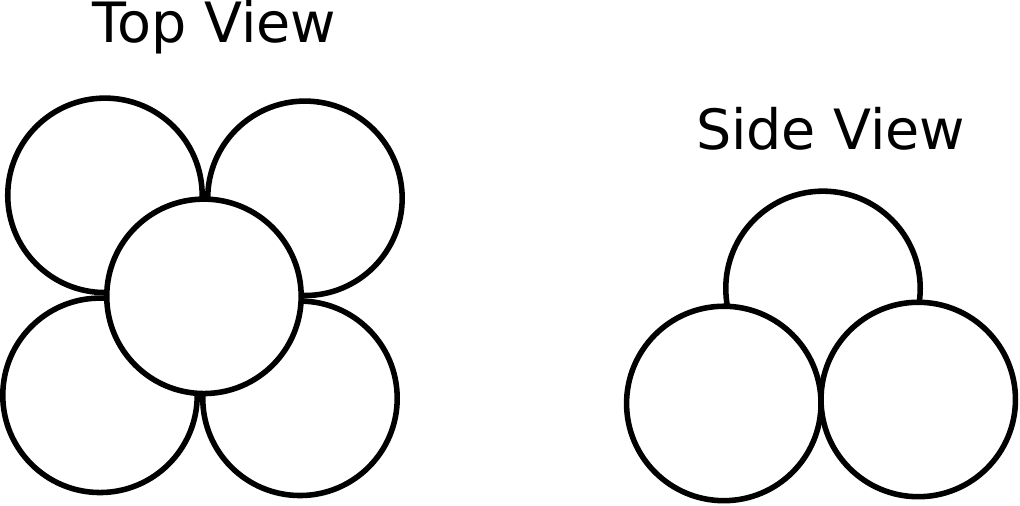}{Sphere Location Diagram}{pyramid_spheres}{0.8}

The lines connecting the centers of the spheres form a pyramid with
sides $2R$, as shown in Fig.~\ref{pyramid_diagram}, where $R$ is the
radius of the spheres. The length of $a$ in the figure is
$\frac{2R}{\sqrt{2}}$, and because $b$ is part of a right triangle,
$(2R)^2 - (\frac{2R}{\sqrt{2}})^2 = b^2 = 4R^2 - \frac{4R^2}{2} =
2R^2$, so $b$ has the same length as $a$, and thus the elevation angle
for all vertexes of the pyramid are $45^\circ$ from horizontal.

\myfigure{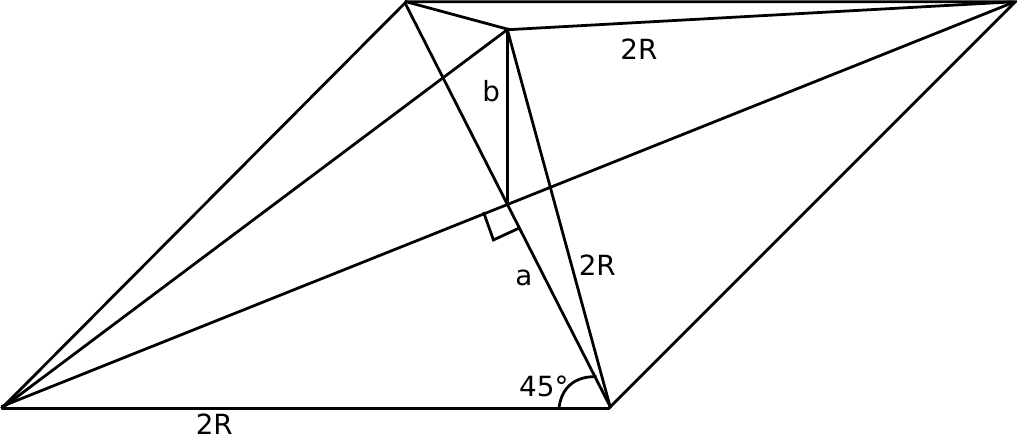}{Pyramid Diagram}{pyramid_diagram}

Taking for origin of the coordinates system the projection of the
pyramid summit onto the ground, the locations (coordinates) of
the sphere centers are given in Table \ref{sphere_location}.

\begin{table}[!htb]
  \centering
  \caption{\bf Sphere location table.}
  \label{sphere_location}
  \begin{tabular}{|c|c|c|}
    \hline
    X & Y & Z \\
    \hline
    $-R$ & $-R$ & $R$ \\
    $R$ & $-R$ & $R$ \\
    $-R$ & $R$ & $R$ \\
    $R$ & $R$ & $R$ \\
    0 & 0 & $R(1+\sqrt{2})$ \\
    \hline
  \end{tabular}
\end{table}

\subsection{Derivation of Minimum Static Frictions}

The initial forces on the base sphere are the force of gravity
$m\mathbf{g}$, and the normal forces $\mathbf{Tn}$ and $\mathbf{Fn}$ as
shown in Fig.~\ref{pyramid_forces}.  This causes initial stuck-slip which
will cause $\mathbf{Fs}$ to develop to counter the slip, and
$\mathbf{Ts}$ to counter the rotation of the base sphere relative to the
top sphere. The top sphere will have no rotation because the forces from the four spheres will be symmetric and counteract each other. 

The forces on the base sphere are:
\begin{quote}
\begin{description}
\item[$\mathbf{Tn}$] -- Normal force from the top sphere
\item[$\mathbf{Ts}$] -- Static friction force from the top sphere
\item[$m\mathbf{g}$] -- Force of gravity on the base sphere
\item[$\mathbf{Fn}$] -- Normal force from floor
\item[$\mathbf{Fs}$] -- Static friction force from the floor
\end{description}
\end{quote}

\myfigures{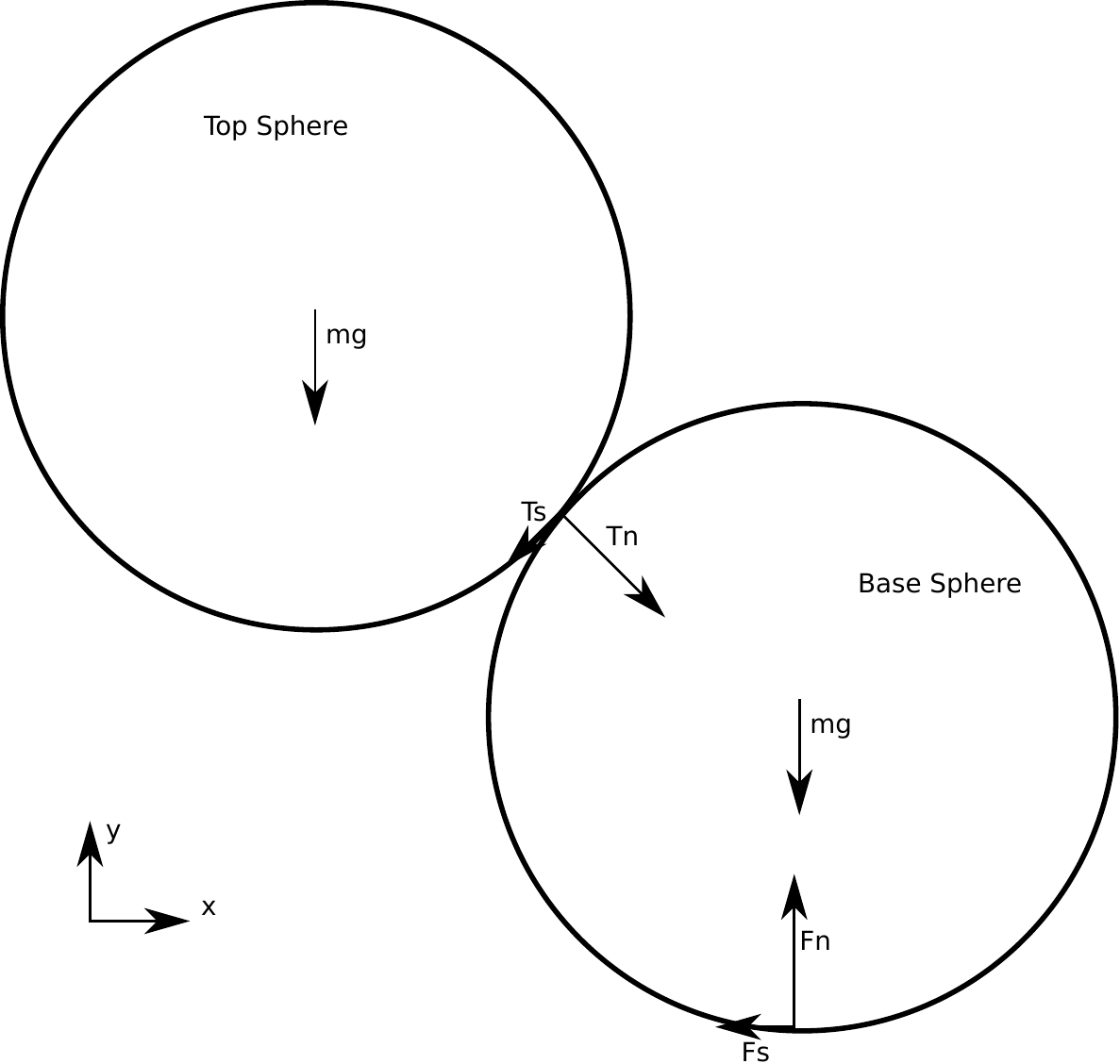}{Force Diagram}{pyramid_forces}{0.9}

Note that $\mathbf{Fn}$ is larger than $\mathbf{Tn}$ since
$\mathbf{Tn}$ is only a portion of the $m\mathbf{g}$ force since the
top sphere transmits (and splits) its force onto all four base
spheres.

There are three requirements for a base sphere to be non-accelerated.

If a base sphere is not rotating than there is no torque, so:

\begin{equation}
|\mathbf{Fs}|=|\mathbf{Ts}|
\label{seq:refText0}
\end{equation}

The resultant of all forces must also be zero in the x and the y direction
(vector notation dropped since they are in one dimension and therefore
scalars) as follows:

\begin{equation}
-Fs-Tsx+Tnx=0
\label{seq:refText1}
\end{equation}

\begin{equation}
-mg-Tsy-Tny+Fn=0
\label{seq:refText2}
\end{equation}

Since the angle of contact between a base sphere and the top sphere is
$45^\circ$, the following two equations hold (where $Ts$ is
the magnitude of $\mathbf{Ts}$ and $Tn$ is the magnitude of
$\mathbf{Tn}$):

\begin{equation}
Tsx=Tsy=\frac{Ts}{\sqrt{2}}
\end{equation}

\begin{equation}
Tnx=Tny=\frac{Tn}{\sqrt{2}}
\end{equation}

This changes equations \ref{seq:refText1} and \ref{seq:refText2} into:

\begin{equation}
-Fs-\frac{Ts}{\sqrt{2}}+\frac{Tn}{\sqrt{2}}=0
\label{seq:refText5}
\end{equation}

\begin{equation}
-mg-\frac{Ts}{\sqrt{2}}-\frac{Tn}{\sqrt{2}}+Fn=0
\label{seq:refText6}
\end{equation}

Combining equation \ref{seq:refText0} and \ref{seq:refText5}
provides:

\begin{equation}
-Ts-\frac{Ts}{\sqrt{2}}+\frac{Tn}{\sqrt{2}}=0
\end{equation}

Which gives the relation:

\begin{equation}
Tn=Ts(\sqrt{2}+1)
\label{seq:refText8}
\end{equation}

By the static friction Equation \ref{static_friction_equation}, 

\begin{equation}
Ts\le \mu_{sphere} Tn
\label{seq:refText9}
\end{equation}

Combining equations \ref{seq:refText8} and \ref{seq:refText9} and
simplifying gives the requirement that

\begin{equation}
\sqrt{2}-1\le \mu_{sphere}
\end{equation}

For use with testing, the static friction program can be tested twice
with a sphere-to-sphere friction coefficient slightly above 0.41421...
and one slightly below 0.41421....  In the first case the pyramid
should be stable, and in the second case the top ball should fall to
the floor.

Since {\textonequarter} of the weight of the top pebble is on one of the
base pebbles, the following holds:

\begin{equation}
Fn=\frac{5}{4}mg
\label{seq:refText11}
\end{equation}

Combining \ref{seq:refText6} and \ref{seq:refText11} provides the
following equation:

\begin{equation}
\frac{mg}{4}-\frac{Ts}{\sqrt{2}}-\frac{Tn}{\sqrt{2}}=0
\label{seq:refText12}
\end{equation}

Equations \ref{seq:refText5} and \ref{seq:refText12} can be added to
produce

\begin{equation}
-Fs-\sqrt{2}Ts+\frac{mg}{4}=0
\end{equation}

Using \ref{seq:refText0} and \ref{seq:refText12} and solving for
\textit{Fs} gives the following value for \textit{Fs}:

\begin{equation}
Fs=\frac{mg}{4(1+\sqrt{2})}
\end{equation}

By the static friction Equation \ref{static_friction_equation}:

\begin{equation}
Fs\le \mu_{surface} Fn.
\end{equation}

Substituting the values for $Fs$ and $Fn$ gives:

\begin{equation}
\frac{mg}{4(1+\sqrt{2})}\le \mu_{surface} \frac{5}{4}mg
\end{equation}

Simplifying provides the following relation for the surface-to-sphere
static friction requirement:

\begin{equation}
\frac{1}{5(1+\sqrt{2})}\le \mu_{surface}.
\end{equation}

This can be used similarly to the other static friction requirement by
setting the value slightly above 0.08284... and slightly below
0.08284... and making sure that it is stable with the higher value and
not stable with the lower value.

This test was inspired by an observation of lead cannon balls stacked
into a pyramid. I tried to stack used glass marbles into a five
ball pyramid and it was not stable. Since lead has a static friction
coefficient around 0.9 and used glass has a much lower static
friction, the physics of pyramid stability was further investigated,
resulting in this benchmark test of static friction modeling.

\subsection{Use of Benchmark}

The benchmark test of two critical static friction coefficients is
defined by the following equations.  If both static friction
coefficients are above the critical values, the spheres will form a
stable pyramid.  If either or both values are below the critical
values the pyramid will collapse.

\begin{equation}
  \mu_{criticalsurface}=\frac{1}{5(1+\sqrt{2})} \approx 0.08284
\end{equation}
\begin{equation}
  \mu_{criticalsphere}=\sqrt{2}-1 \approx 0.41421
\end{equation}

To set up the test cases, the sphere locations from Table
\ref{sphere_location} should be used as the initial locations of the
sphere.  Then, static friction coefficients for the sphere-to-sphere
contact and the sphere-to-surface contact are chosen.  The code is
then run until either the center sphere falls to the surface, or the
pyramid obtains a stable state.  There are three test cases that are run
to test the model.

\begin{enumerate}
\item $\mu_{surface} = \mu_{criticalsurface} + \epsilon$ and $\mu_{sphere} = \mu_{criticalsphere} + \epsilon$ which should result in a stable pyramid.
\item $\mu_{surface} = \mu_{criticalsurface} - \epsilon$ and $\mu_{sphere} = \mu_{criticalsphere} + \epsilon$ which should result in a fall.
\item $\mu_{surface} = \mu_{criticalsurface} + \epsilon$ and $\mu_{sphere} = \mu_{criticalsphere} - \epsilon$ which should result in a fall.
\end{enumerate}

For soft sphere models, there are fundamental limits to how close the
model's results can be to the critical coefficient.  Essentially, as
the critical coefficients are approached, the assumptions become less
valid.  For example, with soft (elastic) spheres, the force from the
center sphere will distort the contact angle, so the actual critical
value will be different.  For the PEBBLES code, an $\epsilon$ value of
0.001 is used.

\section[Janssen's Formula Comparison]{Testing of the Static Friction Model by Comparison with Janssen's Formula}
\label{janssen_test}

The pyramid static friction test is used as a simple test of the
static friction model.  Another test compares the static friction
model against the Janssen formula's behavior \citep{Sperl2006}.  This
formula specifies the expected wall pressure as a function of depth.
This formula only applies when the static friction is fully loaded,
that is when $\mathbf{F}_{s}| = \mu |\mathbf{F}_{\perp }|$.  This
condition is generally not satisfied until some recirculation has
occurred.  Figure \ref{wall_forces} shows the normal force and the
static friction force from a pebble to the wall.  With the PEBBLES
code, this is only satisfied after recirculation with lower values of
the static friction coefficient $\mu$.

\myfigures{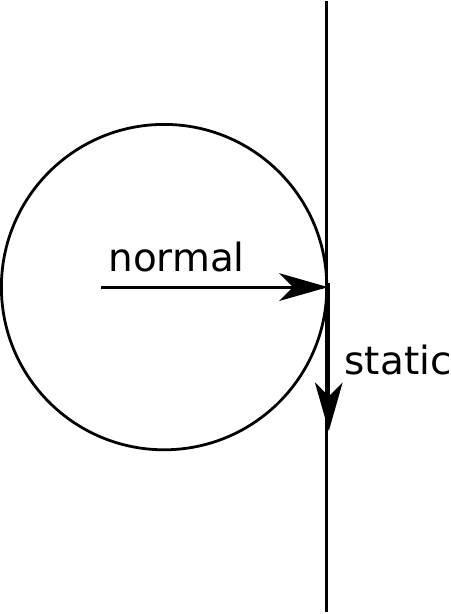}{Relevant Forces on Wall from Pebble}{wall_forces}{0.8}

The equation used to calculate the pressure on the region $R$ from
the normal force in PEBBLES is:

\begin{equation}
p = {1 \over R_h 2 \pi r} \sum_{i \ in \  R} | \mathbf{F}_{\perp ci} |
\end{equation}

where $p$ is the pressure, $R_h$ is the height of the region, and $r$
is the radius of the cylinder.

The equation for calculating the Janssen formula pressure is

\begin{align}
K &= 2\mu_{pp}^2 - 2\mu_{pp}\sqrt{\mu_{pp}^2+1}+1 \\
b &= f \rho g \\
p &= {b 2 r \over 4 \mu_{wall}} \left ( 1 - e^{-{4 \mu_{wall} K z \over 2 r}} \right )
\end{align}

where $\mu_{pp}$ is the pebble to pebble static friction coefficient,
$\mu_{wall}$ is the pebble to wall, $f$ is the packing fraction,
$\rho$ is the density, $g$ is the gravitational acceleration, and $z$
is the depth that the pressure is being calculated.  For the Figures
\ref{janssen_0_05_0_15} and \ref{janssen_0_25_0_9}, a packing fraction
of 0.61 is used and a density of 1760 kg/m$^3$ are used.  There are
20,000 pebbles packed into a 0.5 meter radius cylinder, and 1,000 are
recirculated before the pressure measurement is done.

Figure \ref{janssen_0_05_0_15} compares the Janssen model with the
PEBBLES simulation for static friction values of 0.05 and 0.15.  For
this case, the Janssen formula and the simulated pressures match
closely.  Figure \ref{janssen_0_25_0_9} compares these again.  In this
case, the 0.25 $\mu$ values only approximately match, and the 0.9
static friction pressure values do not match at all.  The static
friction slip vectors were examined, and they are not perfectly
vertical, and they are not fully loaded.  This results in the static
friction force being less than the maximum possible, and thus the
pressure is higher since less of the force is removed by the walls.

\myfigures{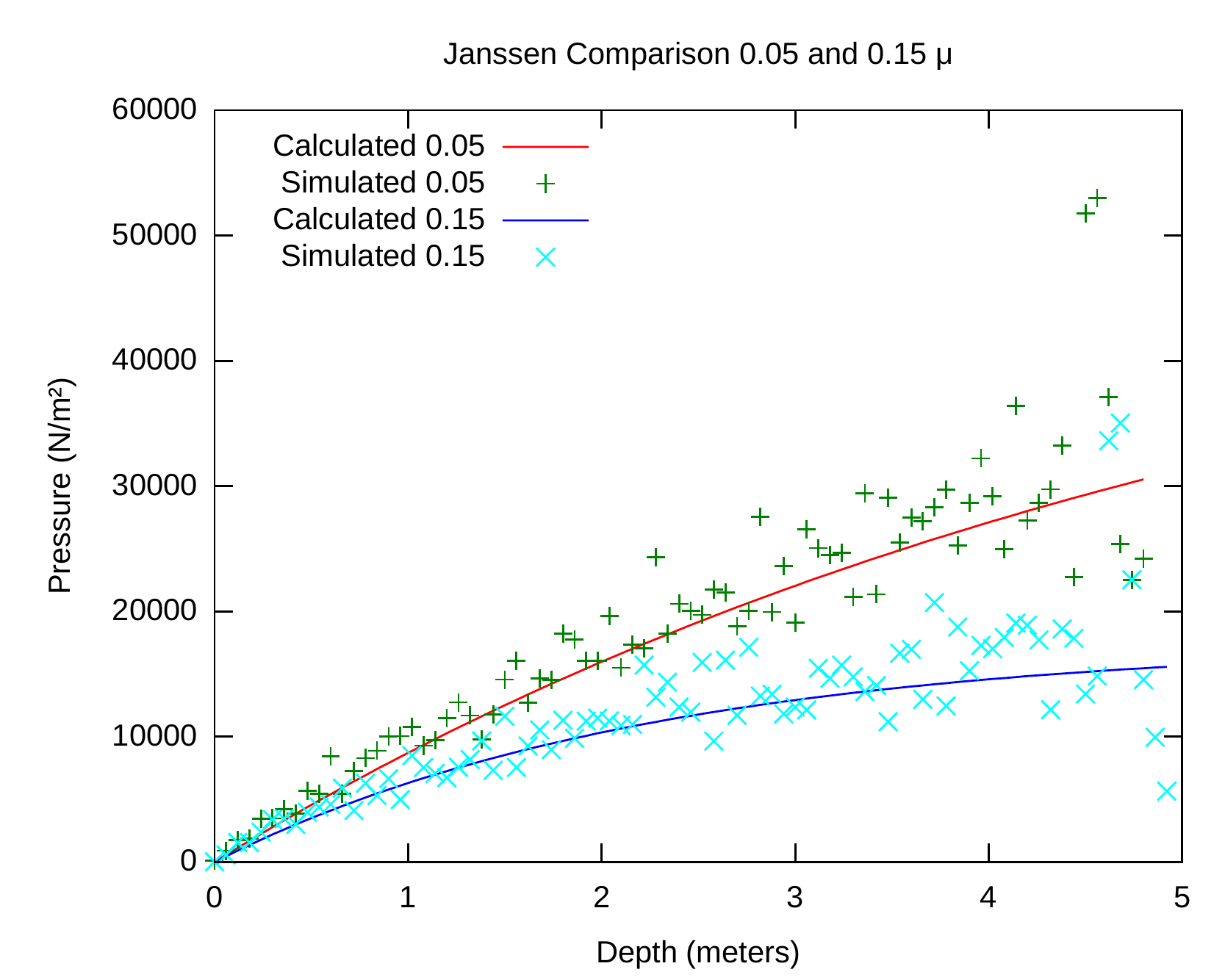}{Comparison With 0.05 and 0.15 $\mu$}{janssen_0_05_0_15}{0.8}
\myfigures{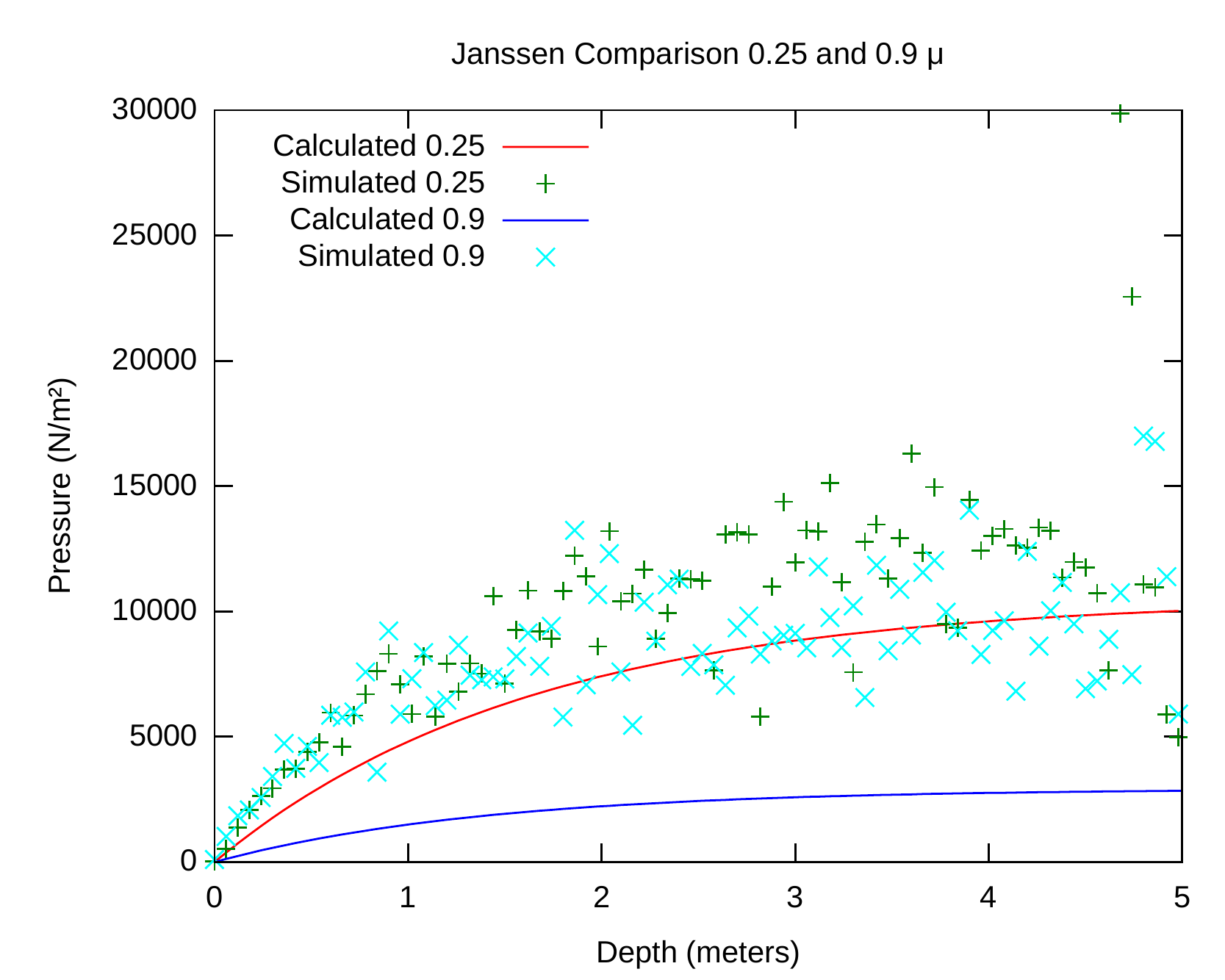}{Comparison With 0.25 and 0.9 $\mu$}{janssen_0_25_0_9}{0.8}

\chapter{Unphysical Approximations}

Modeling the full physical effects that occur in a pebble bed reactor
mechanics is not computationally possible with current computer
resources.  In fact, even modeling all the intermolecular forces that
occur between two pebbles at sufficient levels to reproduce all
macroscopic behavior is probably computationally intractable at the
present time.  This is partially caused by the complexity of effects
such as inter-grain boundaries and small quantities of impurities that
affect the physics and different levels between the atomic effects and
the macroscopic world.  Instead, all attempts at modeling the behavior
of pebble bed reactor mechanics have relied on approximation to make
the task computationally practical.  The PEBBLES simulation has as
high or higher fidelity than past efforts, but it does use multiple
unphysical approximations. This chapter will discuss the
approximations so that future simulation work can be improved, and an
understanding of what limitations exist when applying PEBBLES to
different problems.


In different regions of the reactor, the radioactivity and the fission
will heat the pebbles differently, and the flow of the coolant helium
will distribute this heat around the reactor.  This will change the
temperature of different parts of the reactor.  Since the temperature
will be different, the parameters driving the mechanics of the pebbles
will be different as well.  This includes parameters such as the
static friction coefficients and the size of the pebbles which will
change through thermal expansion.  As well, parameters such as static
friction can also vary depending on the gas in which they currently are in
and in which they were, since some of the gas tends to remain in and
on the carbon surface.  Graphite dust produced by wear may also affect static
friction in downstream portions of the reactor.  


The pebbles in a pebble bed reactor have helium gas flowing around and
past them.  PEBBLES and all other pebble bed simulations ignore
effects of this on pebble movement.  However, the gas will cause both
additional friction when the pebbles are dropping through the reactor,
and the motion of gas will cause additional forces on pebbles.  


Pebble bed mechanics simulations use soft spheres.  Physically, there
will be deflection of spheres under pressure (even the pressure of
just one sphere on the floor), but the true compression is much
smaller than what is actually modeled.  In PEBBLES, the forces are
chosen to keep the compression distance at a millimeter or below.
Another effect related to the physics simulation is that force is
transmitted via contact.  This means the force from one end of the
reactor is transmitted at a speed related to the time-step used for
the simulation, instead of the speed of sound.


Since simulating billions of time-steps is time consuming, two
approximations are made.  First, instead of simulating the physical
time that pebble bed reactors have between pebble additions (on the
order of 2-5 minutes), new pebbles are added at a rate between a
quarter second and two seconds.  This may result in somewhat
unphysical simulations since some vibration that would have dampened
out with a longer time between pebble additions still exists when the
next pebble impacts the bed.  Second, since full recirculation of all
the pebbles is computationally costly, for some simulations, only a
partial recirculation or no recirculation is done.


The physics models do not take into account several physical
phenomena.  The physics do not handle pure spin effects, such as when
two pebbles are contacting and are spinning with an axis around the
contact point.  This should result in forces on the pebbles, but the
physics model does not handle this effect since the contact velocity
is calculated as zero.  In addition, when the pebble is rolling so
that the contact velocity is zero because the pebble's turning axis is
parallel to the surface and at the same rate as the pebble is moving
along the surface, there should be rolling friction, but this effect
is not modeled either.  As well, the equations used assume that the
pebbles are spherically symmetric, but defects in manufacturing and
slight asymmetries in the TRISO particle distribution mean that there
will be small deviations from being truly spherically symmetric.


The physics model does not match classical Hertzian or Mindlin and
Deresiewicz elastic sphere behavior.  The static friction model is a
simplification and does not capture all the hysteretic effects of true
static friction.  Effectively, this means that $h_s$, the coefficient
used to calculate the force from slip, is not a constant. In order to
fully discuss this, some features of these models will be discussed in
the following paragraphs.


Since closed-form expressions exist for elastic contact between
spheres, they will be used, instead of a more general case which lacks
closed-form expressions.  Spheres are not a perfect representation of
the effect of contact between shapes such as a cone and a sphere, but
should give an approximation of the size of the effect of curvature.

The amount of contact area and displacement of distant points for two
spheres or one sphere and one spherical hole (that is negative
curvature) for elastic spheres can be calculated via Hertzian
theory\citep{contactmechanics}.  For two spherical surfaces the
following variables are defined:

\begin{equation}
\frac{1}{R} = \frac{1}{R_1} + \frac{1}{R_2}
\end{equation}

and 

\begin{equation}
\frac{1}{E^*} = \frac{1-\nu_1^2}{E_1} + \frac{1-\nu_2^2}{E_2}
\end{equation}

with $R_i$ the $i$th's sphere's radius, $E_i$ the Young's modulus,
$\nu_i$ the Poisson's ration of the material.  For a concave sphere,
the radius will be negative.  Then, via Hertzian theory, the contact
circle radius will be:

\begin{equation}
a = \left( \frac{3NR}{4E^*} \right)^{1/3}
\end{equation}

where $N$ is the normal force.  The mutual approach of distant points
is given by:

\begin{equation}
\delta = \frac{a^2}{R} = \left( \frac{9N^2}{ 16 R E^{*2}} \right) ^{1/3}
\end{equation}

Notice that the above differs compared to the Hooke's Law formulation
that PEBBLES uses.  The maximum pressure will be:

\begin{equation}
p_0 = \frac{3N}{2\pi a^2}
\end{equation}

So as a function of the radii $R_1$ and $R_2$, the circle radius of the
contact will be:

\begin{equation}
a = \left( \frac{3N}{4E^*} \left[ \frac{1}{R_1} + \frac{1}{R_2} \right]^{-1} \right)^{1/3}
\end{equation}

If $m$ is used for the multiple of negative curvature sphere of the
radius of the other, then the equation becomes:

\begin{equation}
a = \left( \frac{3N}{4E^*} \left[ \frac{1}{R_1} - \frac{1}{m R_1} \right]^{-1} \right)^{1/3}
\end{equation}

which can be rearranged to:

\begin{equation}
a = \left( \frac{3NR_1}{4E^*} \right)^{1/3} \left( 1- \frac{1}{m} \right)^{-1/3}
\end{equation}

From this equation, as $m$ increases, it has less effect on contact
area, so if $R_2$ is much greater than $R_1$, the contact area will
tend to be dominated by $R_1$ rather than $R_2$.  For example, typical
radii in PEBBLES might be 18 cm outlet chute and a 3 cm pebble, which
would put $m$ at 6, so the effect on contact area radius would be
about 33\% difference compared to pebble to pebble contact area
radius, or 6\% compared to a flat surface.\footnote{ Sample values of
  $k = \left( 1- \frac{1}{m} \right)^{-1/3}$: $m = -1, k = 1.26$ for
  sphere to sphere, $m = 6, k = 0.94$ sphere to outlet chute and $m =
  \infty, k = 1$ sphere to flat plane.} 

To some extent, the actual contact area is irrelevant for calculating
the maximum static friction force as long as some conditions are
met. Both surfaces need to be of a uniform material.  The basic
macroscopic description $|F_S| <= \mu |N|$ needs to hold, so changing
the area changes the pressure $P = N/a$, but not the maximum static
friction force.  If the smaller area causes the pressure to increase
enough to cause plastic rather than elastic contact, then through that
mechanism, the contact area would cause actual differences in
experimental values. PEBBLES also does not calculate plastic contact
effects.

The contact area causes an effect through another mechanism.  The
tangential compliance in the case of constant normal and increasing
tangential force, that is the slope of the curve relating displacement
to tangential force, is given in Mindlin and Deresiewicz as:

\begin{equation}
\frac{2-\nu}{8\mu a}
\end{equation}

Since the contact area radius, $a$, is a function of curvature, the
slope of the tangential compliance will be as well, which is another
effect that PEBBLES' constant $h_s$ does not capture.

In summary for the static friction using a constant coefficient for
$h_s$ yields two different approximations. First, using the same
constants for wall contact when there is different curvatures is an
approximation that will give somewhat inconsistent results.  Since the
equations for spherical contact are dominated by the smaller radius
object, this effect is somewhat less but still exists.  Second, using
the same constant coefficient for different loading histories is a
approximation. For a higher fidelity, these effects need to be taken
into account.

\chapter{Code Speedup and Parallelization}


Planned and existing pebble bed reactors can have on the order of
100,000 pebbles.  For some simulations, these pebbles need to be
followed for long time periods, which can require computing billions
of time-steps.  Multiplying the time-steps required by the number of
pebbles being computed over leads to the conclusion that large numbers
of computations are required.  These computations should be as fast as
possible, and should be as parallel as possible, so as to allow
relevant calculations to be done in a reasonable amount of time.  This
chapter discusses the process of speeding up the code and
parallelizing it.

The PEBBLES program has three major portions of calculation.  The
first is determining which pebbles are in contact with other
pebbles. The second computational part is determining the time
derivatives for all the vectors for all the pebbles.  The third
computational part is using the derivatives to update the
values. Overall, for calculation of a single time-step, the
algorithm's computation time is linearly proportional to the number of
pebbles, that is O(n)\footnote{O(n): The algorithm scales linearly (n)
  with increasing input size.  So if it runs with 100 pebbles it takes
  roughly 10 times as long as when it runs it only 10 pebbles.  Or if
  it goes from 10 pebbles to 20 pebbles it will take twice as long to
  run. }.

\section{General Information about profiling}

There are four different generic parts of the complete calculation
that need to considered for determining the overall speed.  The first
consideration is the time to compute arithmetic operations.  Modern
processors can complete arithmetic operations in nanoseconds or
fractions of nanoseconds.  In the PEBBLES code, the amount of time
spent on arithmetic is practically undetectable in wall clock changes.
The second consideration is the time required for reading memory and
writing memory.  For main memory accesses, this takes hundreds of CPU
clock cycles, so these times are on the order of fractions
of microseconds \citep{drepper_memory}.  Because of the time required to
access main memory, all modern CPUs have on-chip caches, that contain
a copy of the recently used data. If the memory access is in the
CPU's cache, the data can be retrieved and written in a small number
of CPU cycles.  Main memory writes are somewhat more expensive than
main memory reads, since any copies of the memory that exist in other
processor's caches need to be updated or invalidated.  So for a
typical calculation like $a + b \to c$ the time spent doing the
arithmetic is trivial compared to the time spent reading in $a$ and
$b$ and writing out $c$.

The third consideration is the amount of time required for parallel
programming constructs.  Various parallel synchronization tools such
as atomic operations, locks and critical sections take time.  These
take an amount of time on the same order of magnitude as memory
writes. However, they typically need a read and then a write without
any other processor being able to access that chunk of memory in
between which requires additional overhead, and a possible wait if the
memory address is being used by another process.  Atomic operations on
x86\_64 architectures are faster than using locks, and locks are
generally faster than using critical sections. The fourth
consideration is network time.  Sending and receiving a value can
easily take over a millisecond for the round trip time.  These four
time consuming operations need to be considered when choosing
algorithms and methods of calculation.

There are a variety of methods for profiling the computer code.  The
simplest method is to use the FORTRAN 95 intrinsics \texttt{CPU\_TIME}
and \texttt{DATE\_\-AND\_\-TIME}.  The \texttt{CPU\_TIME} subroutine
returns a real number of seconds of CPU time.  The
\texttt{DATE\_AND\_TIME} subroutine returns the current wall clock
time in the \texttt{VALUES} argument.  With \texttt{gfortran} both
these times are accurate to at least a millisecond.  The difference
between two different calls of these functions provide information on
both the wall clock time and the CPU time between the calls.  (For the
\texttt{DATE\_AND\_TIME} subroutine, it is easiest if the days, hours,
minutes, seconds and milliseconds are converted to a real seconds past
some arbitrary time.)  The time methods provide basic information and
a good starting point for determining which parts of the program are
consuming time.  For more detailed profiling the
\texttt{oprofile} \citep{oprofile} program can be used on Linux.  This
program can provide data at the assembly language level which makes it
possible to determine which part of a complex function is consuming
the time.  Non-assembly language profilers are difficult to accurately
use on optimized code, and profiling non-optimized code is
misrepresentative.

\section{Overview of Parallel Architectures and Coding}

Parallel computers can be arranged in a variety of ways. Because of
the expense of linking shared memory to all processors, a common
architecture is a cluster of nodes with each node having multiple
processors.  Each node is linked to other nodes via a fast network
connection.  The processors on a single node share memory.  Figure
\ref{cluster_architecture} shows this arrangement.  For this
arrangement, the code can use both the OpenMP (Open Multi-Processing)
\citep{openmp} and the MPI (Message Passing Interface) \citep{mpi}
libraries.  MPI is a programming interface for transferring data
across a network to other nodes.  OpenMP is a shared memory
programming interface.  By using both programming interfaces high
speed shared memory accesses can be used on memory shared on the node
and the code can be parallelized across multiple nodes.

\myfigures{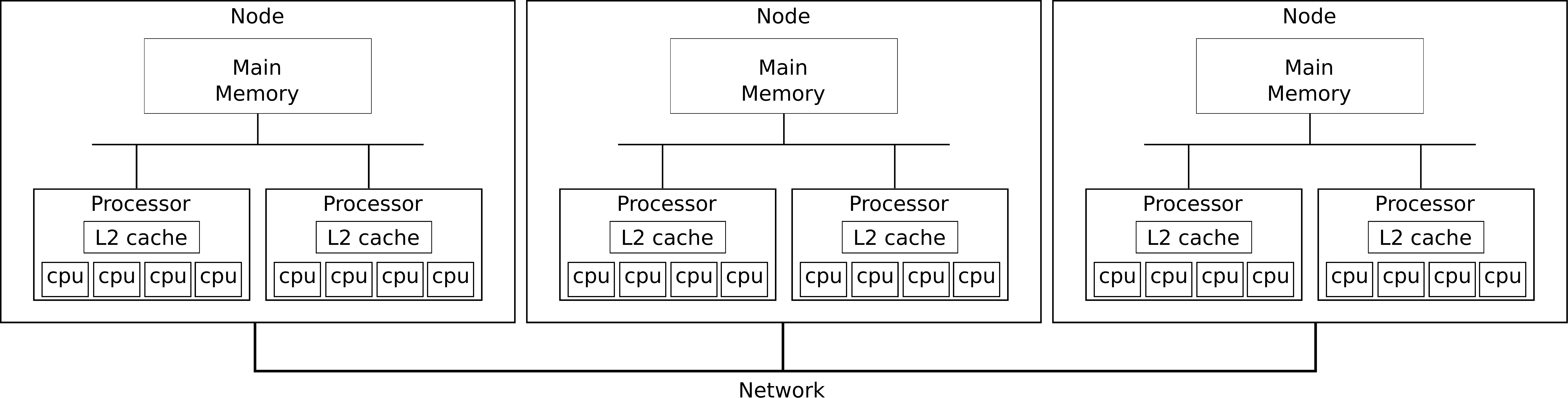}{Sample Cluster Architecture}{cluster_architecture}{0.15}

\section{Lock-less Parallel O(N) Collision Detection}


For any granular material simulation, which particles are in contact
must be determined quickly and accurately for each time-step.  This is
called collision detection, though for pebble simulations it might be
more accurately labeled contact detection.  The simplest algorithm for
collision detection is to iterate over all the other objects and
compare each one to the current object for collision.  To determine
all the collisions using that method, $O(N^2)$ time is required.

An improved algorithm by \citet{cohen_collide} uses six sorted lists
of the lower and upper bounds for each object. (There is one upper
bound list and one lower bound list for each dimension.)  With this
algorithm, to determine the collisions for a given object, the bounds
of the current objects are compared to bounds in the list---only
objects that overlap the bounds in all three dimensions will
potentially collide.  This algorithm typically has approximately $O(N
\log(N))$ time,\footnote{In order from slowest to fastest (for
  sufficiently big N): $O(N^2)$,$O(N\log(N)$,$O(N)$,$O(1)$.} because
of the sorting of the bounding lists \citep{cohen_collide}.

A third, faster method, grid collision detection, is available if the
following requirements hold: (1) there is a maximum diameter of
object, and no object exceeds this diameter, and (2) for a given
volume, there is a reasonably small, finite, maximum number of objects
that could ever be in that volume.  These two constraints are easily
satisfied by pebble bed simulations, since the pebbles are effectively
the same size (small changes in diameter occur due to wear and thermal
effects).  A three-dimensional parallelepiped grid is used over the
entire range in which the pebbles are simulated.  The grid spacing
$gs$ is set at the maximum diameter of any object (twice the maximum
radius for spheres).

Two key variables are initialized, $grid\_count(x,y,z)$, the number of
pebbles in grid locations x,y,z; and $grid\_ids(x,y,z,i)$, the pebble
identification numbers ($ids$) for each x,y,z location.  The $id$ is a
unique number assigned to each pebble in the simulation.  The spacing
between successive grid indexes is $gs$, so the index of a given x
location can be determined by $\lfloor (x-x_{min})/gs \rfloor$ where
$x_{min}$ is the zero x index's floor; similar formulas are used for y
and z.

The grid is initialized by setting $grid\_count(:,:,:) = 0$, and then
the x,y,z indexes are determined for each pebble.  The $grid\_count$
at that location is then atomically\footnote{In this chapter, atomic
  means uncutable, that is the entire operation is done in one action
  without interference from other processors.} incremented by one and
fetched. Because OpenMP 3.0 does not have a atomic add-and-fetch, the
\texttt{lock xadd} x86\_64 assembly language instruction is put in a
function.  The $grid\_count$ provides the fourth index into the
$grid\_ids$ array, so the pebble $id$ can be stored into the $ids$
array.  The amount of time to zero the $grid\_count$ array is a
function of the volume of space, which is proportional to the number
of pebbles.  The initialization iteration over the pebbles can be done
in parallel because of the use of an atomic add-and-fetch function.
Updating the grid iterates over the entire list of pebbles so the full
algorithm for updating the grid is $O(N)$ for the number of pebbles.

Once the grid is updated, the nearby pebbles can be quickly
determined.  Figure \ref{grid_nearby_pebbles} illustrates the general
process.  First, index values are computed from the pebble and used to
generate $xc$, $yc$, and $zc$. This finds the center grid location,
which is shown as the bold box in the figure.  Then, all the possible
pebble collisions must have grid locations (that is, their centers are
in the grid locations) in the dashed box, which can be found by
iterating over the grid locations from $xc-1$ to $xc+1$ and repeating
for the other two dimensions.  There are $3^3$ grid locations to
check, and the number of pebbles in them are bounded (maximum 8), so
the time to do this is bounded.  Since this search does not change any
grid values, it can be done in parallel without any locks.

\myfigures{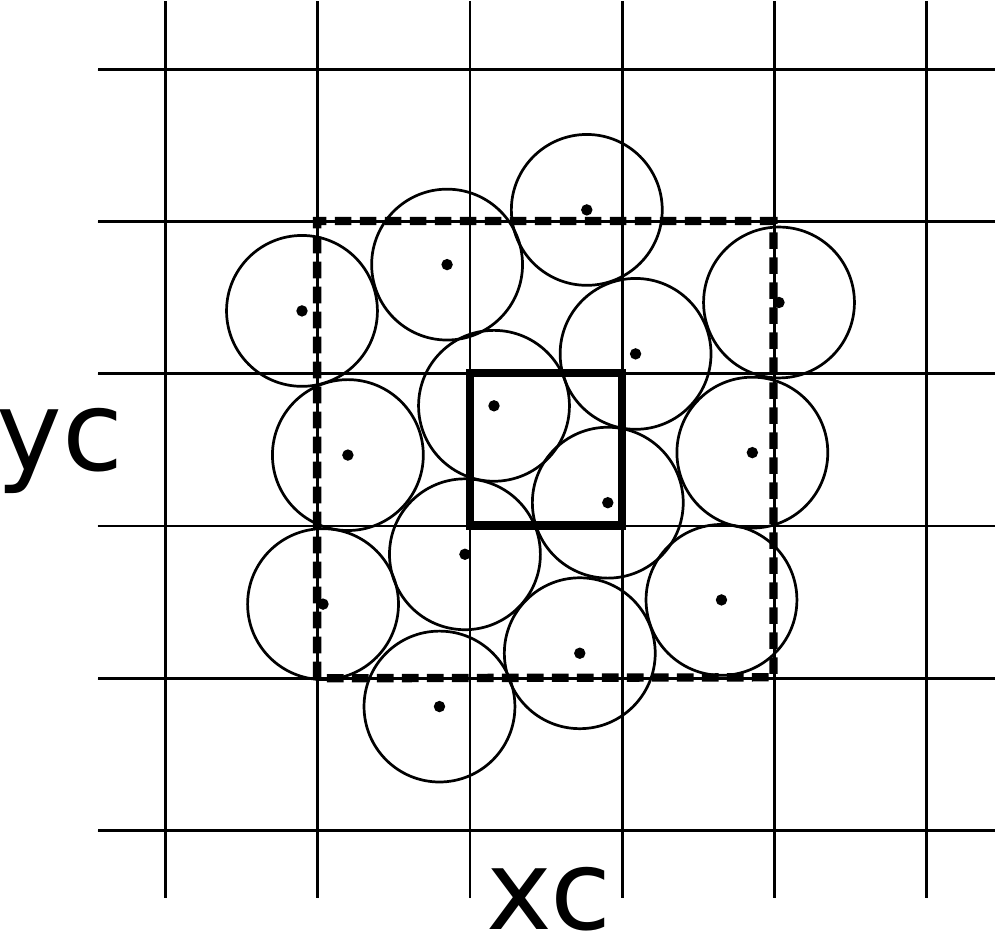}{Determining Nearby Pebbles from Grid}{grid_nearby_pebbles}{0.45}

Therefore, because of the unique features of pebble bed pebbles
simulation, a parallel lock-less $O(N)$ algorithm for determining the
pebbles in contact can be created.

\section{MPI Speedup}

The PEBBLES code uses MPI to distribute the computational work across
different nodes. The MPI/OpenMP hybrid parallelization splits the
calculation of the derivatives and the new variables geometrically and
passes the data at the geometry boundaries between nodes using
messages. Each pebble has a primary node and may also have various
boundary nodes. The pebble-primary-node is responsible for
updating the pebble position, velocity, angular velocity, and
slips. The pebble-primary-node also sends data about the pebble to
any nodes that are the pebble boundary nodes and will transfer the
pebble to a different node if the pebble crosses the geometric
boundary of the node. Boundary pebbles are those close enough to a
boundary that their data needs to be present in multiple nodes so that
the node's primary pebbles can be properly updated. Node 0 is the
master node and does processing that is simplest to do on one node,
such as writing restart data to disk and initializing the pebble
data. The following steps are used for initializing the nodes and then
transferring data between them:

\begin{enumerate}
\item Node 0 calculates or loads initial positions of pebbles.
\item Node 0 creates the initial domain to node mapping.
\item Node 0 sends domain to node mapping to other nodes.
\item Node 0 sends other nodes their needed pebble data.
\end{enumerate}

Order of calculation and data transfers in main loop: 

\begin{enumerate}
\item Calculate derivatives for node primary and boundary pebbles.
\item Apply derivatives to node primary pebble data.
\item For every primary pebble, check with the domain module to
  determine the current primary node and any boundary nodes.
\begin{enumerate}
\item If the pebble now has a different primary node, add the pebble id to the
transfer list to send to the new primary node.
\item If the pebble has any boundary nodes, add the pebble id to the
  boundary send list to send it to the node for which it is a
  boundary.
\end{enumerate}
\item If this is a time step where Node 0 needs all the pebble data
  (such as when restart data is being written), add all the primary
  pebbles to the Node 0 boundary send list.
\item Send the number of transfers and the number of boundary sends that this
node has to all the other nodes using buffered sends.
\item Initialize three Boolean lists of other nodes that this node has:
\begin{enumerate}
\item data-to-send-to with ``true'' if the number of transfers or boundary
  sends is nonzero, and ``false'' otherwise
\item received-data-from to ``false''
\item received-the-number-of-transfers and the
  number-of-boundary-sends with ``false.''
\end{enumerate}
\item While this node has data to send to other nodes and other nodes have
data to send to this node loop:
\begin{enumerate}
\item Probe to see if any nodes that this node needs data from have data
available.
\begin{enumerate}
\item If yes, then receive the data and update pebble data and the Boolean
lists as appropriate
\end{enumerate}
\item If there are any nodes that this node has data to send to, and
  this node has received the number of transfers and boundary sends
  from, then send the data to those nodes and update the Boolean data
  send list for those nodes.
\end{enumerate}
\item Flush the network buffers so any remaining data gets received.
\item Node 0 calculates needed tallies.
\item If this is a time to rebalance the execution load:
\begin{enumerate}
\item Send wall clock time spent computing since last rebalancing to node 0
\item Node 0 uses information to adjust geometric boundaries to move work
towards nodes with low computation time and away from nodes with high
computation time
\item Node 0 sends new boundary information to other nodes, and needed data to
other nodes.
\end{enumerate}
\item Continue to next time step and repeat this process until all
  time-steps have been run.
\end{enumerate}

All the information and subroutines needed to calculate the primary
and boundary nodes that a pebble belongs to are calculated and stored
in a FORTRAN 95 module named \texttt{network\_domain\_module}. The
module uses two derived types: \texttt{network\_domain\_type} and
\texttt{network\_domain\_location\_type}. Both types have no public
components so the implementation of the domain calculation and the
location information can be changed without changing anything but the
module, and the internals of the module can be changed without
changing the rest of the PEBBLES code.  The location type stores the
primary node and the boundary nodes of a pebble. The module contains
subroutines for determining the location type of a pebble based on its
position, primary and boundary nodes for a location type, and
subroutines for initialization, load balancing, and transferring of
domain information over the network.  

The current method of dividing the nodes into geometric domains uses a
list of boundaries between the z (axial) locations. This list is
searched via binary search to find the nodes nearest to the pebble
position, as well as those within the boundary layer distance above
and below the zone interface in order to identify all the boundary
nodes that participate in data transfers. The location type resulting
from this is cached on a fine grid, and the cached value is returned
when the location type data is needed. The module contains a
subroutine that takes a work parameter (typically, the computation
time of each of the nodes) and can redistribute the z boundaries up or
down to shift work towards nodes that are taking less time computing
their share of information.  If needed in the future, the z-only
method of dividing the geometry could be replaced with a full 3-D
version by modifying the network domain module.

\section{OpenMP Speedup}

The PEBBLES code uses OpenMP to distribute the calculation over
multiple processes on a single node. OpenMP allows directives to be
given to the compiler that direct how portions of code are to be
parallelized. This allows a single piece of code to be used for both
the single processor version and the OpenMP version. The PEBBLES
parallelization typically uses OpenMP directives to cause loops that
iterate over all the pebbles to be run in parallel. 

Some details need to be taken into consideration for the
parallelization of the calculation of acceleration and torque. The
physical accelerations imposed by the wall are treated in parallel,
and there is no problem with writing over the data because each
processor is assigned a portion of the total zone inventory of
pebbles. For calculating the pebble-to-pebble forces, each processor
is assigned a fraction of the pebbles, but there is a possibility of
the force addition computation overwriting another calculation because
the forces on a pair of pebbles are calculated and then the calculated
force is added to the force on each pebble. In this case, it is
possible for one processor to read the current force from memory and
add the new force from the pebble pair while another processor is
reading the current force from memory and adding its new force to that
value; they could both then write back the values they have
computed. This would be incorrect because each calculation has only
added one of the new pebble pair forces. Instead, PEBBLES uses an
OpenMP ATOMIC directive to force the addition to be performed
atomically, thereby guaranteeing that the addition uses the latest
value of the force sum and saves it before a different processor has a
chance to read it. 

For calculating the sum of the derivatives using Euler's method,
updating concurrently poses no problem because each individual pebble
has derivatives calculated. The data structure for storing the
pebble-to-pebble slips (sums of forces used to calculate static
friction) is similar to the data structure used for the collision
detection grid. A 2-D array exists where one index is the from-pebble
and the other index is for storing $ids$ of the pebbles that have slip
with the first pebble. A second array exists that contains the number
of $ids$ stored, and that number is always added and fetched
atomically, which allows the slip data to be updated by multiple
processors at once. These combine to allow the program to run
efficiently on shared memory architectures.

\section{Checking the Parallelization}

The parallelization of the algorithm is checked by running the test
case with a short number of time steps (10 to 100).  Various summary
data are checked to make sure that they match the values computed with
the single processor version and between different numbers of nodes
and processors.  For example, with the NGNP-600 model used in the
results section, the average overlap of pebbles at the start of the
run is 9.665281e-5 meters. The single processor average overlap at the
end of the 100 time-step run is 9.693057e-5 meters, the 2 nodes
average overlap is 9.693043e-5 meters, and the 12 node average overlap
is 9.693029e-5 meters.  The lower order numbers change from run to
run. The start-of-run values match each other exactly, and the
end-of-run values match the start of run values to two significant
figures. However, the three different end-of-run values match to five
significant digits.  In short, the end values match each other more
than they match the start values. The overlap is very sensitive to
small changes in the calculation because it is a function of the
difference between two positions.  During coding, multiple defects
were found and corrected by checking that the overlaps matched closely
enough between the single processor calculation and the multiple
processor calculations.  The total energy or the linear energy or
other computations can be used similarly since the lower significant
digits also change frequently and are computed over all the pebbles.


\section{Results}

The data in Table \ref{openmp_results} and Table \ref{mpi_results}
provide information on the time used with the current version of
PEBBLES for running 80 simulation time steps on two models. The
NGNP-600 model has 480,000 pebbles. The AVR model contains 100,000
pebbles. All times are reported in units of wall-clock seconds. The
single processor NGNP-600 model took 251 seconds and the AVR
single processor model took 48 seconds when running the current
version.  The timing runs were carried out on a cluster with two Intel
Xeon X5355 2.66 GHz processors per node with a DDR 4X InfiniBand
interconnect network.  The nodes had 8 processors per node.  The
gfortran 4.3 compiler was used.

\begin{table}[!htb]
  \centering
  \caption{\bf OpenMP speedup results}
  \label{openmp_results}
  \begin{tabular}{c|c|c|c|c|c|c}
    Processes	&AVR &Speedup	&Efficiency	&NGNP-600&Speedup	&Efficiency\\
\hline
1	&47.884	&1	&100.00\%	&251.054	&1	&100.00\%\\
1	&53.422	&0.89633	&89.63\%	&276.035	&0.90950	&90.95\%\\
2	&29.527	&1.6217	&81.09\%	&152.479	&1.6465	&82.32\%\\
3	&21.312	&2.2468	&74.89\%	&104.119	&2.4112	&80.37\%\\
4	&16.660	&2.8742	&71.85\%	&80.375	&3.1235	&78.09\%\\
5	&13.884	&3.4489	&68.98\%	&68.609	&3.6592	&73.18\%\\
6	&12.012	&3.98635	&66.44\%	&61.168	&4.1043	&68.41\%\\
7	&10.698	&4.4760	&63.94\%	&54.011	&4.6482	&66.40\%\\
8	&9.530	&5.0246	&62.81\%	&49.171	&5.1057	&63.82\%\\
  \end{tabular}
\end{table}

\begin{table}[!htb]
  \centering
  \caption{\bf MPI/OpenMP speedup results}
  \label{mpi_results}
{\small
  \begin{tabular}{c|c|c|c|c|c|c|c}
Nodes	&Procs	&AVR &Speedup	&Efficiency	&NGNP-600 &Speedup	&Efficiency\\
\hline
1	&1	&47.884	&1	&100.00\%	&251.054	&1	&100.00\%\\
1	&8	&10.696	&4.4768	&55.96\%	&55.723	&4.5054	&56.32\%\\
2	&16	&6.202	&7.7207	&48.25\%	&30.642	&8.1931	&51.21\%\\
3	&24	&4.874	&9.8244	&40.93\%	&23.362	&10.746	&44.78\%\\
4	&32	&3.935	&12.169	&38.03\%	&17.841	&14.072	&43.97\%\\
5	&40	&3.746	&12.783	&31.96\%	&16.653	&15.076	&37.69\%\\
6	&48	&3.534	&13.550	&28.23\%	&15.928	&15.762	&32.84\%\\
7	&56	&3.285	&14.577	&26.03\%	&15.430	&16.271	&29.05\%\\
8	&64	&2.743	&17.457	&27.28\%	&11.688	&21.480	&33.56\%\\
9	&72	&2.669	&17.941	&24.92\%	&11.570	&21.699	&30.14\%\\
10	&80	&2.657	&18.022	&22.53\%	&11.322	&22.174	&27.72\%\\
11	&88	&2.597	&18.438	&20.95\%	&11.029	&22.763	&25.87\%\\
12	&96	&2.660	&18.002	&18.75\%	&11.537	&21.761	&22.67\%\\
  \end{tabular}
}
\end{table}



Significant speedups have resulted with both the OpenMP and MPI\-/Open\-MP
versions.  A basic time step for the NGNP-600 model went from 3.138
seconds to 146 milliseconds when running on 64 processors.  Since a
full recirculation would take on the order of 1.6e9 time steps, the
wall clock time for running a full recirculation simulation has gone
from about 160 years to a little over 7 years.  For smaller simulation
tasks, such as simulating the motion of the pebbles in a pebble bed
reactor during an earthquake, the times are more reasonable, taking
about 5e5 time steps. Thus, for the NGNP-600 model, a full earthquake
can be simulated in about 20 hours when using 64 processors.  For the
smaller AVR model, the basic time step takes about 34 milliseconds
when using 64 processors.  Since there are less pebbles to
recirculate, a full recirculation would take on the order of 2.5e8
time steps, or about 98 days of wall clock time.

\chapter{Applications}


The knowledge of the packing and flow patterns (and to a much lesser
extent the position) of pebbles in the pebble bed reactor is an
essential prerequisite for many in-core fuel cycle design activities
as well as for safety assessment studies.  Three applications have
been done with the PEBBLES code.  The major application is the
computation of pebble positions during a simulated earthquake.  Two
other applications that have been done are calculation of space
dependent Dancoff factors and calculation of the angle of repose for a
HTR-10 simulation.


\myfigures{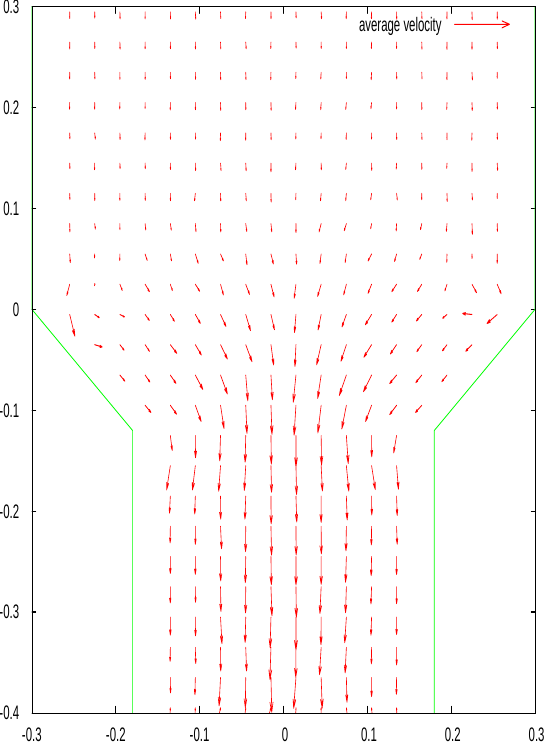}{Flow Field Representation (arrow lengths are proportional to local average pebble velocity)}{flow_lines}{1.0}



\section{Applications in Support of Reactor Physics}

\subsection{Dancoff Factors}
\label{dancoff_factors_ss}

The calculation of Dancoff factors is an example application that
needs accurate pebble position data.  The Dancoff factor is used for
adjusting the resonance escape probability for neutrons.  There are
two Dancoff factors that use pebble position data.  The first is the
inter-pebble Dancoff factor that is the probability that a neutron
escaping from the fuel zone of a pebble crosses a fuel particle in
another pebble.  The second is the pebble-pebble Dancoff factor, which
is the probability that a neutron escaping one fuel zone will enter
another fuel zone without interacting with a moderator nuclide.
\citet{KloostermanandOugouag2005} use pebble location information to
calculate the probability by ray tracing from fuel lumps until another
is hit or the ray escapes the reactor.  The PEBBLES code has been used
for providing position information to  J. L. Kloosterman and
A. M. Ougouag's PEBDAN program.  This program calculate these factors as
shown in Figure \ref{dancoff_factors} which calculates them for the
AVR reactor model. 

\myfigures{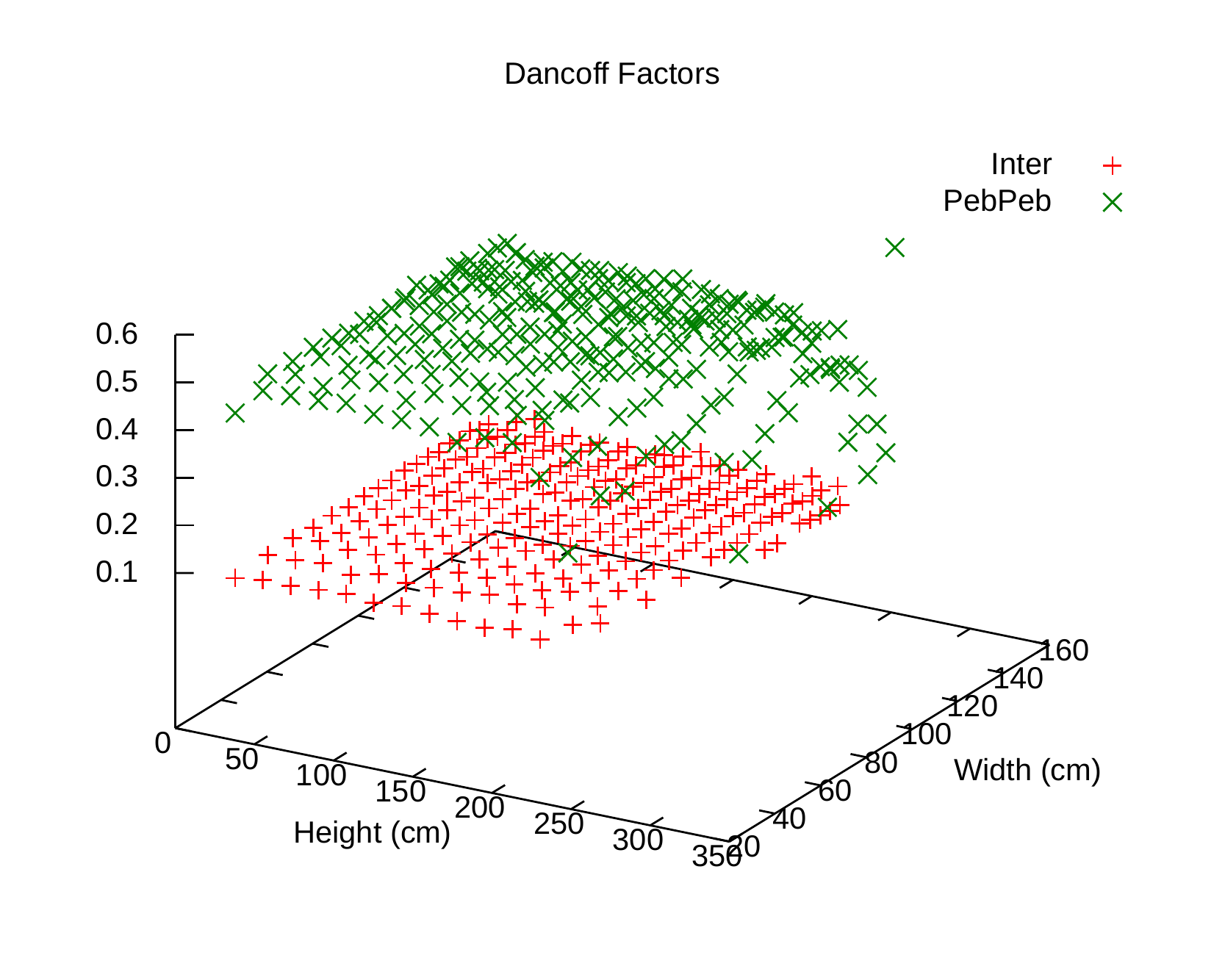}{Dancoff Factors for AVR}{dancoff_factors}{0.8}
\myfigures{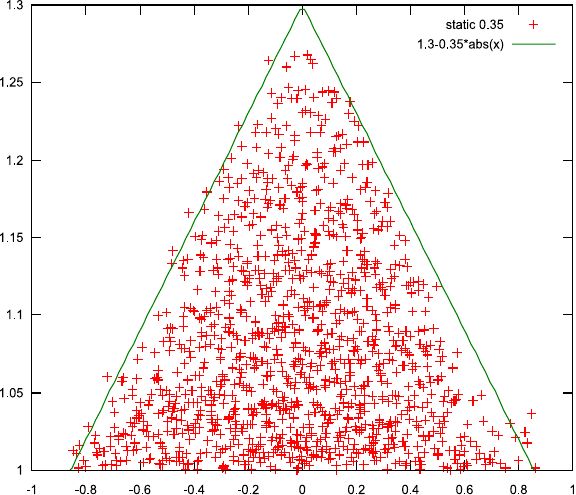}{2-D Projection of Pebble Cone on HTR-10 (crosses represent centers of pebbles)}{cone_angle}{1.0}

\subsection{Angle of Repose}
\label{angle_of_repose}

The PEBBLES code was used for calculating the angle of repose for an
analysis of the HTR-10 first criticality \citep{Terryetal2006}.  The
pebble bed code recirculated pebbles to determine the angle at which
the pebbles would stack at the top of the reactor as shown in Figure
\ref{cone_angle}, since this information was not provided, but was
needed for the simulation of the reactor\citep{AVR1990}.

\subsection{Pebble Ordering with Recirculation}

During experimental work before the construction of the AVR, it was
discovered that when the pebbles were recirculated, the ordering in
the pebbles increased.  Figures \ref{before_recirc} and
\ref{after_recirc} show that this effect occurs in the PEBBLES
simulation as well.  The final AVR design incorporated indentations in
the wall to prevent this from occurring.

\myfigures{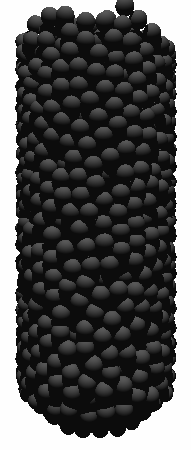}{Pebbles Before Recirculation}{before_recirc}{0.5}
\myfigures{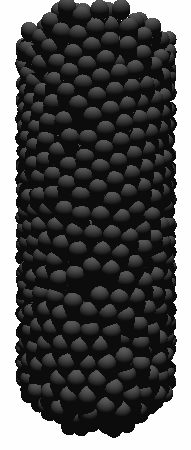}{Pebbles After Recirculation}{after_recirc}{0.5}




\section{Application to Earthquake modeling}
\label{earthquake}

The packing fraction of the pebbles in a pebble bed reactor can vary
depending on the method of packing and the subsequent history of the
packing.  This packing fraction can affect the neutronics behavior of
the reactor, since it translates into an effective fuel density.
During normal operation, the packing fraction will vary only slowly,
over the course of weeks and then stabilize. During an earthquake,
the packing fraction can increase suddenly.  This packing fraction
change is a concern since packing fraction increase can increase the
neutron multiplication and cause criticality concerns as shown by
\citet{OugouagAndTerry2001}.

The PEBBLES code can simulate this increase and determine the rate of
change and the expected final packing fraction, thus allowing the
effect of an earthquake to be simulated.

\subsection{Movement of Earthquakes}

The movement of earthquakes has been well studied in the past. The
magnitude of the motion of earthquakes is described by the Mercalli
scale, which describes the maximum acceleration that a given
earthquake will impart to structures. For a Mercalli X earthquake, the
maximum acceleration is about 1 g. The more familiar Richter scale
measures the total energy release of an earthquake
\citep{Lamarsh1983}, which is not useful for determining the effect on
a pebble bed core. For a given location, the soil properties can be
measured, and using soil data and the motion that the bedrock will
undergo, the motion on the surface can be simulated. The INL site had
this information generated in order to determine the motion from the
worst earthquake that could be expected over a 10,000 years period
\citep{Payne2003}. This earthquake has roughly a Mercalli IX
intensity. The data for such a 10,000 year earthquake are used for the
simulation in this dissertation.

\subsection{Method Of Simulation}

The code simulates earthquakes by adding a displacement to the walls
of the reactor.  As well, the velocity of the walls needs to be
calculated.  The displacement in the simulation can be specified
either as the sum of sine waves, or as a table of displacements that
specifies the x, y, and z displacements for each time. At each time
step both the displacement and the velocity of the displacement are
calculated.  When the displacement is calculated by a sum of sine
functions, the current displacement is calculated by adding vector
direction for each wave and the velocity is calculated from the sum of
the first derivative of all the waves. When the displacement is
calculated from a table of data, the current displacement is a linear
interpolation of the two nearest data points in the table, and the
velocity is the slope between them.  The walls are then assigned the
appropriate computed displacement and velocity. Figure
\ref{total_displacement} shows the total displacement for the INL
earthquake simulation specifications that were used in this paper.

\myfigures{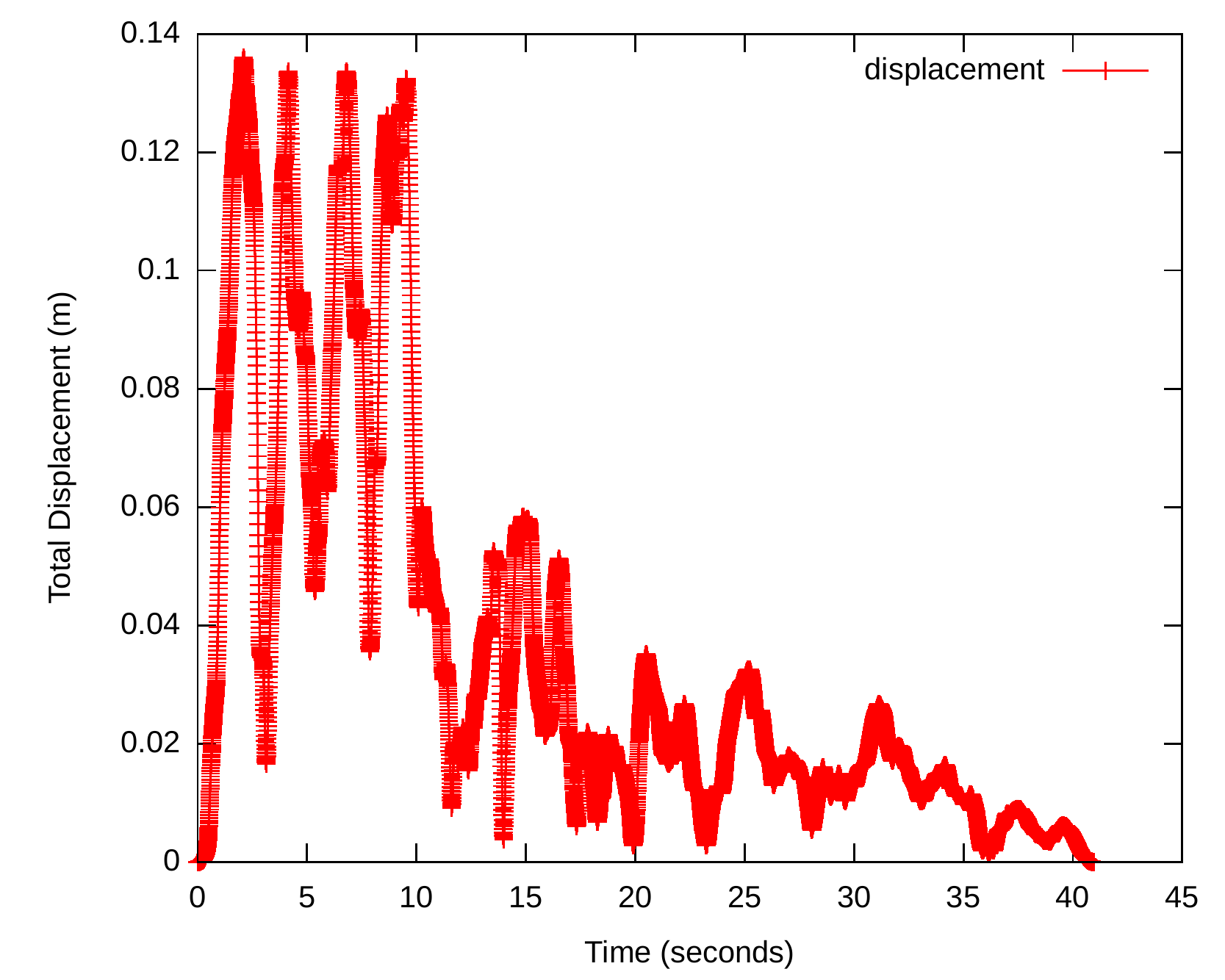}{Total Earthquake Displacement}{total_displacement}{0.8}

\subsection{Earthquake Results}

The results of two simulations carried out here show a substantially
safer behavior than the \citet{OugouagAndTerry2001} bounding
calculations.  The methodology was applied to a model of the PBMR-400
model and two different static friction coefficients were tested, 0.65
and 0.35.  The packing fraction increased from 0.593 to 0.594 over the
course of the earthquake with the 0.65 static friction model, with the
fastest increase was from 0.59307 to 0.59356 and took place over 0.8
seconds. With the 0.35 static friction model, the overall increase was
from 0.599 to 0.601.  The fasted increase was from 0.59964 to 0.60011
in 0.8 seconds.  This is remarkably small when compared to the
bounding calculation packing fraction increase rate of 0.129
sec$^{-1}$ in free fall.\footnote{The free fall rate is determined by
  calculating the packing fraction increase if the pebbles were in
  gravitational free fall.} Both computed increases and packing
fraction change rates are substantially below the free fall bounding
rate and packing fraction change of a transition from 0.60 to 0.64 in
0.31 seconds.  The computed rate and the total packing fraction
increase are in the range that can be offset by thermal feedback
effects for uranium fueled reactors.

\myfigures{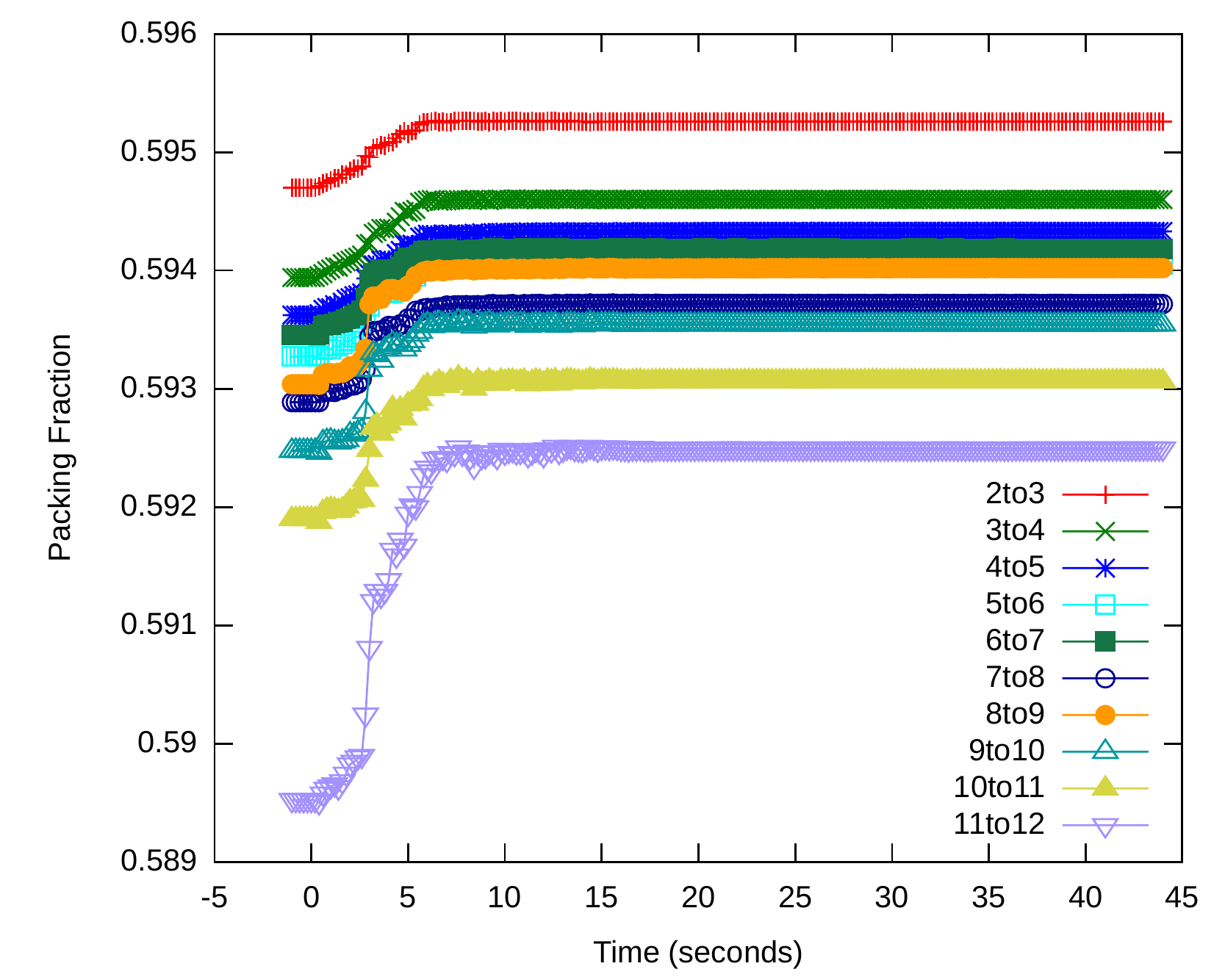}{0.65 Static Friction Packing over Time}{0_65_packing}{0.8}
\myfigures{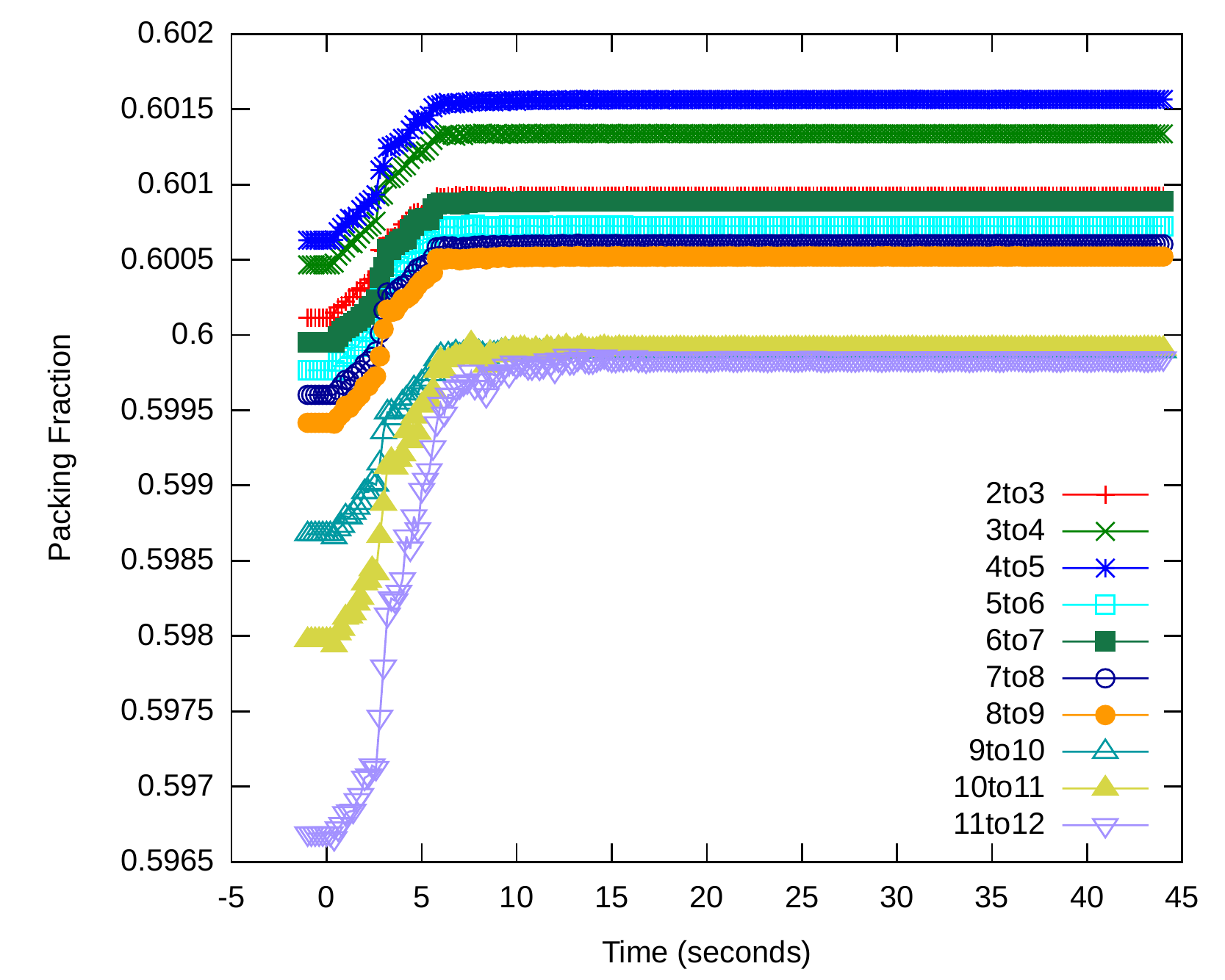}{0.35 Static Friction Packing over Time}{0_35_packing}{0.8}

During the course of the earthquake, the boundary density fluctuations
(that is the oscillations in packing fraction near a boundary) are
observed to increase in amplitude. Figure \ref{radial_eq_packings}
shows the packing fraction before the earthquake and after the
earthquake in the radial direction. These were taken from 4 to 8
meters above the fuel outlet chute in the PBMR-400 model. All the
radial locations have increased packing compared to the packing
fraction before the earthquake, but the points that are at boundary
density fluctuation peaks increase the most. This effect can be seen
in figure \ref{radial_eq_packing_delta}, which shows the increase in
packing fraction before the earthquake and after

\myfigures{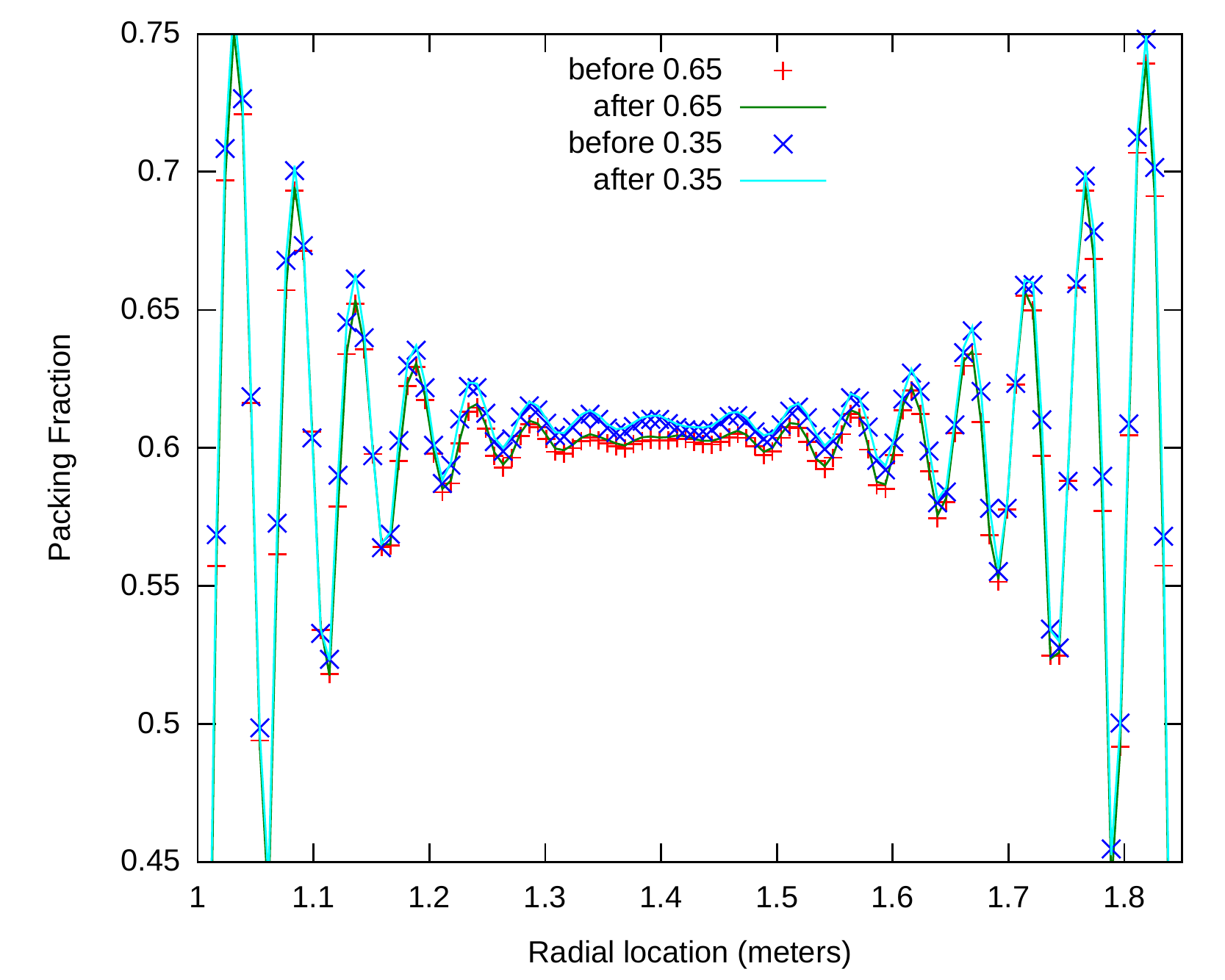}{Different Radial Packing Fractions}{radial_eq_packings}{0.8}
\myfigures{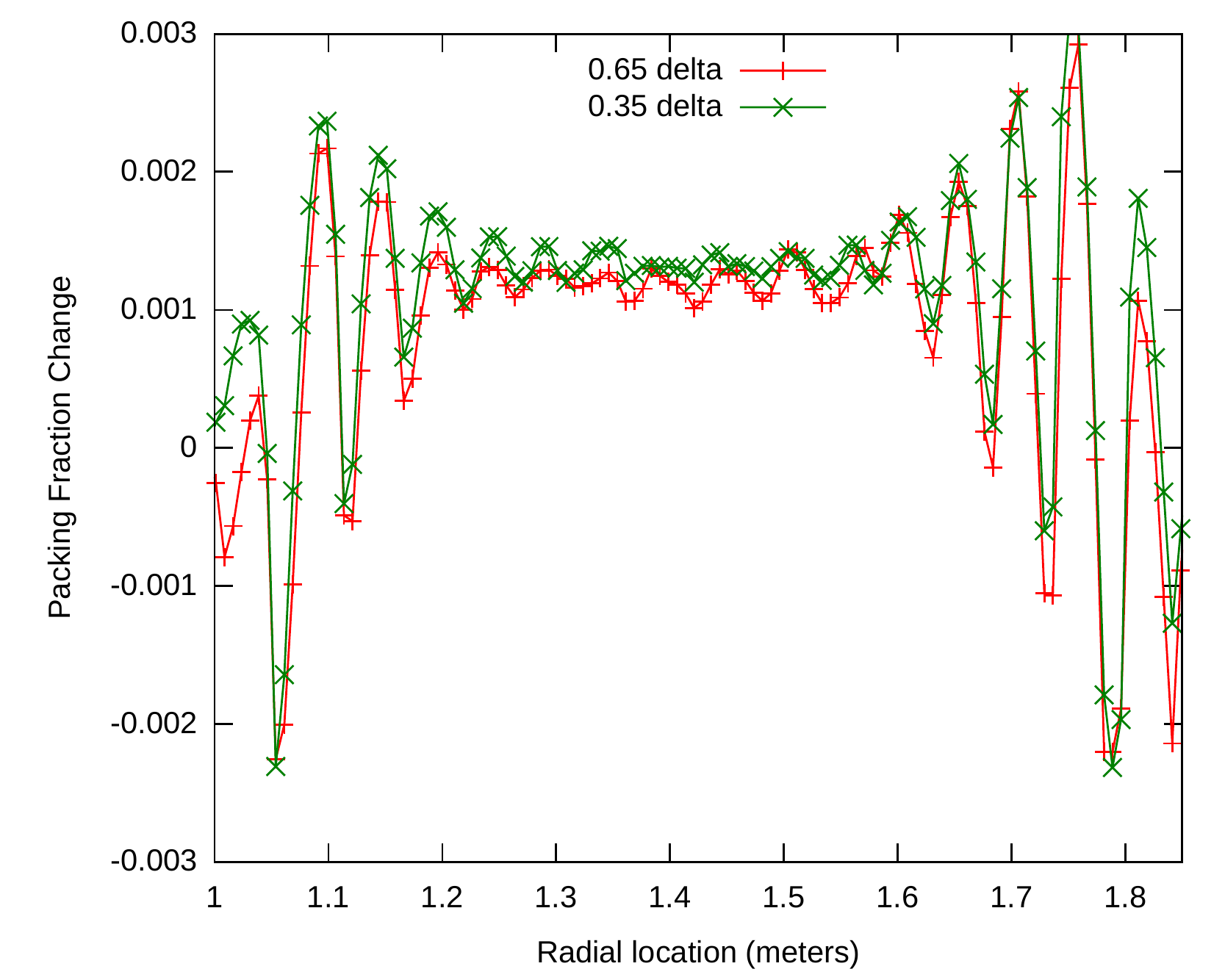}{Changes in Packing Fraction}{radial_eq_packing_delta}{0.8}

A previous version of the positional data from the earthquake
simulation was provided to J. Ortensi.  This data was used by him to
simulate the effects of an earthquake on a pebble bed
reactor\citep{Ortensi2009}.  Essentially, two factors cause an
increase in reactivity.  The first is the increased density of the
pebbles and the second is due to the control rods being at the top of
reactor, so when the top pebbles move down the control rod worth
(effect) decreases.  However, the reactivity increase causes the fuel
temperature to rise, which causes Doppler broadening and more neutrons
are absorbed by the $^{238}$U, which causes the reactivity to fall.
Figure \ref{ortensi_graph} shows an example of this.

\myfigures{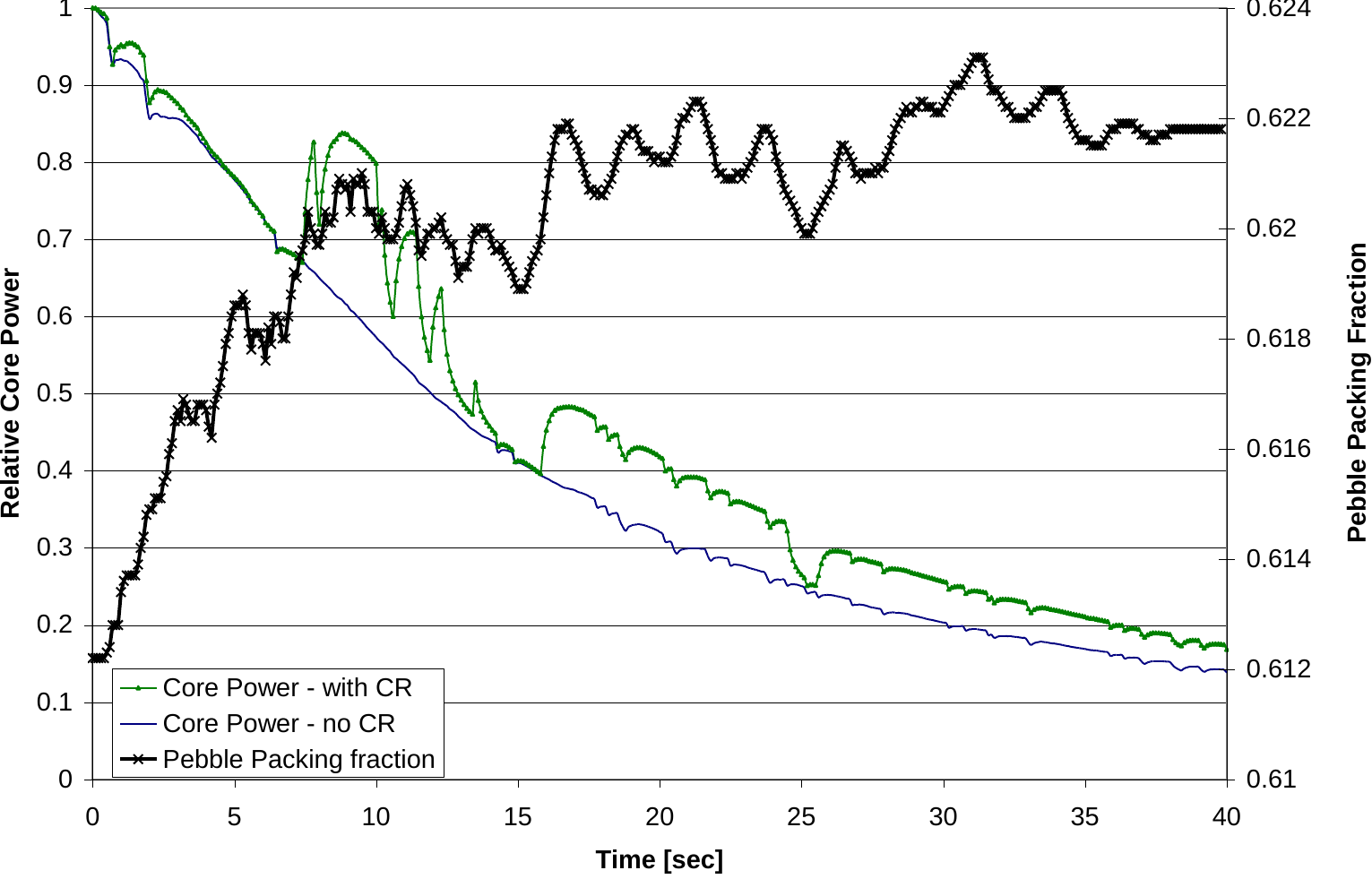}{Neutronics and Thermal Effects from J. Ortensi}{ortensi_graph}{0.9}

\subsection{Earthquake Equations}

For each time-step, the simulation calculates both a displacement and
a wall velocity.

For the sum of waves method, the displacement is calculated by:

\begin{equation}
\mathbf{d} = \sum_i \mathbf{D} \left [ sin \left((t - S){2.0\pi \over p} + c  \right ) + o \right]
\end{equation}

where $t$ is the current time, $S$ is the time the wave starts, $p$ is
the period of the wave, $c$ is the initial cycle of the wave, $o$ is
the offset, and $\mathbf{D}$ is the maximum displacement vector.

The velocity is calculated by:

\begin{equation}
\mathbf{m} = \sum_i {2 \pi \mathbf{D}\over p} cos \left((t - S){2\pi \over p}+c \right ) 
\end{equation}

For the tabular data, the displacement and velocity are calculated by:

\begin{align}
\mathbf{d} &= (1 - o)T_k + o T_{k+1} \\
\mathbf{m} &= {1 \over \delta} (T_{k+1} - T_k )
\end{align}

where $T_i$ is the displacement at the $i$th time-step, $o$ is a
number between 0 and 1 that specifies where 0 is the start of the
time-step and 1 is the end, and $\delta$ is the time in seconds between
time-steps.

With these displacements, the code then uses:

\begin{align}
\mathbf{p'} &= \mathbf{p} + \mathbf{d} \\
\mathbf{v'} &= \mathbf{v} + \mathbf{m}
\end{align}

as the adjusted position and velocity.

\chapter{Construction of a Dust Production Framework}


With the creation of the PEBBLES simulation, one issue that was
examined was using the simulation to attempt to predict the volume of
dust that would be produced by an operating pebble bed reactor.  This
is an important issue that could affect the choice of a pebble bed
reactor versus a prismatic reactor for process heat applications.
However, as this chapter and Appendix \ref{dust_coefficients} will discuss,
while the PEBBLES code has the force and motion data required for this
simulation, the coefficients that would allow this information to be
used have not been sufficiently robustly experimentally determined
yet.

\myfigures{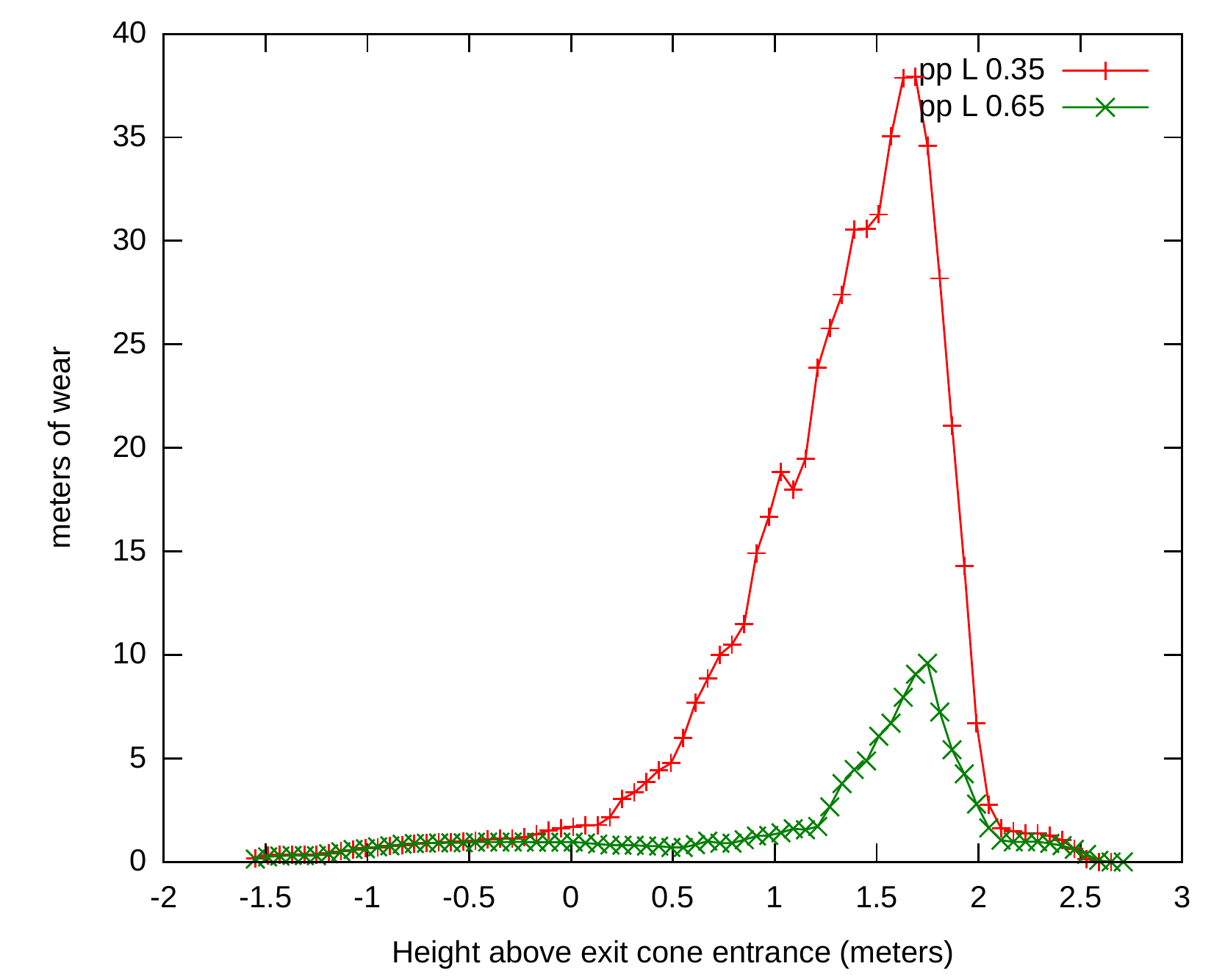}{Pebble to Pebble Distance Traveled}{wear_l_pp}{0.8}

\myfigures{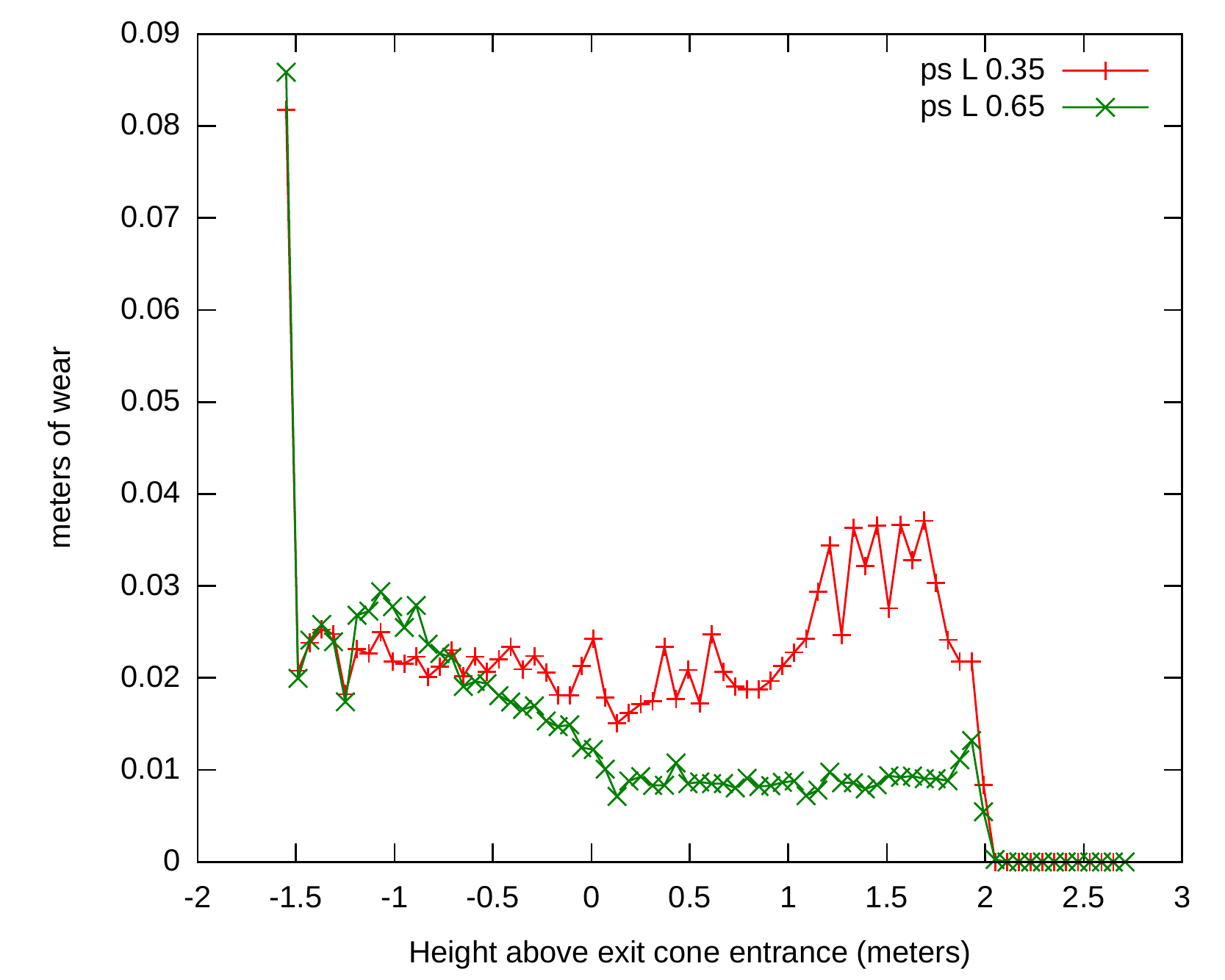}{Pebble to Surface Distance Traveled}{wear_l_ps}{0.8}

\myfigures{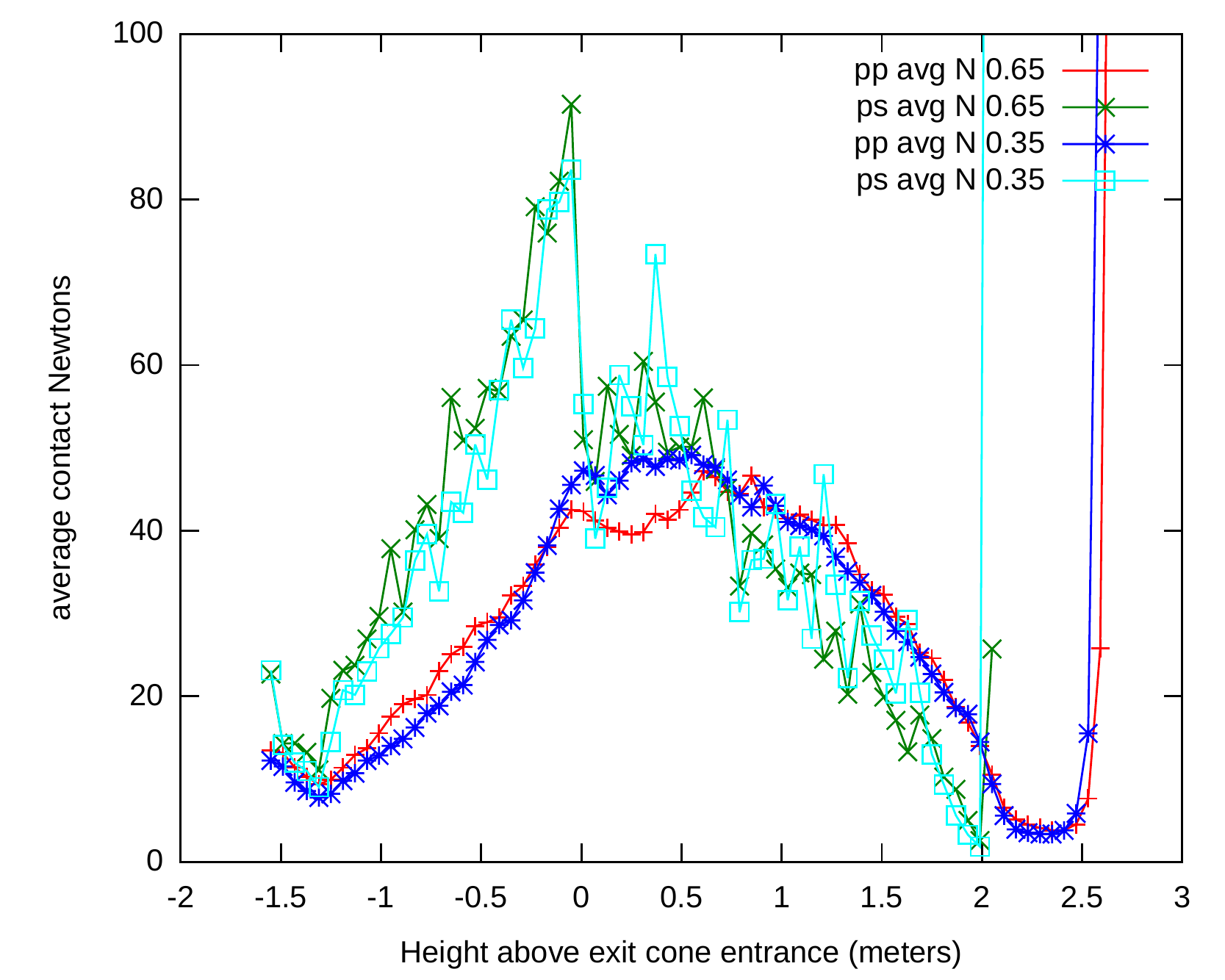}{Average Normal Contact Force}{wear_avg_n}{0.8}

\myfigures{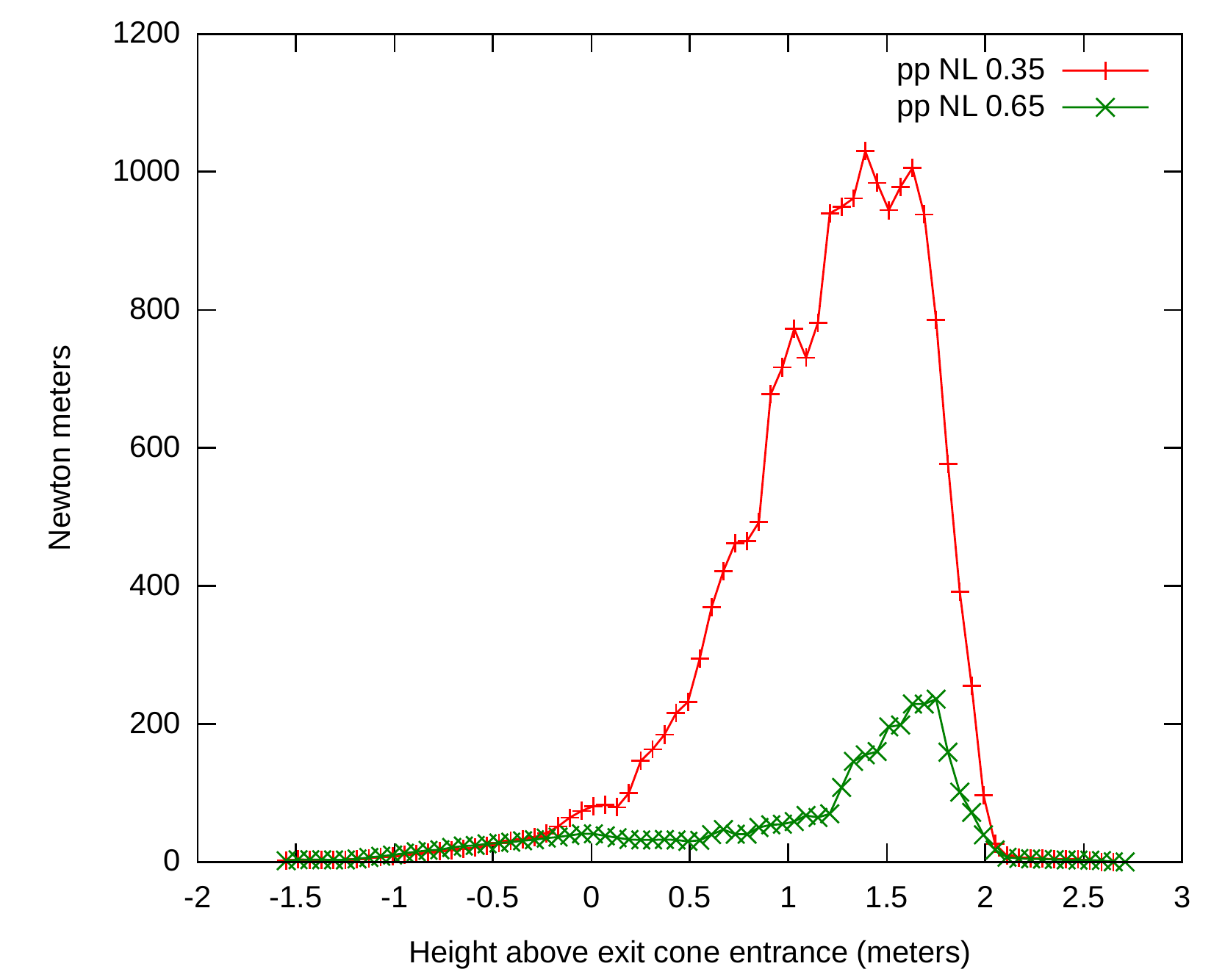}{Pebble to Pebble Wear}{wear_nl_pp}{0.8}

\myfigures{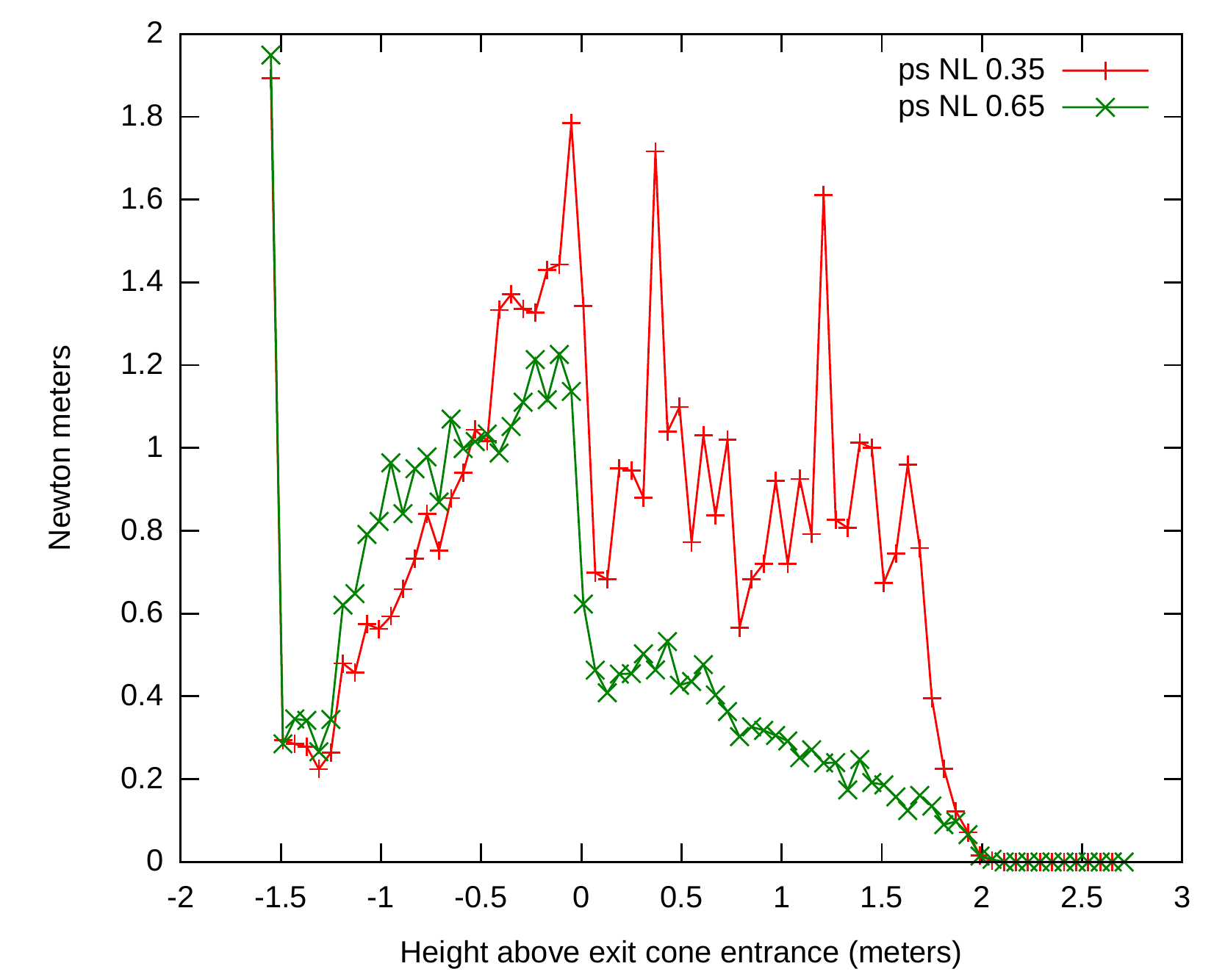}{Pebble to Surface Wear}{wear_nl_ps}{0.8}

With the data provided by PEBBLES, equations to link the dust
production to PEBBLES calculated quantities were examined.  As shown
in equation \ref{wear_volume_approximation}, the volume of dust
produced can be approximated if the normal force of contact, the
length slide and the wear coefficients are known.  The force of
contact and the length slide are calculated as part of the PEBBLES
simulation, so this method was used to calculate dust production for
the AVR reactor.  This resulted in an estimate of four grams of
graphite dust produced per year as compared to the measured value of
three kilograms of dust produced per year.  Several possible causes of
this were identified in the paper documenting this work
\citep{CogliatiandOugouag2008}.  A key first issue as described by
this dissertation's previous work section is that there are no good
measurements of graphite wear rates in pebble bed reactor relevant
conditions (especially for a reactor run at AVR temperatures).  A
second issue is that the previous model of AVR was missing features
including the reactor control rod nose cones and wall indentations.  A
third issue, identified after the paper's publication, is that
significant portions of the length traveled were due not to motion
down through the reactor.  Instead, much of the length that was
tallied was due to pebbles vibrating.  In the model used in the paper,
this problem was about four times more severe than the current model,
due to the new addition of slip correction via equation \ref{dls_eqn}.

As an illustration of the general framework for dust production, a
simple cylindrical vat the size of AVR was simulated.  In this model,
the outlet chute starts shrinking at height zero.  The length (L), and
length times normal force (NL) were tallied on intervals of 6 cm and
the figures \ref{wear_l_pp} to \ref{wear_nl_ps} show the results after
recirculating 400 pebbles.  The results are on a per pebble-pass
basis.  Two sets of static friction and kinetic friction pairs are
used, one with a static friction coefficient of 0.35 and kinetic of
0.25, and the other with static of 0.65 and kinetic of 0.4.  Figure
\ref{wear_l_pp} shows the calculated pebble-to-pebble lengths.  Notice
that for the 0.65 static friction simulation, about 10 meters of
length traveled is occurring at the peak in a 6 cm long tally. Since
the pebble is traveling about 0.06 m and has at most 12 pebble to
pebble contacts, essentially all but a small portion of this length
traveled is due to when the pebbles vibrate relative to each other.
This vibration is caused by the impact of the pebbles coming from the
inlet chutes and hitting the top of the bed.  This likely is a true
physical effect, which has not been discussed in literature this
author is aware of.  However, in order to obtain the correct magnitude
of this vibrational effect, two things must be correct.  First, the
simulation must dissipate the vibration at the correct rate, and
second, the pebbles must impact the bed at the correct velocity.
Quality estimates of both will need to be made to finish the dust
production work.  Note that with the lower static friction value, the
effect is even more pronounced.

Figure \ref{wear_l_ps} is also expected to have a vibrational
component, since only a small portion of pebbles should contact the
wall, and therefore for a 6 cm tally, the length traveled by the
average pebble should be much lower than 6 cm.  As the chute is
entered, the distance the average pebble travels increases in the 0.65
static friction case.  Figure \ref{wear_avg_n} shows the average
normal contact force.  The peak value for the pebble to surface values
is due to the base of the `arch' formed by pebbles.  The curves for
both the 0.65 static friction coefficient and the 0.35 static friction
coefficient are approximately the same because the static friction
force is not reaching the full Coulomb limit, so both have the same
effective $\mu$.

Figure \ref{wear_nl_pp} shows the normal times force sums.  For the
0.65 case, the peak is due to the vibrational impact.  For the 0.35
case, the vibration travels deep into the reactor bed, producing dust
throughout the reactor.  Figure \ref{wear_nl_ps} shows the peak dust
production coming in the base of the reactor, where the forces are the
highest, and the greatest lengths are traveled next to the wall.

\chapter{Future Work}

The dust production simulation requires both proper dust production
wear coefficients, and properly determining the correct method of
dealing with vibrational issues.  It would be useful to determine the
number of pebbles that need to be simulated to provide a correct
representation of a full NGNP-600 sized reactor.  Since the middle
portions are geometrically similar, determining the amount of
recirculation that is required to reach a geometrically asymptotic
state might allow only a portion of the recirculation to be done.
Those two changes might allow quicker simulation of full sized
reactors.  Finally, in order to allow sufficiently fast simulations on
today's computer hardware many approximations to the true behavior are
done.  In the future, some of these approximations maybe relaxed.

\chapter{Summary and Conclusions} 

Research results presented in this dissertation demonstrates a
distinct element method that provides high fidelity and yet has
reasonable run-times for many pebble fuel element flow simulations.
The new static friction test will be useful for evaluating any
implementation of static friction for spheres.  The PEBBLES code
produced for this dissertation has been able to provide data for
multiple applications including Dancoff factor calculation, neutronics
simulation and earthquake simulation.  The new static friction model
provides expected static friction behavior in the reactor including
partial matching of the Janssen model predictions and correctly
matching stability behavior in a pyramid. The groundwork has been
created for predicting the dust production from wear in a pebble bed
reactor once further experimental data is available.  Future work
includes potentially relaxing some of the physical approximations made
for speed purposes when faster computing hardware exists, and
investigating new methods for allowing faster simulations.  This
dissertation has provided significant enhancements in simulation of
the mechanical movement of pebbles in a pebble bed reactor.



\let\oldbibsection\bibsection
\renewcommand{\bibsection}{\oldbibsection\addcontentsline{toc}{chapter}{Bibliography}}
\bibliography{dissertation}{}

\begin{thebibliography}{58}
\providecommand{\natexlab}[1]{#1}
\providecommand{\url}[1]{\texttt{#1}}
\expandafter\ifx\csname urlstyle\endcsname\relax
  \providecommand{\doi}[1]{doi: #1}\else
  \providecommand{\doi}{doi: \begingroup \urlstyle{rm}\Url}\fi

\bibitem[mpi(2009)]{mpi}
Mpi: A message-passing interface standard, version 2.2.
\newblock 2009.
\newblock URL \url{http://www.mpi-forum.org/docs/mpi-2.2/mpi22-report.pdf}.

\bibitem[ope(2008)]{openmp}
Openmp application program interface, version 3.0.
\newblock 2008.
\newblock URL \url{http://www.openmp.org/mp-documents/spec30.pdf}.

\bibitem[opr(2009)]{oprofile}
Oprofile - a system profiler for linux.
\newblock 2009.
\newblock URL \url{http://oprofile.sourceforge.net}.

\bibitem[tht({\natexlab{a}})]{thtr_report}
The commissioning of the thtr 300. a performance report, {\natexlab{a}}.
\newblock D HRB 129288E.

\bibitem[tht({\natexlab{b}})]{thtr_webpage}
Hkg hochtemperatur-kernkraftwerk gesellschaft mit beschr\"a{}nkter haftung,
  {\natexlab{b}}.
\newblock \url{http://www.thtr.de/} Accessed Oct 27, 2009. Technical data on:
  \url{http://www.thtr.de/technik-tht.htm}.

\bibitem[Atomwirtschaft-Atomtechnik-atw(1966)]{avr_atw}
Atomwirtschaft-Atomtechnik-atw.
\newblock Avr-atomversuchskraftwerk mit kugelhaufen-hochtemperatur-reaktor in
  juelich.
\newblock \emph{Atomwirtschaft}, May 1966.
\newblock Sonderdruck aus Heft (Available as part of NEA-1739: IRPhE/AVR
  \url{http://www.nea.fr/abs/html/nea-1739.html}. Pages particularly useful are
  230 and 240).

\bibitem[B{\"a}umer et~al.(1990)B{\"a}umer, Barnert, Baust, Bergerfurth,
  Bonnenberg, B{\"u}lling, Burger, von~der Decken, Delle, Gerwin, Hackstein,
  Hantke, Hohn, Ivens, Kirch, Kirjushin, Kr{\"o}ger, Kr{\"u}ger, Kuzavkov,
  Lange, Marnet, Nickel, Pohl, Scherer, Sch{\"o}ning, Schulten, Singh,
  Steinwarz, Theymann, Wahlen, Wawrzik, Weisbrodt, Werner, Wimmers, and
  Ziermann]{AVR1990}
R.~B{\"a}umer, H.~Barnert, E.~Baust, A.~Bergerfurth, H.~Bonnenberg,
  H.~B{\"u}lling, St. Burger, C.-B. von~der Decken, W.~Delle, H.~Gerwin, K.-G.
  Hackstein, H.-J. Hantke, H.~Hohn, G.~Ivens, N.~Kirch, A.~I. Kirjushin,
  W.~Kr{\"o}ger, K.~Kr{\"u}ger, N.~G. Kuzavkov, G.~Lange, C.~Marnet, H.~Nickel,
  P.~Pohl, W.~Scherer, J.~Sch{\"o}ning, R.~Schulten, J.~Singh, W.~Steinwarz,
  W.~Theymann, E.~Wahlen, U.~Wawrzik, I.~Weisbrodt, H.~Werner, M.~Wimmers, and
  E.~Ziermann.
\newblock \emph{AVR: Experimental High Temperature Reactor; 21 years of
  sucessful operation for a future energy technology}.
\newblock Association of German Engineers (VDI), The Society for Energy
  Technologies, 1990.
\newblock ISBN 3-18-401015-5.

\bibitem[Bedenig et~al.(1968)Bedenig, Rausch, and
  Schmidt]{BedenigRauschandSchmidt1968}
D.~Bedenig, W.~Rausch, and G.~Schmidt.
\newblock Parameter studies concerning the flow behaviour of a pebble with
  reference to the fuel element movement in the core of the thtr 300 mwe
  prototype reactor.
\newblock \emph{Nuclear Engineering and Design}, 7:\penalty0 367--378, 1968.

\bibitem[Benenati and Brosilow(1962)]{BenenatiandBrosilow1962}
R.F. Benenati and C.~B. Brosilow.
\newblock Void fraction distribution in beds of spheres.
\newblock \emph{A. I. Ch. E. Journal}, 8:\penalty0 359--361, 1962.

\bibitem[Bernal et~al.(1960)Bernal, Mason, and Scott]{BernalMasonandScott1960}
J.~D. Bernal, J.~Mason, and G.~David Scott.
\newblock Packing of spheres.
\newblock \emph{Nature}, 188:\penalty0 908--911, 1960.

\bibitem[Bhushan(2000)]{moderntribology}
Bharat Bhushan.
\newblock \emph{Modern Tribology Handbook}.
\newblock CRC Press, Boca Raton, Florida, USA, 2000.
\newblock Chap. 7.5.

\bibitem[Bratberg et~al.(2005)Bratberg, Mal\o{}y, and Hansen]{narrowcolumns}
I.~Bratberg, K.J. Mal\o{}y, and A.~Hansen.
\newblock Validity of the janssen law in narrow granular columns.
\newblock \emph{The European Physical Journal E}, 18:\penalty0 245--252, 2005.

\bibitem[Cogliati and Ougouag(2008)]{CogliatiandOugouag2008}
J.~J. Cogliati and A.~M. Ougouag.
\newblock Pebble bed reactor dust production model.
\newblock Proceedings of the 4th International Topical Meeting on High
  Temperature Reactor Technology, 2008.
\newblock Washington, D.C., USA, September 28 -- October 1.

\bibitem[Cohen et~al.(1995)Cohen, Lin, Manocha, and Ponamgi]{cohen_collide}
Jonathan~D. Cohen, Ming~C. Lin, Dinesh Manocha, and Madhav~K. Ponamgi.
\newblock I–collide: an interactive and exact collision detection system for
  large scale environments.
\newblock Proceedings of the 1995 Symposium on Interactive 3D Graphics, 1995.
\newblock Monterey, CA, April 9-12, pp. 19-24.

\bibitem[Cundall and Strack(1979)]{cundall_and_strack}
P.~A. Cundall and O.~D.~L. Strack.
\newblock A discrete numerical model for granular assemblies.
\newblock \emph{G\'eotechniqe}, 29:\penalty0 47--65, 1979.

\bibitem[Drepper(2007)]{drepper_memory}
Ulrich Drepper.
\newblock What every programmer should know about memory.
\newblock 2007.
\newblock URL \url{http://people.redhat.com/drepper/cpumemory.pdf}.

\bibitem[Duran(1999)]{Duran1999}
Jacques Duran.
\newblock \emph{Sands, Powders, and Grains: An Introduction to the Physics of
  Granular Materials}.
\newblock Springer, New York, New York, USA, 1999.
\newblock ISBN 978-0387986562.

\bibitem[Freund et~al.(2003)Freund, Zeiser, Huber, Klemm, Brenner, Durst, and
  Emig]{Freundetal2003}
Hannsj{\"o}rg Freund, Thomas Zeiser, Florian Huber, Elias Klemm, Gunther
  Brenner, Franz Durst, and Gerhand Emig.
\newblock Numerical simulations of single phase reacting flows in randomly
  packed fixed-bed reactors and experimental validation.
\newblock \emph{Chemical Engineering Science}, 58:\penalty0 903--910, 2003.

\bibitem[Goodjohn(1991)]{SummaryGascooledReactorPrograms}
A.~J. Goodjohn.
\newblock Summary of gas-cooled reactor programs.
\newblock \emph{Energy}, 16:\penalty0 79--106, 1991.

\bibitem[Gotoh et~al.(1997)Gotoh, Masuda, and Higashitanti]{powder}
Keishi Gotoh, Hiroaki Masuda, and Ko~Higashitanti.
\newblock \emph{Powder Technology Handbook, 2nd ed.}
\newblock Marcel Dekker, Inc, New York, New York, 1997.

\bibitem[Gougar et~al.(2004)Gougar, Ougouag, and Terry]{GougarExternal}
Hans~D. Gougar, Abderrafi~M. Ougouag, and William~K. Terry.
\newblock Advanced core design and fuel management for pebble-bed reactors.
\newblock 2004.
\newblock INEEL/EXT-04-02245.

\bibitem[Haile(1997)]{Haile1997}
J.~M. Haile.
\newblock \emph{Molecular Dynamics Simulation}.
\newblock John Wiley {\&} Sons, Inc, New York, 1997.

\bibitem[Johnson(1985)]{contactmechanics}
K.L. Johnson.
\newblock \emph{Contact Mechanics}.
\newblock Cambridge University Press, 1985.
\newblock ISBN 0-521-34796-3.
\newblock Section 4.2.

\bibitem[Jullien et~al.(1992)Jullien, Pavlovitch, and
  Meakin]{JullienPavlovitchandMeakin1992}
R{\'e}mi Jullien, Andr{\'e} Pavlovitch, and Paul Meakin.
\newblock Random packings of spheres built with sequential models.
\newblock \emph{Journal Phys. A: Math. Gen.}, 25:\penalty0 4103--4113, 1992.

\bibitem[Kadak and Bazant(2004)]{KadakandBazant2004}
Andrew~C. Kadak and Martin~Z. Bazant.
\newblock Pebble flow experiments for pebble bed reactors.
\newblock Bejing, China, September 22-24 2004. 2nd International Topical
  Meeting on High Temperature Reactor Technology.

\bibitem[Kloosterman and Ougouag(2005)]{KloostermanandOugouag2005}
J.~L. Kloosterman and A.~M. Ougouag.
\newblock Computation of dancoff factors for fuel elements incorporating
  randomly packed triso particles.
\newblock Technical report, 2005.
\newblock INEEL/EXT-05-02593, Idaho National Laboratory.

\bibitem[Kohring(1995)]{Kohring1995}
G.~A. Kohring.
\newblock Studies of diffusional mixing in rotating drums via computer
  simulations.
\newblock \emph{Journal de Physique I}, 5:\penalty0 1551, 1995.

\bibitem[Lamarsh(1983)]{Lamarsh1983}
J.R. Lamarsh.
\newblock \emph{Introduction to Nuclear Engineering 2nd Ed.}
\newblock Addison-Wesley Publishing Company, Reading, Massaschusetts, 1983.
\newblock pp. 591-593.

\bibitem[Lancaster and Pritchard(1980)]{lancaster}
J.K. Lancaster and J.R. Pritchard.
\newblock On the `dusting' wear regime of graphite sliding against carbon.
\newblock \emph{J. Phys. D: Appl. Phys.}, 13:\penalty0 1551--1564, 1980.

\bibitem[Landry et~al.(2003)Landry, Grest, Silbert, and
  Plimpton]{confined_packings}
James~W. Landry, Gary~S. Grest, Leonardo~E. Silbert, and Steven~J. Plimpton.
\newblock Confined granular packings: Structure, stress, and forces.
\newblock \emph{Physical Review E}, 67:\penalty0 041303, 2003.

\bibitem[Lu et~al.(2001)Lu, Abdou, and Ying]{LuAbdouandYing2001}
Zi~Lu, Mohamed Abdou, and Alice Ying.
\newblock 3d micromechanical modeling of packed beds.
\newblock \emph{Journal of Nuclear Materials}, 299:\penalty0 101--110, 2001.

\bibitem[Luo et~al.(2004)Luo, Zhang, and Suyuan]{loads}
Xiaowei Luo, Lihong Zhang, and Yu~Suyuan.
\newblock The wear properties of nuclear grade graphite ig-11 under different
  loads.
\newblock \emph{International Journal of Nuclear Energy Science and
  Technology}, 1\penalty0 (1):\penalty0 33--43, 2004.

\bibitem[Luo et~al.(2005)Luo, Suyuan, Xuanyu, and Shuyan]{temperature}
Xiaowei Luo, Yu~Suyuan, Sheng Xuanyu, and He~Shuyan.
\newblock Temperature effects on ig-11 graphite wear performance.
\newblock \emph{Nuclear Engineering and Design}, 235:\penalty0 2261--2274,
  2005.

\bibitem[Marion and Thornton(2004)]{MarionandThornton2004}
Jerry~B. Marion and Stephen~T. Thornton.
\newblock \emph{Classical Dynamics of Particles and Systems, 5th Ed.}
\newblock Saunders College Publishing, 2004.
\newblock ISBN 0-534-40896-6.

\bibitem[Miller et~al.(2002)Miller, Petti, and Maki]{MillerPettiMaki2002}
G.~K. Miller, D.~A. Petti, and J.~T. Maki.
\newblock Development of an integrated performance model for triso-coated gas
  reactor particle fuel.
\newblock High Temperature Reactor 2002 Conference, April 2002.
\newblock URL
  \url{http://www.inl.gov/technicalpublications/Documents/3169759.pdf}.
\newblock Chicago, Illinois, April 25-29, 2004, on CD-ROM, American Nuclear
  Society, Lagrange Park, IL.

\bibitem[Mindlin and Deresiewicz(1953)]{MindlinandDeresiewicz1953}
R.~D. Mindlin and H.~Deresiewicz.
\newblock Elastic spheres in contact under varying oblique forces.
\newblock \emph{ASME J. Applied Mechanics}, 20:\penalty0 327--344, 1953.

\bibitem[Moormann()]{moorman}
Rainer Moormann.
\newblock A safety re-evaluation of the avr pebble bed reactor operation and
  its consequences for future htr concepts.
\newblock Berichte des Forschungszentrums J\"u{}lich; 4275 ISSN 0944-2952
  \url{http://hdl.handle.net/2128/3136}.

\bibitem[Ortensi(2009)]{Ortensi2009}
Javier Ortensi.
\newblock \emph{An Earthquake Transient Method for Pebble-Bed Reactors and a
  Fuel Temperature Model for TRISO Fueled Reactors}.
\newblock PhD thesis, Idaho State University, 2009.

\bibitem[Ougouag and Terry(2001)]{OugouagAndTerry2001}
Abderrafi~M. Ougouag and William~K. Terry.
\newblock A preliminary study of the effect of shifts in packing fraction on
  k-effective in pebble-bed reactors.
\newblock Proceeding of Mathemetics \& Computation 2001, September 2001.
\newblock Salt Lake City, Utah, USA.

\bibitem[Ougouag et~al.(2004)Ougouag, Gougar, Terry, Mphahlele, and
  Ivanov]{optimalmoderation}
Abderrafi~M. Ougouag, Hans~D. Gougar, William~K. Terry, Ramatsemela Mphahlele,
  and Kostadin~N. Ivanov.
\newblock Optimal moderation in the pebble-bed reactor for enhanced passive
  safety and improved fuel utilization.
\newblock PHYSOR 2004 – The Physics of Fuel Cycle and Advanced Nuclear
  Systems: Global Developments, April 2004.

\bibitem[Payne(2003)]{Payne2003}
S.~J. Payne.
\newblock Development of design basis earthquake (dbe) parameters for moderate
  and high hazard facilities at tra.
\newblock 2003.
\newblock INEEL/EXT-03-00942,
  http://www.inl.gov/technicalpublications/Documents/2906939.pdf.

\bibitem[Ristow(1998)]{Ristow1998}
Gerald~H. Ristow.
\newblock Flow properties of granual materials in three-dimensional geometries.
\newblock Master's thesis, Philipps-Universität Marburg, Marburg/Lahn, January
  1998.

\bibitem[Rycroft et~al.(2006{\natexlab{a}})Rycroft, Bazant, Grest, and
  Landry]{rycroft}
Chris~H. Rycroft, Martin~Z. Bazant, Gary~S. Grest, and James~W. Landry.
\newblock Dynamics of random packings in granular flow.
\newblock \emph{Physical Review E}, 73:\penalty0 051306, 2006{\natexlab{a}}.

\bibitem[Rycroft et~al.(2006{\natexlab{b}})Rycroft, Crest, Landry, and
  Bazant]{Rycroftetal2006}
Chris~H. Rycroft, Gary~S. Crest, James~W. Landry, and Martin~Z. Bazant.
\newblock Analysis of granular flow in a pebble-bed nuclear reactor.
\newblock \emph{Physical Review E}, 74:\penalty0 021306, 2006{\natexlab{b}}.

\bibitem[Sheng et~al.(2003)Sheng, S., and He]{sheng}
X.~Yu Sheng, X.~S., Luo, and S.~He.
\newblock Wear behavior of graphite studies in an air-conditioned environment.
\newblock \emph{Nuclear Engineering and Design}, 223:\penalty0 109--115, 2003.

\bibitem[Silbert et~al.(2001)Silbert, Ertas, Grest, Halsey, Levine, and
  Plimpton]{granular_flow}
Leonardo~E. Silbert, Deniz Ertas, Gary~S. Grest, Thomas~C. Halsey, Dov Levine,
  and Steven~J. Plimpton.
\newblock Granular flow down an inclined plane: Bagnold scaling and rheology.
\newblock \emph{Physical Review E}, 64:\penalty0 051302, 2001.

\bibitem[Soppe(1990)]{Soppe1990}
W.~Soppe.
\newblock Computer simulation of random packings of hard spheres.
\newblock \emph{Powder Technology}, 62:\penalty0 189--196, 1990.

\bibitem[Sperl(2006)]{Sperl2006}
Matthias Sperl.
\newblock Experiments on corn pressure in silo cells – translation and
  comment of janssen’s paper from 1895.
\newblock \emph{Granular Matter}, 8:\penalty0 59--65, 2006.

\bibitem[Stansfield(1969)]{stansfield}
O.~M. Stansfield.
\newblock Friction and wear of graphite in dry helium at 25, 400, and
  800$^\circ$ c.
\newblock \emph{Nuclear Applications}, 6:\penalty0 313--320, 1969.

\bibitem[Terry et~al.(2006)Terry, Kim, Montierth, Cogliati, and
  Ougouag]{Terryetal2006}
William~K. Terry, Soon~Sam Kim, Leland~M. Montierth, Joshua~J. Cogliati, and
  Abderrafi~M. Ougouag.
\newblock Evaluation of the htr-10 reactor as a benchmark for physics code qa.
\newblock ANS Topical Meeting on Reactor Physics, 2006.
\newblock URL
  \url{http://www.inl.gov/technicalpublications/Search/Results.asp?ID=INL/CON-%
06-11699}.

\bibitem[Vu-Quoc et~al.(2000)Vu-Quoc, Zhang, and Walton]{vu_quoc}
L.~Vu-Quoc, X.~Zhang, and O.~R. Walton.
\newblock A 3-d discrete-element method for dry granular flows of ellipsoidal
  particles.
\newblock \emph{Computer Methods in Applied Mechanics and Engineering},
  187:\penalty0 483--528, 2000.

\bibitem[Vu-Quoc and Zhang(1999)]{VuQuocandZhang1999}
Loc Vu-Quoc and Xiang Zhang.
\newblock An accurate and effcient tangential force-displacement model for
  elastic frictional contact in particle-flow simulations.
\newblock \emph{Mechanics of Materials}, 31:\penalty0 235--269, 1999.

\bibitem[Wahsweiler(1989)]{thtr_dust}
Dr. Wahsweiler.
\newblock Bisherige erkentnisse zum graphitstaub, 1989.
\newblock HRB BF3535 26.07.1989.

\bibitem[Wait(2001)]{Wait2001}
R.~Wait.
\newblock Discrete element models of particle flows.
\newblock \emph{Mathematical Modeling and Analysis I}, 6:\penalty0 156--164,
  2001.

\bibitem[Walker(1966)]{Walker1966}
D.~M. Walker.
\newblock An approximate theory for pressures and arching in hoppers.
\newblock \emph{Chemical Engineering Science}, 21:\penalty0 975--997, 1966.

\bibitem[Wu et~al.(2002)Wu, Lin, and Zhong]{DesignFeaturesOfHTR10}
Zongxin Wu, Dengcai Lin, and Daxin Zhong.
\newblock The design features of the htr-10.
\newblock \emph{Nuclear Engineering and Design}, 218:\penalty0 25--32, 2002.

\bibitem[Xiaowei et~al.(2005)Xiaowei, Suyaun, Zhensheng, and
  Shuyan]{dust_estimation}
Luo Xiaowei, Yu~Suyaun, Zhang Zhensheng, and He~Shuyan.
\newblock Estimation of graphite dust quantity and size distribution of
  graphite particle in htr-10.
\newblock \emph{Nuclear Power Engineering}, 26, 2005.
\newblock ISSN 0258-0926(2005)02-0203-06.

\bibitem[Xu and Zuo(2002)]{OverviewOfHTR10}
Yuanhui Xu and Kaifen Zuo.
\newblock Overview of the 10 mw high temperature gas cooled reactor—test
  module project.
\newblock \emph{Nuclear Engineering and Design}, 218:\penalty0 13--23, 2002.

\end{thebibliography}
\bibliographystyle{plainnat}

\appendix

\chapter{Calculation of Packing Fractions}

For determining the volume of a sphere that is inside a vertical slice,
the following formula can be used
\begin{align}
a &= max(-r,bot-z) \\
b &= min(r,top-z) \\
v &= \pi\left[r^2(b-a)+\frac{1}{3}(a^3-b^3)\right]
\end{align}
where $r$ is the pebble radius, $bot$ is the bottom of the vertical
slice,$top$ is the top of the vertical slice and $z$ is the vertical
location of the pebble center.

To determine the area inside a vertical and radial slice, two auxiliary
functions are defined, one which has the area inside a radial 2d
slice, and another which has the area outside a radial 2d slice.

\myfigures{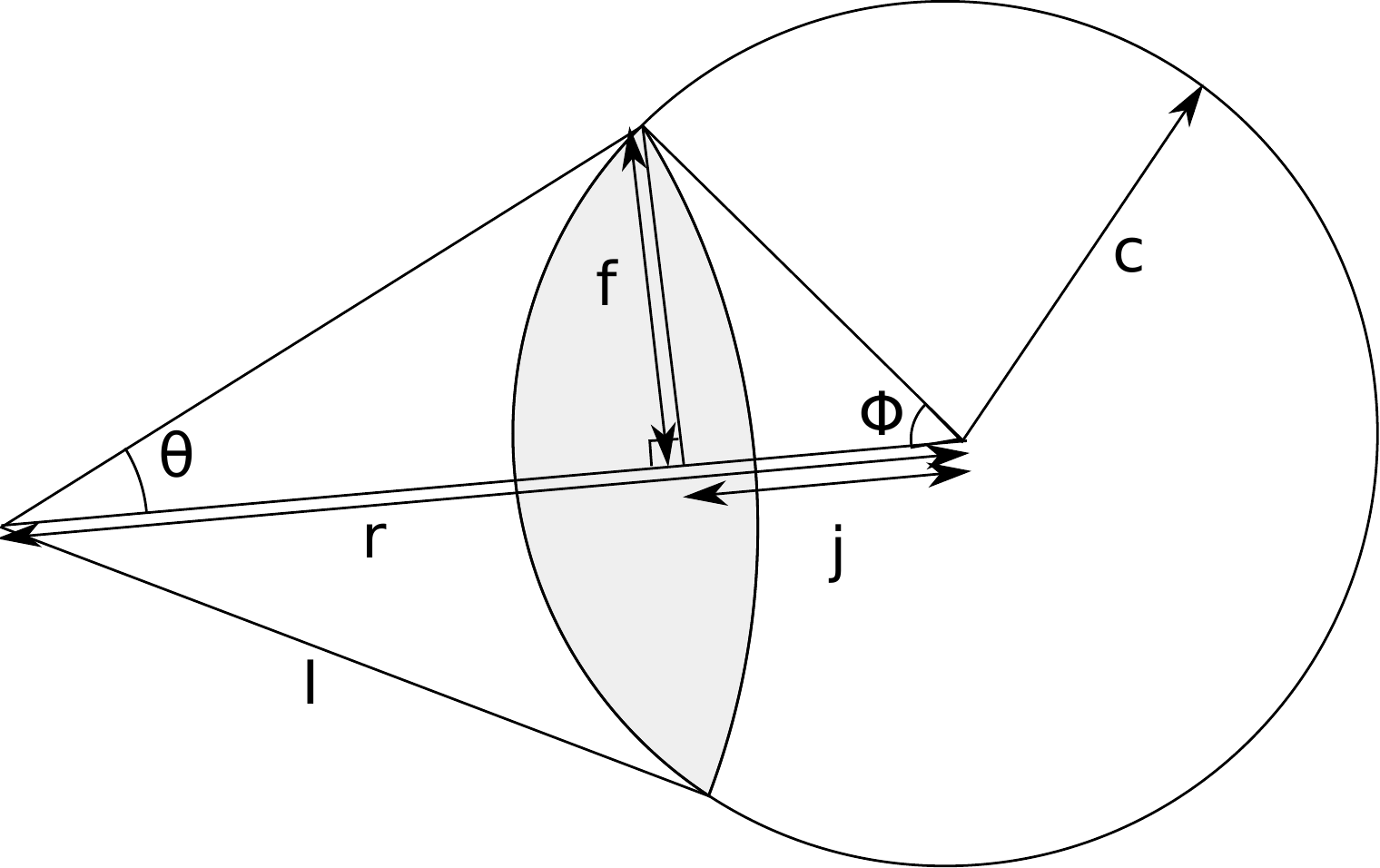}{Area Inside Geometry}{inside_area}{0.5}

\myfigures{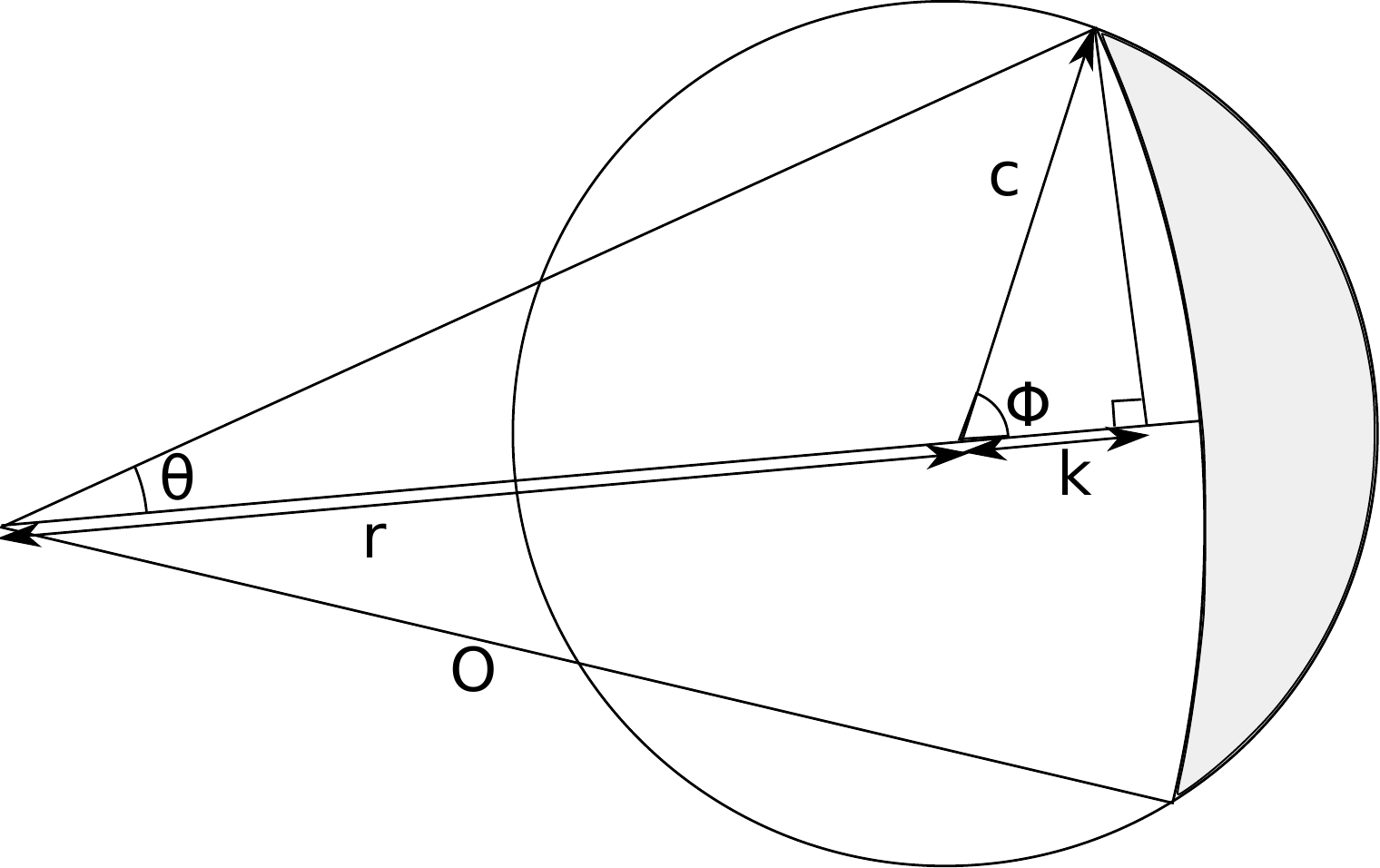}{Area Outside Geometry}{outside_area}{0.5}

Figure \ref{inside_area} shows the area that is inside both a circle
of radius $c$ and a radial slice of $I$.  The circle is $r$ from the
center of the radial circle.  Auxiliary terms are defined, which
include $f$, the distance from the intersection of the segment of the
radial circle perpendicularly to the center line, $j$ the distance to
the intersection of $f$, $\theta$ the angle of segment, and $\phi$ the
angle from the segment intersection on the circle side.  The
area\_inside function has the following definition:

\begin{align}
a_i &= 0.0 \qquad  \mathrm{if}\, I < r - c \\
a_i &= \pi c^2 \qquad  \mathrm{if}\, r + c < I \\
a_i &= \pi I^2 \qquad  \mathrm{if}\, I < r + c \,\mathrm{and}\, I < c - r \\
\mathrm{otherwise} &  \\
j &= \frac{r^2 + c^2 - I^2}{2r} \\
f &= \sqrt{c^2 - j^2} \\
\theta &= 2 \arccos \frac{I^2+r^2-c^2}{2 I r}  \\
\phi &= 2 \arccos \frac{r^2 +c^2 -I^2}{2 r c} \\
a_i &= \frac{1}{2} c^2 \phi + \frac{1}{2} I^2 \theta - f r
\end{align}

Figure \ref{outside_area} shows the area that is outside the radial
slice, but inside the circle.  The radial slice has a radius of $O$.
The new auxiliary term $k$ is the distance from the circle's center to
the perpendicular intercept.  The area\_outside function has the
following definition:

\begin{align}
a_o &= 0.0 \qquad \mathrm{if}\, O > c + r \\
a_o &= \pi c^2 \qquad \mathrm{if}\, c - r > O \\
a_o &= \pi c^2 - \pi O^2 \qquad \mathrm{if}\, O < r + c \,\mathrm{and}\, O < c - r \\
\mathrm{otherwise} & \\
k &= \frac{O^2 - r^2 - c^2}{2r} \\
m &= \sqrt{c^2 - k^2} \\
\theta &= 2 \arccos \frac{k+r}{O} \\
\phi &= 2 \arccos \frac{k}{c} \\
a_o &= ( \frac{1}{2} c^2 \phi - m k ) - ( \frac{1}{2} O^2 \theta - m (k+r) )
\end{align}

Then, the total volume in a radial slice can be determined from the
computation:

\begin{align}
a &= max(-r,bot-z) \\
b &= min(r,top-z) \\
v_t &= \pi\left[R^2(b-a)+\frac{1}{3}(a^3-b^3)\right] \\
v_i &= \int_a^b \mathrm{area\_inside}(c=\sqrt{R^2-z^2}) dz \\
v_o &= \int_a^b \mathrm{area\_outside}(c=\sqrt{R^2-z^2}) dz \\
v &= v_t - v_i - v_o
\end{align}

\chapter{Determination of dust production coefficients}
\label{dust_coefficients}

One potential use of the PEBBLES code is to predict the dust
production of a pebble bed reactor.  This section discusses the
features that make this possible and work that has been done to
determine the necessary coefficients.  Unfortunately, the following
literature review discovered a lack of robust wear coefficients, which
prevents prediction of dust production.

There are essentially four contact wear mechanisms.  Adhesive wear is
from the contacting surfaces adhesively bonding together, and part of
the material is pulled away.  Abrasive wear is when one of the
contacting materials is harder than the other, and plows (or shears)
away material.  Fatigue wear is when the surfaces repeatedly contact
each other causing fracture of the material.  The last mechanism is
corrosive wear, when chemical corrosion causes the surface to behave
with increased wear \citep{moderntribology}.  For pebble bed reactors,
adhesive wear is expected to be the dominate wear mechanism.

As a first order approximation the adhesive dust
production volume is \citep{moderntribology}:

\begin{equation}
\label{wear_volume_approximation}
V = K_{ad} \frac{N}{H} L
\end{equation}

In this equation $V$ is the wear volume, $K_{ad}$ is the wear
coefficient for adhesive wear, $L$ is the length slide and
$\frac{N}{H}$ is the real contact area (with $N$ the normal force and
$H$ the hardness).  Typically, the hardness and the wear coefficient
for adhesive wear are combined with the units of either mass or volume
over force times distance.  For two blocks, the length slide is the
distance that one of the blocks travels over the other while in
contact.  Note that this formula is only an approximation since the
wear volume is only approximately linear with respect to both the
normal force and the distance traveled.  Abrasive wear also can be
approximated by this model, but fatigue and corrosive wear will not be
modeled well by this. To the extent that these wear mechanisms are
present in the pebble bed reactor, this model may be less valid.

The wear coefficient is typically measured by grinding or stroking two
pieces of graphite against each other, and then measuring the decrease
in mass. The details of the experiment such as the contact shape and
the orientation of the relative motion affect the wear coefficient.

The wear that occurs with graphite depends on multiple factors.  A
partial list includes the normal force of contact (load), the
temperature of the graphite and the past wear history (since wear
tends to polish the contact surfaces and remove loose grains).  The
atmosphere that the graphite is in affects the wear rates since some
molecules chemically interact with the carbon or are adsorbed on the
surface.  Neutron damage and other radiation effects can damage the
structure of the graphite and affect the wear.  The type and
processing of the graphite can affect wear rates.  As a related
effect, if harder and softer graphites interact, the harder one can
`plow' into the softer and increase wear rates.

For graphite on graphite, depending on conditions there can be over
three orders of magnitude difference in the wear.  For example
\citet{sheng} experimentally determined graphite on graphite in air at
room temperature can exhibit wear rates of 3.3e-8 g/(Nm) but in the
dusting regime\footnote{In air, above a certain temperature graphite
  wear transitions to dusting wear, which has much greater wear rates.
  Increased water vapor decreases or eliminates the dusting wear.} at
200\degree C the wear coefficient was determined by \citet{lancaster}
to be 2e-5 g/(Nm) which is about a thousand times greater.  For this
reason, conditions as close to the in-core conditions are needed for
determining a better approximation of the wear coefficients.

For tests using nuclear graphite near in-core conditions, the best
data available to the author is from two independent sets of
experiments.  One data-set emerged from the experiments by
\citet{stansfield} and the other is from a series of
experiments performed at the Tsinghua
University\citep{sheng,loads,temperature}.

O.M. Stansfield measured friction and wear with different types of
graphite in helium at different temperatures \citep{stansfield}. In
the experiments, two pieces of graphite were slid against each other
linearly with a 0.32 cm stroke.  Two different loads were used, one
2-kg mass, and another 8-kg mass. The data for wear volumes is only
provided graphically, that is, not tabulated, therefore only order of
magnitude results are available.  The wear values were about an order
of magnitude higher at 25\degree C than at 400\degree C and 800\degree
C.  There was a reduction of friction with increased length slide, but
no explanation was provided\footnote{Possibly this was due to a
  lubrication effect or the removal of rough or loose surfaces.}.
Typical values for the wear rates are 10e-3 cm$^3$/kg for the
25\degree C case and 10e-4 cm$^3$/kg for the 400\degree C and
800\degree C for 12 500 cm distance slide.  With a density of 1.82
g/cm$^3$, these work out to about 1.5e-6 g/(Nm) and 1.5e-7 g/(Nm).
These are only about an order of magnitude above room temperature
wear.


The second set of experiments were done at the Tsinghua university.
The first paper measures the wear coefficient of graphite KG-11 via
pressing a static specimen against a revolving specimen.  The wear is
measured by weighing the difference in mass before the experiment and
after the experiment.  At room temperature in air they measured wear
rates of 7.32e-9 g/(Nm) with 31 N load with surface contact, 3.29e-8
g/(Nm) with 31 N load with line contact and 3.21e-8 g/(Nm) with 62 N
load\citep{sheng}.  The second paper measures the wear coefficient of
graphite IG-11 on graphite and on steel at varying loads\citep{loads}.
Unfortunately, there are inconsistencies in the units used in the
paper.  For example, in Table 2 the mean wear rate for the lower
specimen is listed as 3.0e3 $\mu$g/m, but in the text it is listed as
0.3e-3 $\mu$g/m, seven orders of magnitude different.  The 30 N of
load upper specimen wear coefficient for the first 30 minutes is
listed as 1.4e-3 $\mu$g/m, which works out to 4.7e-10 g/(Nm).  If
1.4e3 $\mu$g/m is used, this works out to 4.7e-4 g/(Nm).  Neither of
these matches the first paper's results.  It seems that the units of
$\mu$g, (or micrograms or 1.0e-6 g) are used where mg (or milligrams
or 1.0e-3 g) should be.  Also, the sign for the exponent is
inconsistent, where sometimes the negative sign is dropped.  These two
mistakes would make the correct exponents 1.0e-3 mg/m and the measured
coefficient 1.4e-3 mg/m or 4.7e-7 g/(Nm), which match reasonably well
to the first paper's values on the order of 1.0e-8 g/(Nm).  For the
rest of this report, it is assumed that these corrections should be
used for the Xiaowei et al.~papers.

The third paper measures the temperature effects in
helium\citep{temperature}.  The experimental setup is similar to the
setup in the second paper, but the atmosphere is a helium atmosphere
and the temperatures used are 100\degree C to 400\degree C with a load
of 30 N.  In Figure 2 of that paper, it can be qualitatively
determined that as the temperature increases, the amount of wear
increases.  As well, the wear tends to have a higher rate initially,
and then decrease.  Since the wear experiment was performed using a 2
mm long stroke, it seems plausible that wear rates in an actual pebble
bed might be closer to the initially higher rates since the pebble
flow might be able to expose more fresh surfaces of the pebbles to
wear.  From the graph, there does not seem to be a clear trend in the
wear as a function of temperature.  This makes it difficult to
estimate wear rates since pebble bed reactor cores can have
temperatures over 1000\degree C in normal operation.  The highest wear
rate in Table 2 of the paper is 31.3e-3 mg/m at 30 N, so the highest
wear rate measured is 1.04e-6 g/(Nm).  This is about 20 times lower
than wear in the dusting regime.  Since the total amount of wear (from
Fig.~2) between 200\degree C and 400\degree C roughly doubles in the
upper specimen and increases by approximately 35\% in the lower
specimen, substantially higher wear rates in over 1000\degree C
environments are hard to rule out.  Note, however, that the opposite
temperature trend was observed in the Stansfield paper.

\section{Calculation of Force in Reactor Bed}

In order to calculate the dust produced in the reactor, the force
acting on the pebbles is needed.  Several different approximations can
be used to calculate this with varying accuracy.  The simplest (but
least accurate) method of approximating the pressure in the reactor is
using the hydrostatic pressure, or

\begin{equation}
P = \rho f g h
\label{hydro_pressure}
\end{equation}
where $P$ is the pressure at a point, $\rho$ is the density of the
pebbles, $f$ is the packing fraction of the pebbles (typical values
are near 0.61 or 0.60), $g$ is the gravitational acceleration and $h$ is
the height below the top of the pebble bed.  With knowledge of how
many contacts there are per unit area or per unit volume, this can be
converted into pebble to surface or pebble to pebble contact forces.
This formula is not correct when static friction occurs since the
static friction allows forces to be transferred to the walls. Therefore,
Equation \ref{hydro_pressure} over-predicts the actual pressures in the pebble bed.

In the presence of static friction, more complicated calculations are required.  The fact
that static friction transfers force to the wall was observed by the
German engineer H.A.~Janssen in 1895 \citep{Sperl2006}.  Formulas for the
pressure on the wall for cylindrical vessels with conical exit chutes
were derived by \citet{Walker1966}.  Essentially, when the upward
force on the wall from static friction for a given segment matches the
downward gravitational force from the additional pebbles in that
segment, the pressure stops increasing.  

For a cylinder, the horizontal pressure equation is \citep{powder}:

\begin{equation}
P_h = \frac{\gamma D}{4 \mu_w} \left[ 1 - \exp \left(\frac{-4\mu_w K}{D}x \right) \right]
\end{equation}
where $\gamma$ is the bulk weight (or $f \rho g$), $D$ is the diameter
of the cylinder, $\mu_w$ is the static friction coefficient between the
pebbles and the wall, $K$ is the Janssen Coefficient, and $x$ is the
distance below the top of the pile.

The Janssen coefficient is dependent upon the pebble to pebble static
friction coefficient and can be calculated from:

\begin{equation}
K=\frac{1-\sin \phi}{1+\sin \phi}
\end{equation}
where $\tan \phi = \mu_p$ and $\mu_p$ is the pebble to pebble static
friction.  Since $\tan^{-1} \mu = \sin^{-1} \left (
\frac{\mu}{\sqrt{\mu^2+1}} \right )$ then $K$ can also be written as:

\begin{equation}
K=2\mu_p^2-2\mu_p\sqrt{\mu_p^2+1}+1
\end{equation}

The Janssen formula derivations make assumptions that are not
necessarily true for granular materials.  These include assuming the
granular material is a continuum and that the shear forces on the wall
are at the Coulomb limit \citep{narrowcolumns}.  The static friction
force ranges from zero at first contact up to $\mu N$ (the Coulomb
limit) when sufficient shear force has occurred. If the force is not
at the Coulomb limit, then an effective $\mu$ may be able to be found
and used instead.  In general, this assumption will not be true when
the pebbles are freshly loaded as they will not have slid against the
wall enough to fully load the static friction.  Even after the pebbles
have been recirculated, they may not reach the Coulomb limit and
effective values for the static friction constant may be needed
instead for predicting the wall pressure.  Finally, real reactors have
more complicated geometries than a smooth cylinder above a cone exit
chute.

\section{Prior data on dust production}

\myfigures{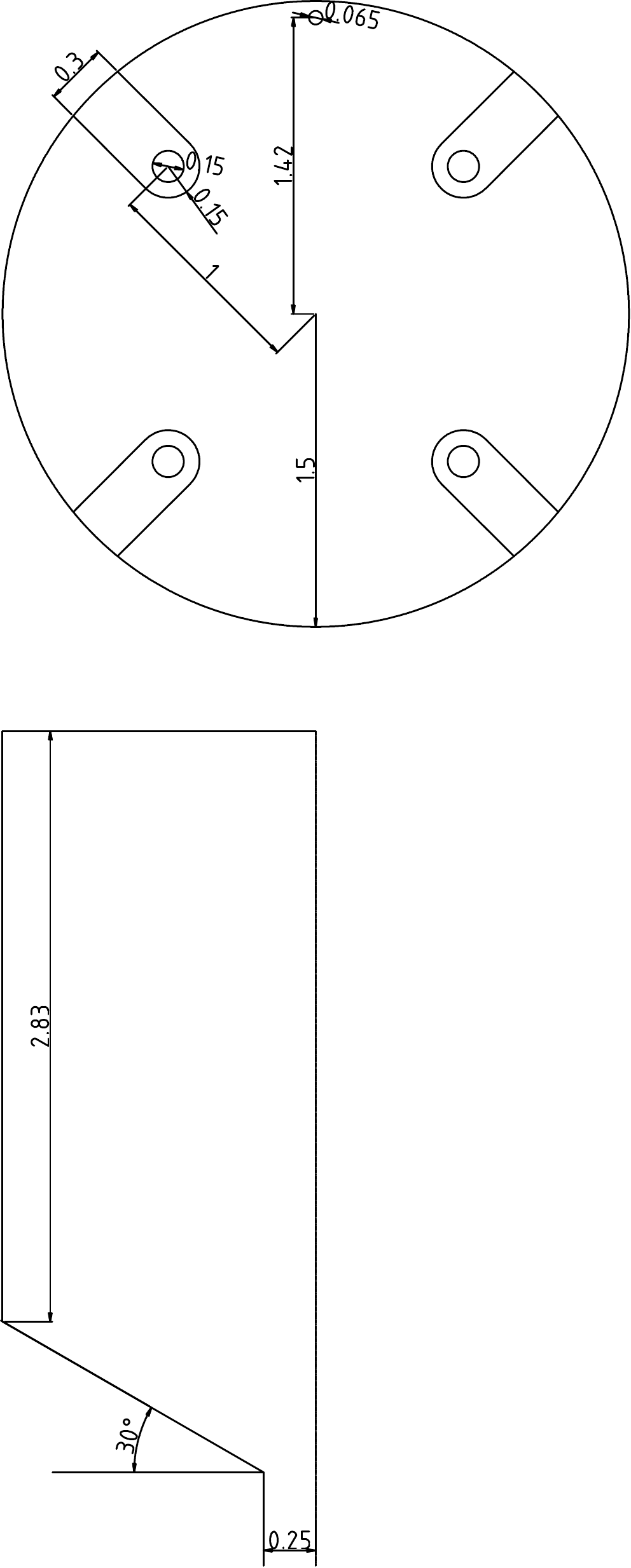}{AVR Dimensions}{avr_dimensions}{0.6}

The 46 MW thermal pebble bed reactor Arbeitsgemeinschaft
VersuchsReaktor (AVR) was created in the 1960s in Germany and operated
for 21 years.  The pebbles were added into the reactor through four
feeding tubes spaced around the reactor and one central feeding tube
at the top of the reactor.  There was one central outlet chute below.
Into the reactor cavity there were four noses of U shaped graphite
with smooth sides for inserting the control rods.  The cylinder walls
contained dimples about 1/2 a pebble diameter deep and that alternated
location periodically.  All the structural graphite was a needle coke
graphite.  Dimensions are shown in Figure \ref{avr_dimensions} and
design and measured data is provided in Table \ref{avr_data}.  The measured dust
production rate was 3 kg per year. No real conclusions were inferred 
because of a water ingress, an oil ingress, the uncertainty in the composition 
of the dust (i.e., metallic components) and the uncertainty of the location of
dust production\citep{AVR1990,avr_atw}.  The interior of the AVR
reactor reached over 1280\degree C as determined by melt wire
experiments\citep{moorman}.

\begin{table}
  \caption{\bf AVR Data}
  \label{avr_data}
  \begin{tabular}{l|r}
    Name & Value\\
    \hline
    Average Inlet Temperature & 250\degree C\\
    Average Outlet Temperature & 950\degree C\\
    Pebble Circulation Rate & 300-500 per day\\
    Dust Produced & 3 kg per year\\
    Pebbles in Reactor Core & 100,000\\
    Reactor Radius & 1.5 m\\
    Outlet Chute Radius & 0.25 m\\
    Angle of Outlet Cone & 30\degree\\
    Control Rod Nose Thickness & 0.3 m\\
    Radius of Control Rod Nose & 0.15 m\\
    Feed tube to outlet chute & 2.83 m\\
  \end{tabular}
\end{table}

The THTR-300 reactor was a thorium and uranium powered pebble bed
reactor that first went critical in 1983 and ran through 1988.
THTR-300 produced 16 kg of dust per Full Power Year (FPY), and an
estimated 6 kg of that was produced in the core of the
reactor\citep{thtr_dust}.  The control rods in the THTR-300 actually
pushed into the pebble bed.  On a per pebble basis, the amount of dust
produced in the THTR-300 is lower than in the AVR.  Further data on
the THTR-300 is summarized in Table
\ref{thtr_data}\citep{thtr_report,thtr_webpage}.

\begin{table}
  \caption{\bf THTR Data}
  \label{thtr_data}
  \begin{tabular}{l|r}
    Name & Value\\
    \hline
    Average Inlet Temperature & 250\degree C\\
    Average Outlet Temperature & 750\degree C\\
    Core Height & 6.0 m\\
    Pebbles Circulated  & 1,300,000 per FPY\\
    Core Diameter & 5.6 m\\
    Pebbles in Full Core & 657,000\\
    Total Dust Produced & 16 kg per FPY\\
    Estimated In-core Dust & 6 kg per FPY\\
  \end{tabular}
\end{table}

\section{Prior Prediction Work}

There are two papers published that attempt to predict the in-core
pebble dust production.  The first paper is ``Estimation of Graphite
Dust Quantity and Size Distribution of Graphite Particle in HTR-10''
\citep{dust_estimation} and was created to estimate the dust
production that the core of the HTR-10 reactor would produce.  The
second is co-authored by this author and attempts to estimate the dust
that the AVR reactor produced.


The HTR-10 paper started by calculating from the hydrostatic pressure
the force between the pebbles at the bottom of the reactor. The force
was approximated to be 30N.  The remainder of the paper uses 30N as
the force for conservatism.  Note that the HTR-10 paper is in Chinese,
so this review may contain mistakes in understanding due to language
differences.

The dust production is calculated in three regions, the core of the
reactor, the outlet chute of the reactor and the fuel loading pipe.
As with the other papers, the assumption is made that $\mu$g should
actually be mg.

For the core of the reactor the temperature used is 550\degree C with
pebble to pebble wear rates of 4.2e-3 mg/m extrapolated from
400\degree C data. The pebble to wall wear rates are extrapolated to
480\degree C to 12.08e-3 mg/m from the 400\degree C data.  The pebble
to pebble wear is estimated to occur for\footnote{This is the length
  slide and is multiplied by 4.2e-3 mg/m to get per pass dust
  production} 2.06 m and 3.85\% of pebbles are estimated to wear
against the wall.  From this data the average pebble dust production
per pass in the core is determined to be 8.65e-3 mg for pebble to
pebble wear and 0.99e-3 mg from pebble to wall.  The total in-core
graphite dust produced per pebble pass is 9.64e-3 mg.

The outlet chute wear is estimated to occur for 2.230 m in the
graphite portion and 1.530 m in the stainless steel portion, and that
44.16\% of the pebbles wear against the chute.  Both these portions
are estimated to be at 400\degree C.  Wear rates of 3.5e-3 mg/m are
used for the pebble to pebble wear, and 10.4e-3 mg/m for the pebble to
graphite chute and 9.7e-3 mg/m for pebble to steel.  Thus for the
outlet chute the upper portion has 18.05e-3 mg of dust produced per
average pebble and the lower portion has 11.91e-3 mg produced for a
total outlet chute amount of 29.96e-3 mg.

The fuel loading pipe is approximately 25 m long and the temperature
is 200\degree C which gives a wear value of 2.1e-3 mg/m and 52.50e-3
mg.  Thus, for an estimated average pebble pass, 10.5\% of the dust is
produced in core, 32.5\% is produced in the outlet chute and 57.0\% is
produced in the loading pipes. The paper estimates that 50\% of the
outlet chute graphite dust enters the core and that 75\% of the
graphite dust produced in the fuel loading pipes enters the reactor
core, for a total amount of graphite dust entering the core of 64.0e-3
mg per pebble pass.  Since there are 125 pebbles entering the reactor
a day, and 365 days in a year, this works out to 2.92 g/year of pebble
dust per year (reported in the paper as 2.74 kg/year due to a
precision loss and unit errors)\citep{dust_estimation}.

HTR-10 has 27 thousand pebbles compared to AVR's 100 thousand
and a rate of 125 pebbles per day compared to about 400 pebbles per
day.  A crude scaling factor estimate of 35 grams of dust
per year would be produced per year in AVR.  Measured values of dust
generation rates from HTR-10 would provide valuable information on
pebble bed reactor dust production but appear to be unavailable.


\end{document}